\documentclass[runningheads]{llncs}

\usepackage[top=2cm,bottom=2cm,left=2.5cm,right=2.5cm,includehead,includefoot]{geometry}

\makeatletter
\renewcommand\subsubsection{\@startsection{subsubsection}{3}{\z@}%
                       {-18\p@ \@plus -4\p@ \@minus -4\p@}%
                       {8\p@ \@plus 4\p@ \@minus 4\p@}%
                       {\normalfont\normalsize\bfseries\boldmath
                        \rightskip=\z@ \@plus 8em\pretolerance=10000 }}
\renewcommand\paragraph{\@startsection{paragraph}{4}{\z@}%
                       {-18\p@ \@plus -4\p@ \@minus -4\p@}%
                       {8\p@ \@plus 4\p@ \@minus 4\p@}%
                       {\normalfont\normalsize\bfseries\boldmath
                        \rightskip=\z@ \@plus 8em\pretolerance=10000 }}

\makeatother
\usepackage{graphicx}

\usepackage[ruled,english,onelanguage,linesnumbered,vlined]{algorithm2e} 
\usepackage{amsmath}
\usepackage{amsfonts}
\usepackage{amssymb}
\usepackage{array}
\usepackage[title]{appendix}
\usepackage[french,english]{babel}
\usepackage{capt-of}
\usepackage{chngcntr}
\usepackage{color}
\usepackage{easybmat}
\usepackage{lineno}
\usepackage{listings}
\usepackage{multirow,hhline}
\usepackage{nicefrac}
\usepackage{setspace}
\usepackage{tabularx}
\usepackage{ulem}
\usepackage[dvipsnames]{xcolor}

\renewcommand{\lstlistingname}{Snippet}

\SetCommentSty{mycommfont}

\makeatletter
\newcommand{\biggg}{\bBigg@{4}}
\newcommand{\Biggg}{\bBigg@{5}}
\makeatother

\renewcommand{\v}{\vspace{5mm}}
\newcommand{\h}{\hspace{5mm}}
\newcommand{\hsd}{\hspace{2.5mm}}

\newcommand{\dsp}{\displaystyle}
\newcommand{\dspfrac}[2]{\dsp{\frac{#1}{#2}}}
\newcommand{\ie}{\textit{id est} }

\newcommand{\etc}{\textit{et c\ae tera} }
\newcommand{\ite}{\item[$\bullet$]}

\newcommand{\xinit}{x^{\text{init}}}
\newcommand{\tinit}{t^{\text{init}}}
\newcommand{\tend}{t^{\text{end}}}
\newcommand{\tN}{t^{[N]}}
\newcommand{\tNp}{t^{[N+1]}}
\newcommand{\dt}{\delta t}
\newcommand{\dtN}{\delta t^{[N]}}

\newcommand{\nst}{n_{st}}
\newcommand{\nin}{n_{in}}
\newcommand{\nout}{n_{out}}
\newcommand{\nstk}[1]{n_{st,#1}}
\newcommand{\nink}[1]{n_{in,#1}}
\newcommand{\noutk}[1]{n_{out,#1}}

\newcommand{\rnst}{\mathbb{R}^{\nst}}
\newcommand{\rnin}{\mathbb{R}^{\nin}}
\newcommand{\rnout}{\mathbb{R}^{\nout}}
\newcommand{\rnstk}[1]{\mathbb{R}^{\nstk{k}}}
\newcommand{\rnink}[1]{\mathbb{R}^{\nink{k}}}
\newcommand{\rnoutk}[1]{\mathbb{R}^{\noutk{k}}}

\newcommand{\Inst}{[\![1, \nst]\!]}
\newcommand{\Inin}{[\![1, \nin]\!]}
\newcommand{\Inout}{[\![1, \nout]\!]}

\newcommand{\Nmax}{N_{\text{max}}}
\newcommand{\mmax}{m_{\text{max}}}
\newcommand{\m}{{}^{[m]}}

\newcommand{\padUp}{\vphantom{\LARGE A}}
\newcommand{\padDown}{\vspace{0.6mm}}
\newcommand{\PadUp}{\vphantom{\huge A}}
\newcommand{\PadDown}{\vspace{1.5mm}}
\newcommand{\Xid}{\mathrm{\Xi}}
\newcommand{\half}{\frac{1}{2}}
\newcommand{\fourth}{\frac{1}{4}}
\newcommand{\tildexN}{\tilde{x}^{[N]}}
\newcommand{\tildeuNplus}{\tilde{u}^{[N]+}}
\newcommand{\tildexdNplus}{\tilde{\dot{x}}^{[N]^+}}
\newcommand{\tildeudNplus}{\tilde{\dot{u}}^{[N]^+}}
\newcommand{\tildexddNplus}{\tilde{\ddot{x}}^{[N]^+}}
\newcommand{\tildeuddNplus}{\tilde{\ddot{u}}^{[N]^+}}
\newcommand{\halftildexddNplus}{\frac{\tildexddNplus}{2}}
\newcommand{\halftildeuddNplus}{\frac{\tildeuddNplus}{2}}
\newcommand{\tildefNplus}{\tilde{f}^{[N]^+}}
\newcommand{\tildepfNplus}[1]{\widetilde{\partial_{#1} f}^{[N]^+}}
\newcommand{\halftildepfNplus}[1]{\frac{\tildepfNplus{#1}}{2}}
\newcommand{\ex}{\varepsilon_{x}}
\newcommand{\ext}{\varepsilon_{xt}}
\newcommand{\taumtn}{\tau\!-\!\tN}
\newcommand{\ed}{\dot{e}}
\newcommand{\DeltaN}{\Delta^{[N]}}

\newcolumntype{P}[1]{>{$}>{\centering\arraybackslash}p{#1}<{$}}

\newcommand{\figuresdir}{.}

\def\wordwrap#1#2{
\let\oldarraystretch\arraystretch
\renewcommand{\arraystretch}{#1}
\begin{tabular}{c}
#2
\end{tabular}
\let\arraystretch\oldarraystretch
}

\def\lwordwrap#1#2{
\let\oldarraystretch\arraystretch
\renewcommand{\arraystretch}{#1}
\begin{tabular}{l}
#2
\end{tabular}
\let\arraystretch\oldarraystretch
}

\AtBeginDocument{\counterwithout{lstlisting}{chapter}}
\begin{document}

\title{COSTARICA estimator for rollback-less systems handling in iterative co-simulation algorithms\thanks{Supported by organization Siemens Digital Industries Software.}}

\titlerunning{COSTARICA estimator for iterative co-simulation methods}

\author{Yohan Eguillon\inst{1,2}\orcidID{0000-0002-9386-4646} \and
Bruno Lacabanne\inst{2}\orcidID{0000-0003-1790-3663} \and
Damien Tromeur-Dervout\inst{1}\orcidID{0000-0002-0118-8100}}

\authorrunning{Y. Eguillon et al.}

\institute{
Institut Camille Jordan, Université de Lyon, UMR5208 CNRS-U.Lyon1, Villeurbanne, France\\
\email{\{yohan.eguillon,damien.tromeur-dervout\}@univ-lyon1.fr}\\
\and
Siemens Digital Industries Software, Roanne, France
\email{\{yohan.eguillon,bruno.lacabanne\}@siemens.com}\\
}

\maketitle


\begin{abstract}
Co-simulation is widely used in the industry due to the emergence of modular dynamical models made up of interconnected, black-boxed systems. Several co-simulation algorithms have been developed, each with different properties and different levels of accuracy and robustness. Among them, the most accurate and reliable ones are the iterative ones, although they have a main drawback in common: the involved systems are required to be capable of rollback. The latter denotes the ability of a system to integrate over a co-simulation time step that has already been simulated. Non-rollback-capable system can only go forward in time and every integrated step is definitive. In practice, the industrial modelling and simulation platforms rarely produce rollback-capable systems. This paper proposes a solution that slightly changes the co-simulation methodology and that enables to use iterative co-simulation methods on a modular model which contains non-rollback-capable systems in case the latter represent ordinary differential equations. The idea is to replace such a system by a simplified version, which is used to estimate the results of the integrations instead of integrating the real system. Once the co-simulation method's surrogate iterations on these estimators predict the convergence on the co-simulation step, the non-rollback-capable systems genuinely integrate the step using the estimated solution on the other systems before moving forward, transforming the iterative co-simulation method into a non-iterative one.

\end{abstract}


\section{Introduction}
\label{section:introduction}

Co-simulation is an area of research that now attracts more and more interest in the industry \cite{Gomes2018survey}. Also called simulators coupling or solvers coupling, a co-simulation involves two or more interconnected systems. Each of the latter contains its own solver. The set of interconnected systems is called a modular model.

The ability to connect such systems makes it possible to assemble modular models out of black-boxed systems. In the case of multiphysics simulation, this enables each system to embed a solver tailored to the physics it represents. For instance, electrical systems can use a solver method dedicated to the electrical simulation, fluid systems can benefit from simulation algorithms preserving conservation laws, etc. In addition, the intellectual property of the system manufacturer can be protected even though the system can be simulated thanks to a simple set of interactions. The industrial interest of co-simulation lies in this possibility: simulating a modular model made of systems which do not need to disclose their know-how.

A minimal set of possible interactions is required from such systems. In case one of them is not supported, no co-simulation can occur. Otherwise, basic co-simulation algorithms can be used to simulate the modular model. Many co-simulation algorithms have been developed \cite{Kubler2000} \cite{Arnold2001} \cite{Gu2004} \cite{Bartel2013} \cite{Sicklinger2014} \cite{Busch2016} \cite{Benedikt2013NEPCE} and analyzed \cite{Li2014} \cite{Schweizer2015ImplicitExplicitStabAndConv} \cite{Schweizer2016} over the past few years. Indeed, the most basic ones are usually not sufficient to generate accurate enough results. The main trade-off that is tackled by the advanced co-simulation algorithms is the balance between accuracy and computational time. A high accuracy can usually be reached when additional information can be retrieved from the systems. Among others, the model-based methods \cite{Benedikt2013NEPCE} \cite{Stettinger2014ECC} \cite{Stettinger2014IEEEConfControl} use structural information in order to adapt the co-simulation so that it leads to very accurate results, usually through preservation of some quantities (for instance: energy on a physical coupling \cite{Sadjina2017} \cite{Sadjina2020}). The main drawback of such methods is that the systems must be disclosed.

Other methods do not require systems to be disclosed: modular models can be made of interconnected, black-boxed systems. Such methods, when they correspond to advanced co-simulation algorithms \cite{Ochel2019} \cite{Gomes2019} \cite{Eguillon2019Ifosmondi} \cite{Eguillon2021IfosmondiJFM} \cite{Kraft2021}, usually require the systems to perform advanced actions in addition to the minimal set of possible ones. Indeed, depending on the modeling and simulation platform that generates a system, the latter might be able to perform more than just the basic actions. These advanced actions are called "capabilities".

Some of these advanced capabilities are namely formalized in the FMI standard \cite{FMIStandard}, with a dedicated mechanism for each system to notify the supported and unsupported ones. Some capabilities are well-known and lots of co-simulation methods use them, some other are exotic, and some capabilities are very rare in practice. Among the latter, the \textbf{rollback} is one of the most promising, yet scarce. The rollback is the ability of a system to re-integrate itself on a time slice on which it has already undergone integration. When every system of a modular model has this capability, an iterative co-simulation algorithm can be used \cite{Kubler2000} \cite{Arnold2001} \cite{Bartel2013} \cite{Eguillon2019Ifosmondi} \cite{Eguillon2021IfosmondiJFM} \cite{Kraft2021} \cite{Sicklinger2014} \cite{Schweizer2014PredCorr} or the methods referred to as ICSs (implicit coupling schemes) in \cite{Viot2018}. Iterative methods, when they converge, are a good way to reach a required accuracy on a wide range of models while supporting black-boxed systems. The only problem of these methods is the scarcity, in practice, of rollback-capable systems.

This paper introduces an alternative that mimics the rollback on rollback-free systems corresponding to ODEs (ordinary differential equations) so that iterative co-simulation algorithms can be adapted into a version that can be applied on modular models even if the latter involve rollback-free systems. This adaptation consists in replacing the rollback-free systems by a simplified version which is used to estimate the results of the integrations instead of integrating them for real. These simplified systems are estimators which require advanced capabilities that are less rare than the rollback on most of the black-boxed systems embedding a tailored solver. Once the co-simulation method's surrogate iterations on these estimators predict the convergence on the co-simulation step, the non-rollback-capable systems genuinely integrate the step using the estimated solution on the other systems before moving forward, thus transforming the iterative co-simulation method into a non-iterative adaptation of it.

On each non-rollback-capable system, the associated estimator used in the surrogate iterative stage of the co-simulation method is designed to imitate the integration as if it was done for real. The predicted quantities are the data of the system that will be used by other systems: the output coupling variables (and eventually related data such as their time-derivatives). This estimation depends, among other things, on its input coupling variables (determined by the co-simulation method). Such an estimator, possible on most of the systems given very common capabilities that do not require system disclosure, lead to a surrogate system on the coupling variables. The basics of this estimator are the following: the ODE of the system is linearized at the most recent reached time, and thanks to a Laplace transform of this linearized system, a relation can be established between the input coupling variables and the output coupling variables. In case the input coupling variables can be expressed as polynomials (which is the case in the overwhelming majority of cases), the output values at the time to reach have a linear expression in terms of the coefficients of the polynomial of the input coupling variables. This linear expression involves transfer matrices that can be obtained using Patel \& Misra method \cite{Misra1987} and their inverse Laplace transform that can be computed with the Gaver-Stehfest algorithm \cite{Jacquot1983}.

This paper is structured as follows: the motivation of a step estimator and the associated formalism will be described first, then the COSTARICA itself will be detailed both formally (mathematically) and practically (pseudo-code). A quick analysis with corroborative examples will follow the COSTARICA description as the latter is based on a partial linearization. Finally, a few examples will be presented in order to convince the reader about the practical aspect of the COSTARICA process.

\section{Framework and motivations}
\label{section:framework_and_motivation}

Elements of formalism about ODE systems for co-simulation will be given in this section. Starting from the equations inside of the systems, the notions of co-simulation step (or macro-step) will be defined as well as the simulation function and its iterative version.

The purpose is to explain how a step estimator can, approximately or exactly, solve the problem of the application of an iterative co-simulation algorithm to a modular model.

\subsection{System and macro-step}
\label{subsection:system_and_macro_step}

Let's consider an abstraction of an iterative co-simulation algorithm. The latter can be any method that needs to proceed several times an integration of one or more systems on the same time slice \cite{Kubler2000} \cite{Eguillon2021IfosmondiJFM} \cite{Schweizer2014PredCorr} \cite{Sicklinger2014}.

We will formerly define what an iterative co-simulation method implies and how it works.

First of all, as mentioned in the introduction, each system is supposed to represent an ODE. As the systems may be (and, usually, are) connected with other systems of the modular model, we consider the inputs $u$ and the outputs $y$ being vectorial functions of the time, respectively of dimension $\nin$ and $\nout$. With $x$ being the state vector of dimension $\nst$, a given system represents the ODE equation as follows:

\begin{equation}
\label{eq:ODE}
\begin{array}{ccc}
	\left\{
		\begin{array}{lcl}
			\dspfrac{dx}{dt} & = & f(t, x, u) \\
			y & = & g(t, x, u)
		\end{array}
	\right.
	&
	\phantom{abc}
	\text{where}
	\phantom{abc}
	&
	\begin{array}{l}
		t \in [\tinit, \tend] \\
		x \in L([\tinit, \tend], \rnst) \\
		u \in L([\tinit, \tend], \rnin) \\
		y \in L([\tinit, \tend], \rnout)
	\end{array}
\end{array}
\end{equation}

The $f$ and $g$ functions of a system are respectively called the \textbf{derivatives} and the \textbf{outputs} functions of the ODE. The initial time $\tinit$ and the final time $\tend$ are supposed to be the same on every system of a modular model, so that the co-simulation occurs on the time domain $[\tinit, \tend]$.

Moreover, a co-simulation system embeds a solver. Abstractly, a solver embedded in a system enables to get the output response of the latter to a certain stimulus of its inputs (which can be seen like a vectorial command), and on a small time domain that starts at a time where the states have an initial value. This small time domain is either called a macro-step or a co-simulation step, as opposed to the micro-step, also called solver step. In practice, the output response can rarely be retrieved on the whole macro-step, but only its final value can be retrieved (as well as its time derivative, in some cases). The ability to retrieve intermediate values might exist in some cases, as mentioned in \cite{Benedikt2013NonIter}.

The system (ODE and solver) can thus be represented by a discretized system over any macro-step $[t^{[N]}, t^{[N+1]}[$ where $\tinit \leqslant t^{[N]} < t^{[N+1]} \leqslant \tend$.

\begin{equation}
\label{eq:ODE_cosim}
\begin{array}{ccc}
	\left\{
		\renewcommand{\arraystretch}{1.5}
		\begin{array}{lcl}
			\dspfrac{dx^{[N]}}{dt} & = & f(t, x^{[N]}, u^{[N]}) \\
			y^{[N]} & = & g(t, x^{[N]}, u^{[N]})
		\end{array}
		\renewcommand{\arraystretch}{1.0}
	\right.
	&
	\phantom{abc}
	\text{where}
	\phantom{abc}
	&
	\begin{array}{l}
		t \in [t^{[N]}, t^{[N+1]}[ \\
		x^{[N]} \in L([t^{[N]}, t^{[N+1]}[, \rnst) \\
		u^{[N]} \in \left(\mathbb{R}_n[t]\right)^{\nin} \\
		y^{[N]} \in L([t^{[N]}, t^{[N+1]}[, \rnout)
	\end{array}
\end{array}
\end{equation}

In \eqref{eq:ODE_cosim}, $N \in [\![0, \Nmax]\!]$ denotes the index of the macro-step, and $\Nmax$ the number of macro-steps on the total co-simulation time-domain $[\tinit, \tend]$.

Please note that we are assuming that the inputs are polynomial in time in this paper. In \eqref{eq:ODE_cosim}, we have $u^{[N]} \in \left(\mathbb{R}_n[t]\right)^{\nin}$ with $n\in\mathbb{N}$ denoting the maximum polynomial degree among all the $\nin$ inputs. In most of the co-simulation methods, the inputs are not known on $[t^{[N]}, t^{[N+1]}[$ when the integration of \eqref{eq:ODE_cosim} is being performed, so an extrapolation has to be made on this interval. Most of these extrapolations (or interpolations) used in practice are covered by the polynomial form assumption: zero-order hold \cite{Sicklinger2014}, first-order hold, Hermite entries \cite{Eguillon2022F3ornits} \cite{Eguillon2021IfosmondiJFM}, smooth polynomial extrapolations \cite{Busch2019}, ...

\newpage 

As a co-simulation system interacts through its inputs and outputs, the initialization of the state values at each macro-step is done with respect to their ending values at the end of the previous macro-step. In other words, the initial condition of \eqref{eq:ODE_cosim} is:

\begin{equation}
\label{eq:ODE_cosim_initialization_1}
x^{[0]}(\tinit) = \tilde{x}^{[0]} = \xinit
\end{equation}

\noindent for the first macro-step, where the system embeds the information $\xinit$, and

\begin{equation}
\label{eq:ODE_cosim_initialization_2}
\forall N \in [\![1, \Nmax]\!],\
x^{[N]}(t^{[N]}) = \tilde{x}^{[N]} = \lim_{\substack{t\to t^{[N]}\\t<t^{[N]}}} x^{[N-1]}(t)
\end{equation}

\noindent for the other macro-steps.

The retrievable output at the end of a macro-step $N$ will be denoted by $\tilde{y}^{[N+1]}$ and defined by:

\begin{equation}
\label{eq:ODE_cosim_output}
\tilde{y}^{[N+1]} = \lim_{\substack{t\to t^{[N+1]}\\t<t^{[N+1]}}} y^{[N]}(t)
\end{equation}

Moreover, we let $\tilde{y}^{[0]}$ be defined as the initial outputs of the system. These initial outputs are supposed to be a known data of the co-simulation model.

Throughout this paper, quantities corresponding to evaluation of vectorial or scalar time-dependent function at a given time will be denoted by the name of the function with a tilde symbol ${}^{\tilde{}}$ added to it. In particular: $\tilde{x}^{[N]} \in \rnst$ because $x^{[N-1]}\in L([t^{[N-1]}, t^{[N]}[, \rnst)$ (see \eqref{eq:ODE_cosim_initialization_2}) and $\tilde{y}^{[N+1]}\in\rnout$ because $y^{[N]}\in L([t^{[N]}, t^{[N+1]}[, \rnout)$ (see \eqref{eq:ODE_cosim_output}).

Finally, a call to a system on a given co-simulation step $[t^{[N]}, t^{[N+1]}[$ can be seen as a call to the following function called \textbf{simulation function}, \textbf{co-simulation step function} or simply \textbf{step function}:

\begin{equation}
\label{eq:step_function}
S^{[N]}:
\left\{
	\begin{array}{lcl}
		\rnst \times L([t^{[N]}, t^{[N+1]}[, \rnin) & \rightarrow & \rnst \times \rnout \\
		\left(\tilde{x}^{[N]}, u^{[N]}\right) & \mapsto & \left(\tilde{x}^{[N+1]}, \tilde{y}^{[N+1]}\right)
	\end{array}
\right.
\end{equation}

Less formally, yet more comprehensive, a call to the step function acts as follows:

\begin{scriptsize}
$$
\left(
	\begin{array}{c}
		states\ at\ the\ end\ of\ the\ step \\
		outputs\ at\ the\ end\ of\ the\ step
	\end{array}
\right)^T
=
S^{
	\left[
		\begin{array}{c}
			step \\
			index
		\end{array}
	\right]
}
\left(
	\left(
		\begin{array}{c}
			states\ at\ the\ beginning\ of\ the\ step \\
			input\ command\ over\ the\ step
		\end{array}
	\right)^T
\right)
$$
\end{scriptsize}

In practice, the successive calls to the $S^{[N]}$ function on a given system on successive macro-steps $[t^{[N]}, t^{[N+1]}[$, $[t^{[N+1]}, t^{[N+2]}[$, $[t^{[N+2]}, t^{[N+3]}[$, ... are done with only control on the inputs $u^{[N]}$, $u^{[N+1]}$, $u^{[N+1]}$, ... and with only retrievable outputs $y^{[N+1]}$, $y^{[N+2]}$, $y^{[N+3]}$, ...\ .

Indeed, the system keeps its states from a call to the other, at the corresponding time both being the end of a macro-step and the beginning of the upcoming one.

\subsection{Rollback formalism: iteration}
\label{subsection:rollback_formalism_iteration}

The co-simulation algorithm has several roles. Among those, the definition of the time mesh $t^{[1]}$, $t^{[2]}$, ..., $t^{[\Nmax-1]}$ (also called the \textbf{time-stepping}) has been developed in the literature \cite{Schierz2012} \cite{Benedikt2013NonIter} \cite{Meyer2021} \cite{Kraft2019}. The definition of the input variables, another task the co-simulation method is responsible of, is not always a simple dispatching of the corresponding connected output values. Indeed, such a dispatching might not be trivial on asynchronous cases \cite{Muller2016} \cite{Eguillon2022F3ornits}, and sometimes the inputs might be reconstructed (\textit{e.g.} by using extrapolation on the past values \cite{Kubler2000} or with other methods \cite{Eguillon2022F3ornits} \cite{Busch2011}).

A particular range of methods called the \textbf{iterative} co-simulation algorithms are namely designed to find very accurate input commands by iterating on the set of interconnected systems until a satisfactory result is found. For instance, the outputs and corresponding inputs can be compared at the end of each macro-step so that a given coupling relationship is satisfied \cite{Kubler2000} \cite{Sicklinger2014} \cite{Schweizer2014PredCorr}. This is namely the case for co-simulation methods based on the fixed-point method \cite{Kubler2000} \cite{Eguillon2019Ifosmondi}, the Newton method \cite{Sicklinger2014} or Newton-like methods \cite{Eguillon2021IfosmondiJFM}.

When such algorithms are used, every system must be able to integrate a step more than once. This is called the \textbf{rollback}. With the formalism introduced in \ref{subsection:system_and_macro_step}, a rollback-capable system is simply a system on which the step function \eqref{eq:step_function} can be called several times on the same macro-step $[t^{[N]}, t^{[N+1]}[$.

Let's denote by a left superscript $\m $ the iteration index of a given call of the step function on a given macro-step. Let also $\mmax(N)$ be the iteration index of the last iteration done on the macro-step $[t^{[N]}, t^{[N+1]}[$. On a given macro-step $[t^{[N]}, t^{[N+1]}[$, at a given iteration $m\in[\![0, \mmax(N)]\!]$, the call to the step function generates the $m$\up{th} output values:

\begin{equation}
\label{eq:step_function_rb}
\left(\m \tilde{x}^{[N+1]}, \m \tilde{y}^{[N+1]}\right) = S^{[N]}\left({}^{[\mmax(N-1)]}\tilde{x}^{[N]}, \m u^{[N]}\right)
\end{equation}

In \eqref{eq:step_function_rb}, we can namely identify ${}^{[\mmax(N-1)]}\tilde{x}^{[N]}$ the state variables at the end of the previous macro-step, $\m u^{[N]}$ the input commands computed by the co-simulation algorithm at the $m$\up{th} iteration of the co-simulation on the step $[t^{[N]}, t^{[N+1]}[$, and $S^{[N]}$ the step function defined in \eqref{eq:step_function}. The co-simulation algorithm can retrieve the outputs $\m \tilde{y}^{[N+1]}$ of this system at $t^{[N+1]}$, corresponding to the final value on this step of the output response to the stimulus ${}^{[\mmax(N-1)]}\tilde{x}^{[N]}$.

\subsection{Rollback formalism: rejection}
\label{subsection:rollback_formalism_rejection}

Iterative co-simulation algorithms might redefine the end of the current co-simulation step $t^{[N+1]}$, for instance when they are based on a numerical iterative method that diverged or that did not converge fast enough. In such cases, the co-simulation step $[t^{[N]}, t^{[N+1]}[$ is said to be \textbf{rejected} \cite{Eguillon2021IfosmondiJFM} \cite{Schierz2012}. In this case, a step restarting from $t^{[N]}$ but with a different ending time $t^{[N+1]}$ is redefined, and the iterative process restarts in the newly defined macro-step. This namely occurs in the context of adaptive step size iterative co-simulation methods.

\begin{center}
\includegraphics[scale=0.65]{\figuresdir/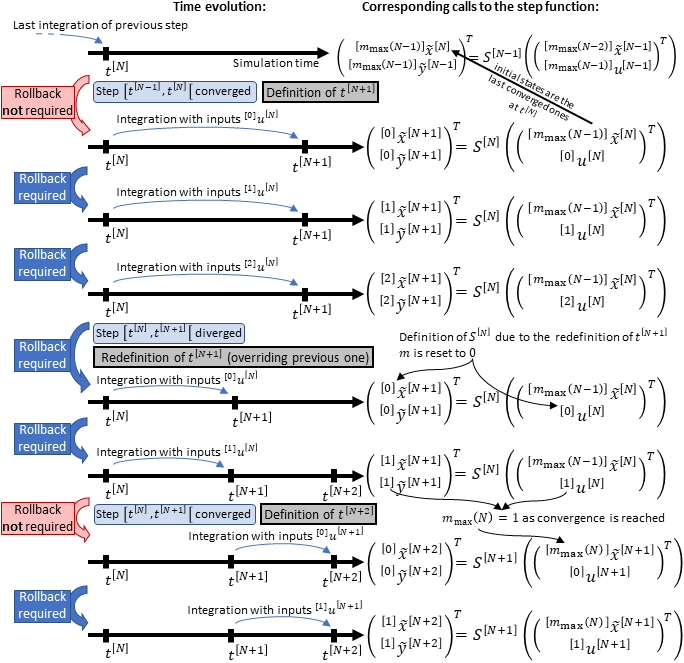}
\captionof{figure}{Successive calls to the step function corresponding to a given scenario (the same one as in figure \ref{fig:Rollback_3loops})}
\label{fig:Rollback_step_function}
\end{center}

For the sake of readability, no supplementary subscript of superscript will be added to the quantities introduced above. We will simply consider that, when the macro-step starting at $t^{[N]}$ is considered, the method can redefine $t^{[N+1]}$. This case will be denoted as \textbf{divergence}. The counter $m$ restarts at $0$ in this case, and at each macro-step $N$ we can consider that $\mmax(N-1)$ corresponds to the iteration that led to convergence and acceptance on the previous macro-step.

Figure \ref{fig:Rollback_step_function} shows a scenario with both iterations described in \ref{subsection:rollback_formalism_iteration} and rejections described in this subsection. The parallel calls to the step function are also given next to each step computation.

\subsection{Iterative co-simulation algorithm}
\label{subsection:iterative_co_simulation_algorithm}

For the sake of genericity, let's consider an abstraction of an iterative co-simulation method. Such method can be seen as a set of two algorithms (also called \textbf{programs}):

\begin{itemize}
\ite an \textbf{orchestrator program}, running in parallel with
\ite as many clones of a \textbf{worker program} as there are systems.
\end{itemize}

Each clone of the worker program is responsible for one system, also called \textbf{simulation unit} in this context (that can, in practice, come from any modelling and simulation tool, for instance Simcenter Amesim, Simulink, ... or anything represented with the FMI co-simulation standard \cite{FMIStandard}). Figure \ref{fig:orchestrator_worker} schematically shows the architecture of such a way to represent a given co-simulation algorithm.

\begin{center}
\includegraphics[scale=0.5]{\figuresdir/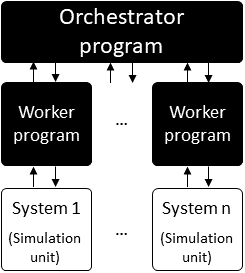}
\vspace{-3mm} 
\captionof{figure}{Orchestator-workers formalism of a co-simulation algorithm}
\label{fig:orchestrator_worker}
\end{center}

\begin{footnotesize}
\begin{algorithm}[H]
\caption{Worker program of an iterative co-simulation method}
\label{alg:single_worker_icsa}
{\setstretch{1.2}
$N:=0$\;
$t^{[0]}:=\tinit$\;
\While(\tcp*[h]{Time loop}){$t^{[N]} < \tend$}
{
	\While(\tcp*[h]{Co-sim. step loop}){Step starting on $t^{[N]}$ is not converged}
	{
		$m:=0$\;
		Method (re)computes $t^{[N+1]}$\;
		
		\While(\tcp*[h]{Internal loop}){True}
		{
			Method computes $\m u^{[N]}$\;
			Compute $\m \tilde{y}^{[N+1]}$ by step integration $S^{[N]}$ with inputs $\m u^{[N]}$\;
			From $\m \tilde{y}^{[N+1]}$ and outputs of other systems on other workers, \hspace{10cm} method decides if the step is \textbf{converged}, \textbf{to-be-redone} or \textbf{rejected}\;
			\uIf{Step $[t^{[N]}, t^{[N+1]}[$ is \textbf{to-be-redone}}
			{
				$m:=m+1$\;
			}
			\Else
			{
				\tcp{either step is converged or has been rejected}
				Break\;
			}
		}
	}
	$\mmax(N)=m$\;
	$N := N+1$\;
}
} 
\end{algorithm}
\end{footnotesize}

\v

An abstraction of the worker program is presented in algorithm \ref{alg:single_worker_icsa}. Please note that, on this algorithm, the computations of $u$ and $t^{[N+1]}$ at each step are not detailed as they depend on the algorithm itself (they might be determined using data received from the orchestrator program, deduced from the past of the connected output, computed from a numerical method, ...). Also, $t^{[N+1]}-t^{[N]}$ might change across $t^{[N]}$ in the case of adaptive step size co-simulation methods \cite{Schierz2012} \cite{Meyer2021}.

As algorithm \ref{alg:single_worker_icsa} is an abstraction of a worker program of an iterative co-simulation method, some iterative co-simulation methods might need to have their formalism slightly adapted in order to fit in with this formalism. For instance, some methods never reject a step. In that case, exiting from the internal loop also exits from the co-simulation loop. In other words, the co-simulation loop might do one iteration only for each macro-step, without loss of generality.

In algorithm \ref{alg:single_worker_icsa}, three nested loops can be identified:

\begin{itemize}
\ite the \textbf{time loop}, denoting the global forward movement of time during the co-simulation,
\ite the \textbf{co-simulation step loop}, denoting the attempts to locally move forward once a time $t^{[N]}$ has been reached (in other words, this loop tries to reach convergence until a time strictly further than the currently farthest time where a convergence has been obtained), and
\ite the \textbf{internal loop}, corresponding to the attempt to validate a given macro-step $[t^{[N]}, t^{[N+1]}[$, either by convergence of an iterative numerical method (like in \cite{Eguillon2019Ifosmondi}, \cite{Eguillon2021IfosmondiJFM} or \cite{Sicklinger2014}), with a given procedure requiring several evaluations of the step function on a macro-step (like in \cite{Schweizer2014PredCorr}), or with any other methodology.
\end{itemize}

A way to visualize how these nested loops correspond to a co-simulation, the scenario presented in figure \ref{fig:Rollback_step_function} is shown together with these different loops in figure \ref{fig:Rollback_3loops}.

\begin{center}
\includegraphics[scale=0.65]{\figuresdir/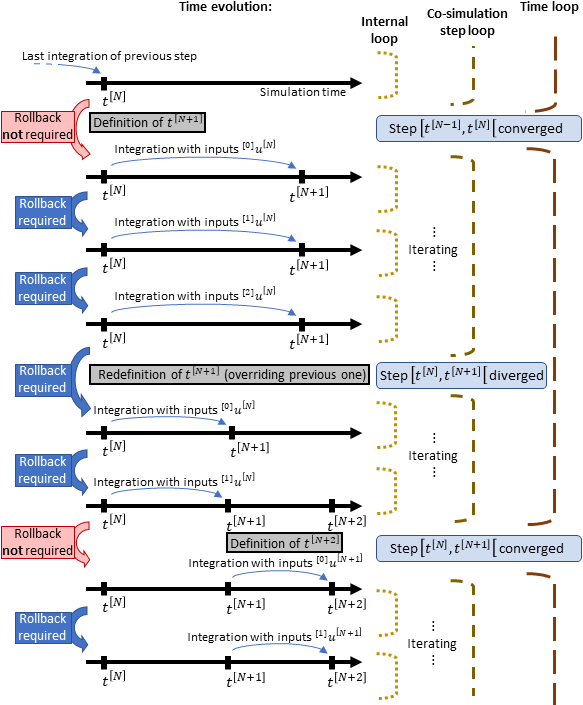}
\vspace{-3mm} 
\captionof{figure}{Nested loops of an iterative co-simulation algorithm and the corresponding scenario (same than in figure \ref{fig:Rollback_step_function})}
\label{fig:Rollback_3loops}
\end{center}

Figure \ref{fig:Rollback_3loops}, as well as figure \ref{fig:Rollback_step_function}, shows that the rollback capability is required on the systems that compose a modular model meant to undergo a co-simulation done with an iterative co-simulation algorithm.

\subsection{Replacing the rollback with a step estimator}
\label{subsection:replacing_the_rollback_with_a_step_estimator}

The idea of this paper is to replace the stages on which the rollback is required. Replacing the integrations by estimations on the non-rollback-capable systems would enable the latter to avoid moving forward in time in a macro-step before convergence.

In other words, the idea is to replace all integrations inside of the co-simulation step loop and the internal loop by an estimation in order to let the co-simulation algorithm find the inputs that lead to a convergence. Once convergence is reached, the forward movement in time is done by a single genuine integration.

Algorithm \ref{alg:single_worker_stepestim_icsa} represents this adaptation applied on algorithm \ref{alg:single_worker_icsa}.

\begin{footnotesize}
\begin{algorithm}[H]
\caption{Worker program of an iterative co-simulation method adapted to a rollback-less system}
\label{alg:single_worker_stepestim_icsa}
{\setstretch{1.2}
$N:=0$\;
$t^{[0]}:=\tinit$\;
\While(\tcp*[h]{Time loop}){$t^{[N]} < \tend$}
{
	\While(\tcp*[h]{Co-sim. step loop}){Step starting on $t^{[N]}$ is not converged}
	{
		$m:=0$\;
		Method (re)computes $t^{[N+1]}$\;
		
		\While(\tcp*[h]{Internal loop}){True}
		{
			Method computes $\m u^{[N]}$\;
			\sout{Compute $\m \tilde{y}^{[N+1]}$ by step integration $S^{[N]}$ with inputs $\m u^{[N]}$}\;
			\fbox{Compute $\m \hat{y}^{[N+1]}$, estimator of $\m \tilde{y}^{[N+1]}$, depending on $\m u^{[N]}$\;}\\
			From $\m y^{[N+1]}$ and outputs of other systems on other workers, \hspace{10cm} method decides if the step is \textbf{converged}, \textbf{to-be-redone} or \textbf{rejected}\;
			\uIf{Step $[t^{[N]}, t^{[N+1]}[$ is \textbf{to-be-redone}}
			{
				$m:=m+1$\;
			}
			\Else
			{
				\tcp{either step is converged or has been rejected}
				Break\;
			}
		}
	}
	$\mmax(N)=m$\;
	\fbox{Compute ${}^{[\mmax(N)]} \tilde{y}^{[N+1]}$ by step integration $S^{[N]}$ with inputs ${}^{[\mmax(N)]} u^{[N]}$\;}\\
	$N := N+1$\;
}
} 
\end{algorithm}
\end{footnotesize}

In algorithm \ref{alg:single_worker_stepestim_icsa}, the strikethrough line has been removed from algorithm \ref{alg:single_worker_icsa}, and the boxed lines added to it.

The real integrations now only occur on successive steps of the time loop: a first one on $[t^{[0]}, t^{[1]}[$, and then a single one on $[t^{[1]}, t^{[2]}[$, and then a single one on $[t^{[2]}, t^{[3]}[$ and so on until the last one on $[t^{[N_{\text{max}}-1]}, t^{[N_{\text{max}}]}]$ where $t^{[N_{\text{max}}]}=\tend$. This way, the system behaves as if it were used by a non-iterative co-simulation method, and the iterative co-simulation method can still iterate thanks to the estimations on the nested loops (co-simulation step loop and internal loop).

The estimator suggested in this paper is the COSTARICA one, yet this process can be used with any estimator (reduced versions of the concerned systems, surrogate models, ...).
Moreover, workers only need to be transformed from algorithm \ref{alg:single_worker_icsa} to algorithm \ref{alg:single_worker_stepestim_icsa} when they are attached to a system that is not capable of rollback (see figure \ref{fig:orchestrator_worker}). Indeed, hybrid configurations both involving rollback-capable and rollback-less systems can be implemented. In this case, only the workers attached to rollback-less systems must be adapted.

\section{COSTARICA estimator}
\label{section:costarica_estimator}

COSTARICA stands for \textbf{Cautiously Obtrusive Solution To Avoid Rollback in Iterative Co-simulation Algorithms}. The way it avoids requiring the rollback has been explained in \ref{subsection:replacing_the_rollback_with_a_step_estimator}, yet the "cautiously obtrusive" part hasn't.

This will be explained in this section, and then the estimator itself will be defined. Finally, a fast version of the update of this estimator will be given.

\subsection{Cautious obtrusiveness}
\label{subsection:cautious_obtrusiveness}

The COSTARICA process has been conceived to be usable in industrial co-simulations. One of the main constraints this specification brings is the need for genericity. Indeed, model-based co-simulation approaches \cite{Stettinger2014ECC} \cite{Stettinger2014IEEEConfControl} \cite{Sadjina2017} \cite{Sadjina2020} usually take advantage of the knowledge about the systems' internal structure, yet industrial co-simulation algorithms don't, for the sake of genericity \cite{Gu2004} \cite{Sicklinger2014} \cite{Busch2019} \cite{Eguillon2021IfosmondiJFM} \cite{Eguillon2022F3ornits}.

Although COSTARICA is not a co-simulation algorithm (it is an estimator, acting as described in \ref{subsection:replacing_the_rollback_with_a_step_estimator}), it must comply with the genericity specification. This implies two things: the required advanced interactions must be standardized (details in \ref{subsubsection:black_box_and_interfacing}), and they must be present on most of the systems (details in \ref{subsubsection:required_capabilities}).

\subsubsection{Black-box and interfacing}
\label{subsubsection:black_box_and_interfacing}

As black-boxed systems must be handled, no assumptions can be done about what is inside of the systems. No physical-based information can be retrieved.

Regarding the generic interactions that are possible without disclosing the systems, there exists a standard on which we can base our method: the FMI standard (functional mock-up interface) \cite{FMIStandard}.

Indeed, a very large majority of modelling and simulation tools offer a way to export a system as an FMU (functional mock-up unit), \textit{id est} a standardized interface for the system. Referring to the features that are available in the FMI interface makes it possible to set up a generic procedure since we \textit{at least} know that these actions are generic.

Please note that, although the considered interactions are among those listed in the FMI standard, they might also be very similar to interactions available with other interfaces.

In this paper, we consider five among the numerous possible interactions defined in the FMI standard \cite{FMIStandard}:

\begin{itemize}
\ite the support of time-dependent (usually polynomial) inputs,
\ite the possibility to retrieve instantaneous values of the internal state variables,
\ite the possibility to retrieve instantaneous derivatives of the latter,
\ite the possibility to retrieve the directional derivatives,
\ite the possibility to retrieve the time-derivative of the outputs, and
\ite the rollback. 
\end{itemize}

These interactions are not possible on any FMU (\textit{id est} on any system with an FMI interface), yet each and every FMU embeds the information about the available interactions it provides. This info is called the \textbf{capability flag}.

\newpage 

\subsubsection{Required capabilities}
\label{subsubsection:required_capabilities}

We distinguish the capabilities that COSTARICA must mimic (in case the algorithm on which this process is used requires them) from the capabilities COSTARICA requires (in order to produce the estimations of the calls to the step function.

\v
\textit{\underline{\underline{Capabilities required by the co-simulation algorithm:}}}
\v

The aim of COSTARICA is to replace the rollback by estimating the results of the step function. Knowing that, a legitimate question could be: does COSTARICA mimics calls to the step function on a system that has advanced capabilities, such as time-depending inputs? The answer is yes for two advanced capabilities: the time-dependent inputs (as far as the latter are polynomial in time), and the ability to retrieve the time-derivatives of the outputs at the end of a macro-step.

In other words, a co-simulation method that requires these two advanced interactions can still use COSTARICA as far as this process can mimic these interactions during the estimated steps.

\v
\textit{\underline{\underline{Capabilities required by COSTARICA:}}}
\v

The COSTARICA estimator requires some advanced capabilities from the system on which the rollback will be mimicked. These capabilities are the possibility for the system to provide its internal state variables and directional derivatives at a given time (in practice: the reached time). Fortunately, these capabilities are way more common than the rollback itself.

\subsubsection{Summary}
\label{subsubsection:summary}

The relation between the above-mentioned interactions, capabilities, and COSTARICA are presented in table \ref{table:capabilities_related_to_COSTARICA}.

\begin{center}
\captionof{table}{Capabilities related to COSTARICA}
\label{table:capabilities_related_to_COSTARICA}
\renewcommand{\arraystretch}{1.2}
\begin{small}
\begin{tabular}{|c||c|c|c|}
\hline
\multirow{2}{*}{Interaction} & Capability name & Required by & COSTARICA \\
 & in FMI standard & COSTARICA & mimics it \\
\hline
\hline
Support of & \multirow{2}{*}{canInterpolateInput} & \multirow{2}{*}{No} & \multirow{2}{*}{It can} \\
polynomial\up{$*$} inputs & & &  \\
\hline
\multirow{3}{*}{\begin{tabular}{c}Getting values and \\derivatives of the \\internal state variables\end{tabular}} & not a capability: internal & \multirow{3}{*}{Yes} & \multirow{3}{*}{No} \\
& state variables and their & & \\
& derivatives must be exposed & & \\
\hline
Getting directional & \multirow{2}{*}{providesDirectionalDerivative} & \multirow{2}{*}{Yes} & \multirow{2}{*}{No} \\
derivatives & & & \\
\hline
Getting output & \multirow{2}{*}{maxOutputDerivativeOrder} & \multirow{2}{*}{No} & \multirow{2}{*}{It can} \\
time-derivative & & & \\
\hline
\multirow{2}{*}{Rolling back} & \multirow{2}{*}{canGetAndSetFMUstate} & \multirow{2}{*}{No} & Yes (this is \\
& & & its purpose) \\
\hline
\end{tabular}
\end{small}
\renewcommand{\arraystretch}{1.0}
\end{center}

{}\up{$*$} In the FMI standard, nothing forces a modelling and simulation platform to reconstruct inputs with a polynomial shape when the capability canInterpolateInputs is active. In this paper, we only consider the case where the inputs are reconstructed with a polynomial shape.

\subsection{Estimator definition}
\label{subsection:estimator_definition}

Let's consider that the system already reached a time $t^{[N]}$. To avoid the loss of generality, at the beginning of the co-simulation we can consider that the system "reached" $t^{[0]}=\tinit$. Let's build the COSTARICA estimator based on quantities that the system can compute without moving forward in time.

The estimator is supposed to estimate the behavior of the system to a certain stimulus. At the iteration $m$, this stimulus is $\m u^{[N]}$ as seen in \ref{subsection:rollback_formalism_iteration}. The COSTARICA estimator only works on polynomial inputs (see \eqref{eq:ODE_cosim}), covering the zero-order hold case among others, as the latter can be seen as  polynomial of degree $0$. Let's define $n$ the maximum polynomial order of the inputs produced by the co-simulation algorithm. We denote by $a$ the coefficients of the polynomials of every coordinate of the polynomial inputs:

\begin{equation}
\label{eq:polynomial_inputs_coeffs}
\forall N\in[\![0, \Nmax]\!],\ \forall m\in[\![0, \mmax(N)]\!],\ \m u^{[N]}:t\mapsto\left(\sum_{k=0}^{n}\m a_{jk}^{[N]}t^k\right)_{j\in\Inin}
\end{equation}

Let's consider the first-order approximation of \eqref{eq:ODE_cosim} (\ie its linearization): $A^{[N]}$, $B^{[N]}$, $C^{[N]}$ and $D^{[N]}$ matrices contain the instantaneous directional derivatives at time $t^{[N]}$. They are recoverable thanks to the provideDirectionalDerivatives capability (see table \ref{table:capabilities_related_to_COSTARICA}).

\begin{equation}
\label{eq:Lin_on_a_step}
\begin{array}{ccc}
	\left\{
	\renewcommand{\arraystretch}{1.5}
	\begin{array}{lcl}
		\dspfrac{d\ \m x^{[N]}}{dt} & = & f\left(\tN, \m x^{[N]}(\tN), \m u^{[N]}(\tN)\right) \\
		& & + A^{[N]}\ (\m x^{[N]} - \m x^{[N]}(\tN)) \\
		& & + B^{[N]}\ (\m u^{[N]} - \m u^{[N]}(\tN)) \\
		\m y^{[N]} & = & y^{[N]}(\tN)\\
		& & + C^{[N]}\ (\m x^{[N]} - \m x^{[N]}(\tN)) \\
		& & + D^{[N]}\ (\m u^{[N]} - \m u^{[N]}(\tN)) \\
	\end{array}
	\renewcommand{\arraystretch}{1.0}
	\right.
	&
	\phantom{a}
	\text{where}
	\phantom{a}
	&
	\begin{array}{lcl}
		A^{[N]} & \in & M_{\nst, \nst}(\mathbb{R}) \\
		B^{[N]} & \in & M_{\nst, \nin}(\mathbb{R}) \\
		C^{[N]} & \in & M_{\nout, \nst}(\mathbb{R}) \\
		D^{[N]} & \in & M_{\nout, \nin}(\mathbb{R}) \\
	\end{array}
\end{array}
\end{equation}

\textbf{Remark:} Please note that, despite the $f$ and $g$ function have $3$ arguments (time, states, inputs), the linearization \eqref{eq:Lin_on_a_step} only takes into account the directional derivatives of these functions in the directions of the states and the inputs. The reason for this is technical: there is no standardized way (in the FMI standard, for instance) to retrieve the partial derivative with respect to the time of $f$ and $g$. The consequence of such a partial linearization is studied in section \ref{section:Relevance_of_the_linearization_over_time}.

\textbf{Remark:} Please note the link with the state-space representation: if $f$ and $g$ are of the form $f:t,x,u\mapsto Ax+Bu$ and $g:t,x,u\mapsto Cx+Du$, equation \eqref{eq:Lin_on_a_step} becomes \eqref{eq:SSR_on_a_step}.

\begin{equation}
\label{eq:SSR_on_a_step}
\left\{
\renewcommand{\arraystretch}{1.5}
\begin{array}{lcl}
	\dspfrac{dx^{[N]}}{dt} & = & A\ \m x^{[N]} + B\ \m u^{[N]} \\
	\m y_L^{[N]} & = & C\ \m x^{[N]} + D\ \m u^{[N]} \\
\end{array}
\renewcommand{\arraystretch}{1.0}
\right.
\end{equation}

In case the time-derivatives of the state variables cannot be retreived (for instance, they haven't been exposed in the FMU), we can only assume \eqref{eq:SSR_erroneous_version} (which is not true in general, and which is a worse approximation than the linearization \eqref{eq:Lin_on_a_step}). Only a down-graded version of COSTARICA, based on \eqref{eq:SSR_on_a_step}, can be implemented in this case. An example is presented in subsection \ref{subsection:lotka_volterra_predation} and illustrates how this lack of capability affects the accuracy obtained with this degradation.

\begin{equation}
\label{eq:SSR_erroneous_version}
f(t, x, u) = \frac{\partial f}{\partial x} (t, x, u) x + \frac{\partial f}{\partial u} (t, x, u) u
\end{equation}

\newpage 

In the general case, system \eqref{eq:Lin_on_a_step} can be re-written under the form \eqref{eq:Lin_on_a_step_alter}:

\begin{equation}
\label{eq:Lin_on_a_step_alter}
\left\{
\renewcommand{\arraystretch}{1.5}
\begin{array}{lcl}
	\dspfrac{d\ \m x^{[N]}}{dt} & = &
		A^{[N]}\ \m x^{[N]} + B^{[N]}\ \m u^{[N]} + \tilde{f}_C^{[N]} \\
	\m y_L^{[N]} & = &
		C^{[N]}\ \m x^{[N]} + D^{[N]}\ \m u^{[N]} \\
	\m y^{[N]} & = & \m y_L^{[N]} + y_C^{[N]} \\
\end{array}
\renewcommand{\arraystretch}{1.0}
\right.
\end{equation}

\noindent where we define $\tilde{f}_C^{[N]} = f(x^{[N]}(t^{[N]})$, and $\m y_L^{[N]}$ and $y_C^{[N]}$ are detailed hereafter.

In \eqref{eq:Lin_on_a_step_alter}, $\m y_L^{[N]} \in L([t^{[N]}, t^{[N+1]}[, \rnout)$ is the linear part of the output response of the system to the stimulus $\m u^{[N]}$ on $[t^{[N]}, t^{[N+1]}[$. Two phenomena can explain the difference between $\m y_L^{[N]}$ and $\m y^{[N]}$:

\begin{itemize}
\ite the non-linearity of the system (in case the system has a non-linear output equation), and
\ite time-dependent outputs, in case of a predefined signal, for instance, or an added offset (converting physical units like degrees Celsius into degrees Fahrenheit, for instance).
\end{itemize}

To take this difference into account, we define the \textbf{control part} of the outputs as the difference between them and their linear part.

\begin{equation}
\label{eq:y_control}
y_C^{[N]} = \m y^{[N]} - (C^{[N]}\ \m x^{[N]}+D^{[N]}\ \m u^{[N]}) \in L([t^{[N]}, t^{[N+1]}[, \rnout)
\end{equation}

Please note that, at the beginning of the step, we have \eqref{eq:yC_at_tn}.

\begin{equation}
\label{eq:yC_at_tn}
y_C^{[N]}(\tN) = \m y^{[N]}(\tN) - (C^{[N]}\ \m x^{[N]}(\tN) + D^{[N]}\ \m u^{[N]}(\tN))
\end{equation}

\textbf{Remark:} In case the co-simulation method guarantees the continuity of the inputs (like in \cite{Busch2016} \cite{Eguillon2019Ifosmondi} \cite{Eguillon2021IfosmondiJFM}\cite{Eguillon2022F3ornits} among others), and as the continuous states are continuous, we can use the known quantities at the end of the converged co-simulation step $[t^{[N-1]}, \tN[$ to compute $y_C^{[N]}(\tN)$ (see \ref{subsubsection:control_part_estimation}) and $\tilde{f}_C^{[N]}$ (see \ref{subsubsection:linear_part_estimation}).

As the system has already reached the time $t^{[N]}$, and as the co-simulation algorithm is supposed to use COSTARICA at the stage where an estimation of the step function is required, the quantities of table \ref{table:known_quantities_at_tn} are known.

The estimator should use the quantities from table \ref{table:known_quantities_at_tn} (and potentially their equivalent at previous communication times, like $t^{[N-1]}$, $t^{[N-2]}$, ...) to compute the estimators of table \ref{table:estimators_at_tnp}.

These estimators will be split into two terms: their control part and their linear part:

\begin{equation}
\label{eq:estimators_terms}
\renewcommand{\arraystretch}{1.5}
\begin{array}{lcl}
	\m \hat{y}^{[N+1]} & = & \hat{y}_C^{[N+1]} + \m \hat{y}_L^{[N]} \\
	\m \hat{\dot{y}}^{[N+1]} & = & \hat{\dot{y}}_C^{[N+1]} + \m \hat{\dot{y}}_L^{[N]} \\
\end{array}
\renewcommand{\arraystretch}{1.0}
\end{equation}

Please note that, in expressions \eqref{eq:estimators_terms}, the control terms have no iteration indices. Indeed, as the linear terms are supposed to be the ones already taking into account the behavior of the (linearized) system with the inputs, the control terms might not depend on these inputs. Moreover, none of the different estimation strategies for such terms presented in \ref{subsubsection:control_part_estimation} depend on the inputs. Consequently, instead of having ${}^{[0]}\hat{y}^{[N+1]}$, ${}^{[1]}\hat{y}^{[N+1]}$, ... being equals, we simply removed the iteration index.

\newpage 

\begin{center}
\captionof{table}{Known quantities at $t^{[N]}$}
\label{table:known_quantities_at_tn}
\renewcommand{\arraystretch}{1.5}
\begin{small}
\begin{tabular}{|c|c|c|c|}
\hline
Quantity & Definition & Source & Domain \\
\hline
\hline
$\tilde{u}^{[N]}$ & $\displaystyle{\lim_{\substack{t\to t^{[N]}\\t<t^{[N]}}}}\!\!\!{}^{[\mmax(\!N\!-\!1\!)]}u^{[N\!-\!1]}(t)$ & \wordwrap{1.0}{\padUp The co-simulation \\algorithm, as it is \\responsible for the \\inputs computation\padDown} & $\rnin$ \\
\hline
$\tilde{y}^{[N]}$ & \wordwrap{1.0}{$\displaystyle{\lim_{\substack{t\to t^{[N]}\\t<t^{[N]}}}}\!\!\!{}^{[\mmax(\!N\!-\!1\!)]}y^{[N\!-\!1]}(t)$\\(iterative version of \eqref{eq:ODE_cosim_output})} & \wordwrap{1.0}{\padUp The system, thanks \\to the basic mandatory \\interaction to provide its \\outputs (last call to $S^{[N-1]}$)\padDown} & $\rnout$ \\
\hline
$\tilde{x}^{[N]}$ & \wordwrap{1.0}{$\displaystyle{\lim_{\substack{t\to t^{[N]}\\t<t^{[N]}}}}\!\!\!{}^{[\mmax(\!N\!-\!1\!)]}x^{[N\!-\!1]}(t)$\\(iterative version of \eqref{eq:ODE_cosim_initialization_2})} & \wordwrap{1.0}{\padUp The system, \\with the "Getting \\internal state variables"\\interaction (see table \ref{table:capabilities_related_to_COSTARICA})\padDown} & $\rnst$ \\
\hline
$\tilde{f}^{[N]}$ & \wordwrap{1.0}{$f\left(\tN, \tilde{x}^{[N]}, \tilde{u}^{[N]}\right)$ \\with $\tilde{x}^{[N]}$ and $\tilde{u}^{[N]}$ \\defined hereabove} & \wordwrap{1.0}{\padUp The system, \\with the "Getting \\internal state derivatives"\\interaction (see table \ref{table:capabilities_related_to_COSTARICA})\padDown} & $\rnst$ \\
\hline
\wordwrap{1.0}{\PadUp$A^{[N]}$\\$B^{[N]}$\\$C^{[N]}$\\$D^{[N]}$\PadDown} & \wordwrap{1.0}{instantaneous\\directional\\derivatives} & \wordwrap{1.0}{The system, \\with the "Getting \\directional derivatives"\\interaction (see table \ref{table:capabilities_related_to_COSTARICA})} & \wordwrap{1.0}{matrices,\\see \eqref{eq:Lin_on_a_step} for\\the sizes} \\
\hline
$\m u^{[N]}$ & \wordwrap{1.0}{\multicolumn{1}{l}{\padUp$t\mapsto$}\\$\left(\displaystyle{\sum_{k=0}^{n}}\m a_{jk}^{[N]}t^k\!\!\right)_{j\in\Inin}$\PadDown} & \wordwrap{1.0}{The co-simulation \\algorithm, as shown \\in algorithm \ref{alg:single_worker_stepestim_icsa}} & \wordwrap{1.0}{vectorial\\polynomial\\$\left(\mathbb{R}_n[t]\right)^{\nin}$} \\
\hline
\end{tabular}
\end{small}
\renewcommand{\arraystretch}{1.0}
\end{center}

\begin{center}
\captionof{table}{Estimators at $t^{[N+1]}$}
\label{table:estimators_at_tnp}
\renewcommand{\arraystretch}{1.5}
\begin{small}
\begin{tabular}{|c|c|c|}
\hline
\wordwrap{1.0}{\padUp Estimator\\notation\padDown} & \wordwrap{1.0}{\padUp Estimated\\quantity\padDown} & Domain \\
\hline
\hline
$\m \hat{y}^{[N+1]}$ & \wordwrap{1.0}{Output response of the system to the stimulus \\$\m u^{[N]}$ at the end of the macro-step} & $\rnout$ \\
\hline
$\m \hat{\dot{y}}^{[N+1]}$ & \wordwrap{1.0}{Time-derivative of the output response of the system \\to the stimulus $\m u^{[N]}$ at the end of the macro-step} & $\rnout$ \\
\hline
\end{tabular}
\end{small}
\renewcommand{\arraystretch}{1.0}
\end{center}

\subsubsection{Control part estimation}
\label{subsubsection:control_part_estimation}

In order to estimate the control parts of the estimators  at time $t^{[N+1]}$, we use the following vectorial sequences $\left(\tilde{y}_C^{[N]}\right)_{N\in[\![0, \Nmax]\!]}$ defined by:

\begin{equation}
\label{eq:control_part_sequence}
\forall N\in[\![0, \Nmax]\!],\
\tilde{y}_C^{[N]} = \tilde{y}^{[N]} - \left(C^{[N]}\tilde{x}^{[N]}+D^{[N]}\tilde{x}^{[N]}\right)
\end{equation}

\noindent where, by convention, we consider that ${}^{[\mmax(-1)]}u^{[-1]}$ is constant and equal to the initial inputs of the system, which are known data on every system of a given co-simulation model.

Once the system reached time $t^{[N]}$, the $\left(\tilde{y}_C^{[N]}\right)_{N\in[\![0, N]\!]}$ are known and the estimators $\hat{y}_C^{[N+1]}$ and $\hat{\dot{y}}_C^{[N+1]}$ need to be computed. Any reconstruction algorithm can be used.

\v
\newpage 

\textit{\underline{\underline{Simplest reconstruction:}}}
\v

The simplest definition of the control terms of the estimators has to be understood as: the one using the least amount of points.

In the case of the output values, this corresponds to a simple zero-order hold:

\begin{equation}
\label{eq:control_part_ZOH_val}
\hat{y}_{C, ZOH}^{[N+1]} = \tilde{y}_C^{[N]}
\end{equation}

As the zero-order hold cannot provide non-zero time-derivatives, it cannot be used for the control term of the output time-derivatives estimator. Therefore, ZOH estimation for control part can only be used when output time-derivatives do not need to be estimated, in other words: when the co-simulation algorithm does not require the time-derivatives of the outputs of the systems.

\v
\textit{\underline{\underline{First order reconstruction:}}}
\v

In case the co-simulation algorithm requires the time-derivative of the outputs, we can increase the number of points to obtain a first-order hold estimator for the outputs

\begin{equation}
\label{eq:control_part_FOH_val}
\hat{y}_{C, FOH}^{[N+1]} = \tilde{y}_C^{[N]} + \frac{t^{[N+1]} - t^{[N]}}{t^{[N]} - t^{[N-1]}}\left(\tilde{y}_C^{[N]} - \tilde{y}_C^{[N-1]}\right)
\end{equation}

\noindent so that the time-derivatives can be estimated analogously:

\begin{equation}
\label{eq:control_part_FOH_der}
\hat{\dot{y}}_{C, FOH}^{[N+1]} = \frac{1}{t^{[N]} - t^{[N-1]}}\left(\tilde{y}_C^{[N]} - \tilde{y}_C^{[N-1]}\right)
\end{equation}

The problem with this method is that the estimations can only occur from $t^{[2]}$ as two previous points need to exist. At $t^{[0]}$, all data are supposed to be known (initialization of the co-simulation model), but for the estimation at $t^{[1]}$, first-order hold cannot be used. During this estimation, only $\hat{y}_{C, ZOH}^{[1]}$ can be used regarding the output values, and (artificially, arbitrary) $0$ regarding the output time-derivatives.

\v
\textit{\underline{\underline{Flexible order reconstruction:}}}
\v

As it might be difficult to know which reconstruction order is relevant (zero, one, or more), auto-adaptive methods can be used. For instance, \cite{Kraft2019} or the flexible order signal reconstruction method presented in \cite{Eguillon2022F3ornits} enable to try to catch the best order at each step and for each coordinate (as outputs might be vectorial), and this is easily adaptable to the output time-derivatives.

This flexible order reconstruction method (the one in \cite{Eguillon2022F3ornits}) is the one that has been used in the examples presented at the end of this paper.

\subsubsection{Linear part estimation}
\label{subsubsection:linear_part_estimation}

In order to estimate the outputs at time $t^{[N+1]}$ using the known data at $t^{[N]}$ as presented in table \ref{table:known_quantities_at_tn}, we will consider a linear ODE problem from $0$ to $\dt^{[N]}$ where the latter denotes the macro-step size.

\begin{equation}
\label{eq:dt_N}
\dt^{[N]} = t^{[N+1]} - t^{[N]}
\end{equation}

In order to express the time-shifted problem, we need to dispose of a time-shifted version of the inputs. This will be denoted with a caron symbol $\check{}\ $.

\begin{equation}
\label{eq:u_shifted}
\forall \check{t}\in[0, \dt^{[N]}[,\
\m \check{u}^{[N]}(t)
=
\left(
	\displaystyle{\sum_{k=0}^{n}} \m \check{a}_{jk}^{[N]}\check{t}^k
\right)_{j\in\Inin}
=
\m u^{[N]}(\check{t} + t^{[N]})
\end{equation}

We will now compute the value of the time-shifted version of the linear system \eqref{eq:Lin_on_a_step}, that is to say:

\begin{equation}
\label{eq:Lin_on_a_step_time_shifted}
\begin{array}{ccc}
	\left\{
	\renewcommand{\arraystretch}{1.5}
	\begin{array}{lcl}
		\dspfrac{d\ \m \check{x}^{[N]}}{dt} & = & A^{[N]}\ \m \check{x}^{[N]} + B^{[N]}\ \m \check{u}^{[N]} + \tilde{f}_C^{[N]} \\
		\m \check{y}_L^{[N]} & = & C^{[N]}\ \m \check{x}^{[N]} + D^{[N]}\ \m \check{u}^{[N]} \\
	\end{array}
	\renewcommand{\arraystretch}{1.0}
	\right.
	&
	\phantom{a}
	\text{with}
	\phantom{a}
	&
	\begin{array}{l}
		\phantom{a} \\
		\m \check{x}^{[N]}(0) = \tilde{x}^{[N]} \\
		\text{as defined in table \ref{table:known_quantities_at_tn}} \\
	\end{array}
\end{array}
\end{equation}

The estimator we are computing is given by:

\begin{equation}
\label{eq:y_L_is_end_of_time_shifted_SSR}
\m \hat{y}_L^{[N+1]} = \lim_{\substack{\check{t}\to\dt^{[N]}\\\check{t}<\dt^{[N]}}}\m \check{y}_L^{[N]}(\check{t})
\end{equation}

The $\tilde{f}_C^{[N]}$ term corresponds to the constant terms of the linearization \eqref{eq:Lin_on_a_step}. These terms have been gathered in the single term  $\tilde{f}_C^{[N]}$ in \eqref{eq:Lin_on_a_step_alter}. This term can simply be computed using the known quantities at the converged iteration of the previous co-simulation step (as introduced in table \ref{table:known_quantities_at_tn}) so that $\tilde{f}_C^{[N]}$ is corresponds to the point around which the linearzation \eqref{eq:Lin_on_a_step} occurred.

\begin{equation}
\label{eq:fCN_computation}
\forall N\in[\![0, \Nmax]\!],\
\tilde{f}_C^{[N]} = \tilde{f}^{[N]} - \left(A^{[N]}\tilde{x}^{[N]}+B^{[N]}\tilde{u}^{[N]}\right)
\end{equation}

Let's compute the polynomial coefficients of the time-shifted inputs based on the polynomial coefficients of the inputs.

\begin{equation}
\label{eq:time_shifted_inputs_coeffs_computation}
\renewcommand{\arraystretch}{1.5}
\begin{array}{lcl}
	\forall j\in\Inin,\ (\m \check{u}^{[N]}(t))_j
	& = &
		(\m u^{[N]}(\check{t} + t^{[N]}))_j \\
	& = &
		\dsp{\sum_{k=0}^{n}} \m a_{jk}^{[N]}(\check{t} + t^{[N]})^k \\
	& = &
		\dsp{\sum_{k=0}^{n}} \m a_{jk}^{[N]} \dsp{\sum_{l=0}^{k}} \left(\begin{smallmatrix}k\\l\end{smallmatrix}\right) \check{t}^l (t^{[N]})^{k-l} \\
	& = &
		\dsp{\sum_{0\leqslant l\leqslant k\leqslant n}} \m a_{jk}^{[N]} \left(\begin{smallmatrix}k\\l\end{smallmatrix}\right) \check{t}^l (t^{[N]})^{k-l} \\
	& = &
		\dsp{\sum_{l=0}^{n}} \dsp{\sum_{k=l}^{n}} \m a_{jk}^{[N]} \left(\begin{smallmatrix}k\\l\end{smallmatrix}\right) \check{t}^l (t^{[N]})^{k-l} \\
	& = &
		\dsp{\sum_{l=0}^{n}} \check{t}^l \underbrace{\dsp{\sum_{k=l}^{n}} \m a_{jk}^{[N]} \left(\begin{smallmatrix}k\\l\end{smallmatrix}\right) (t^{[N]})^{k-l}} \\
	& = &
		\hspace{0.4cm}\dsp{\sum_{l=0}^{n}}\hspace{0.4cm} \check{t}^l \hspace{0.4cm}\m \check{a}_{jl}^{[N]}
\end{array}
\renewcommand{\arraystretch}{1.0}
\end{equation}

By switching $l$ and $k$ in the above computations, we get the expressions of the polynomial coefficients of the vectorial polynomial $\m \check{u}^{[N]}$:

\begin{equation}
\label{eq:a_check_def}
\forall j\in\Inin,\ \forall k\in[\![0, n]\!],\ \m \check{a}_{jk}^{[N]} = \sum_{l=k}^{n} \m a_{jl}^{[N]} \left(\begin{smallmatrix}l\\k\end{smallmatrix}\right) (t^{[N]})^{l-k}
\end{equation}

\textbf{Note:} For the sake of readability, for the following computations in this subsection, the newly introduced quantities won't have any $\m$ and ${}^{[N]}$ left and right superscripts despite the fact that they change depending on the macro-step and the iteration. This will only apply to the computations that will lead to the $\m \check{y}^{[N+1]}$ outputs estimator.

Let $\check{\Xid}\in M_{\nin, (n+1)}(\mathbb{R})$ be the matrix representation of the coefficients of the polynomial of all coordinates of the time-shifted inputs.

\begin{equation}
\label{eq:Xi_check}
\check{\Xid} = \left(\m\check{a}_{jk}^{[N]}\right)_{\substack{j\in\Inin\\k\in [\![0,\phantom{ab} n]\!]}}
\end{equation}

The rows and columns indexing of the $\check{\Xid}$ matrix are intentionally numbered starting from $1$ and starting from $0$ respectively as the rows represent the different coordinates (from $1$ to $\nin$), and the columns represent the successive terms of the polynomials (from $0$ for constant term, to $n$ for the term of maximum degree).

Let's define the Laplace transforms of the inputs, outputs and states of the time-shifted linear system \eqref{eq:Lin_on_a_step_time_shifted}.

\begin{equation}
\label{eq:Laplace_transforms}
\begin{array}{lclclcl}
	\check{X} & = & \check{X}(s) & = & \mathcal{L}(\m\check{x}^{[N]})(s) & \in & L(\mathbb{R}_+, \rnst) \\
	\check{U} & = & \check{U}(s) & = & \mathcal{L}(\m\check{u}^{[N]})(s) & \in & L(\mathbb{R}_+, \rnin) \\
	\check{Y} & = & \check{Y}(s) & = & \mathcal{L}(\m\check{y}_L^{[N]})(s) & \in & L(\mathbb{R}_+, \rnout) \\
\end{array}
\end{equation}

We can now write the Laplace transform of the time-shifted linear system \eqref{eq:Lin_on_a_step_time_shifted}.

\begin{equation}
\label{eq:Lin_on_a_step_time_shifted_Laplace}
\left\{
\begin{array}{ccl}
	s \check{X} - \tilde{x}^{[N]} & = & A^{[N]}\ \check{X} + B^{[N]}\ \check{U} + \frac{1}{s}\tilde{f}_C^{[N]} \\
	\check{Y}  & = & C^{[N]}\ \check{X} + D^{[N]}\ \check{U} \\
\end{array}
\right.
\end{equation}

Let $P$, $G$ and $R$ be the matrix functions of the Laplace domain as defined in \eqref{eq:P_G_R} (in particular, $G$ is the transfer function of the linear system).

\begin{equation}
\label{eq:P_G_R}
\begin{array}{lclclcl}
	P & = & P(s) & = & C^{[N]}\ (sI-A^{[N]})^{-1} & \in & L(\mathbb{R}_+, M_{\nout, \nst}(\mathbb{R})) \\
	G & = & G(s) & = & P(s)\ B^{[N]} + D^{[N]} & \in & L(\mathbb{R}_+, M_{\nout, \nin}(\mathbb{R})) \\
	R & = & R(s) & = & \frac{1}{s}P(s) & \in & L(\mathbb{R}_+, M_{\nout, \nst}(\mathbb{R})) \\
\end{array}
\end{equation}

With $P$, $G$ and $R$ as defined in \eqref{eq:P_G_R}, we can use \eqref{eq:Lin_on_a_step_time_shifted_Laplace} to write:

\begin{equation}
\label{eq:transfer_system_Laplace}
\check{Y} = G\ \check{U} + P\ \tilde{x}^{[N]} + R\ \tilde{f}_C^{[N]}
\end{equation}

The next step consists in splitting $\check{U}$ into two parts: one depending on the coefficients and one depending only on $n$.

\begin{equation}
\label{eq:split_U_and_coeffs}
\renewcommand{\arraystretch}{1.8}
\begin{array}{lcl}
	\check{U}
	& = &
		\left(\mathcal{L}(\m \check{u}^{[N]})(s)\right)_{j\in\Inin} \\
	& = &
		\left(\mathcal{L}\left(\check{t}\mapsto\displaystyle{\sum_{k=0}^{n}}\m \check{a}_{jk}^{[N]}\check{t}^k\right)(s)\right)_{j\in\Inin} \\
	& = &
		\left( \check{t}\mapsto\displaystyle{\sum_{k=0}^{n}} \check{a}_{jk}\mathcal{L}\left(\check{t}^k\right)(s) \right)_{j\in\Inin} \\
	& = &
		\check{\Xid}\ \bar{U}
\end{array}
\renewcommand{\arraystretch}{1.0}
\end{equation}

\noindent where

\begin{equation}
\label{eq:U_bar}
\bar{U}
=
\left(\mathcal{L}\left(\check{t}\mapsto\check{t}^k\right)(s)\right)_{k\in[\![0, n]\!]}
=
\left(\frac{k!}{s^{k+1}}\right)_{k\in[\![0, n]\!]}
\end{equation}

In order to remove ambiguity in the upcoming calculations, let's define the notation $\otimes$ as the outer product (particular case of tensor product). In particular, applied to a matrix $M$ and vector $v$, the outer product gives a $3$\up{rd} order tensor:

\begin{equation}
\label{eq:outer_product_Mv}
\renewcommand{\arraystretch}{1.5}
\begin{array}{c}
	\forall (n_1, n_2, n_3) \in (\mathbb{N}^{*})^3,\
	\forall M \in M_{n_1, n_2}(\mathbb{R}),\
	\forall v \in \mathbb{R}^{n_3},
	\\
	M \otimes v^T
	=
	\left(
		(M)_{i_1, i_2} (v)_{i_3}
	\right)_{\substack
	{
		i_1 \in [\![1, n_1]\!] \\
		i_2 \in [\![1, n_2]\!] \\
		i_3 \in [\![1, n_3]\!]
	}}
\end{array}
\renewcommand{\arraystretch}{1.0}
\end{equation}

In \eqref{eq:outer_product_Mv}, the vector $v$ is transposed so that $v^T$ is a row vector. Indeed, this enables an analogy with the Kronecker product $\otimes_{\text{Kron}}$.

\begin{equation}
\label{eq:outer_produt_and_knonecker}
\renewcommand{\arraystretch}{1.5}
\begin{array}{c}
	\forall (n_1, n_2, n_3) \in (\mathbb{N}^{*})^3,\
	\forall M \in M_{n_1, n_2}(\mathbb{R}),\
	\forall v \in \mathbb{R}^{n_3},\
	\forall Q \in M_{n_2, n_3}(\mathbb{R}),
	\\
	\begin{array}{ll}
		\left(M \otimes v^T\right) Q
		& =
		\left(
			\displaystyle{\sum_{i_2 = 1}^{n_2}}\
			\displaystyle{\sum_{i_3 = 1}^{n_3}}
			\left(
				M \otimes_{\text{Kron}} v^T
			\right)_{i_1, (i_2-1)n_3 + i_3}
			(Q)_{i_2, i_3}\!
		\right)_{i_1 \in [\![1, n_1]\!]}
		\\
		& =
		\left(
			\displaystyle{\sum_{i_2 = 1}^{n_2}}\
			\displaystyle{\sum_{i_3 = 1}^{n_3}}
			(M)_{i_1, i_2} (v)_{i_3} (Q)_{i_2, i_3}
		\right)_{i_1 \in [\![1, n_1]\!]}
		\\
		& =
		\left(M \otimes_{\text{Kron}} v^T\right) \text{vec}(Q^T)
		\\
	\end{array}
\end{array}
\renewcommand{\arraystretch}{1.0}
\end{equation}

\noindent where $\text{vec}(Q^T)$ denotes the vectorization of the matrix $Q^T$, formed by stacking the columns of $Q^T$ into a single column vector. This can also be seen as the concatenated rows of the matrix $Q$, transposed into a single column vector.

Among other properties, we notice the following: the reordering of a matrix-matrix-vector product.

\begin{equation}
\label{eq:reordering_a_MMv_product}
\renewcommand{\arraystretch}{1.5}
\begin{array}{c}
	\forall (n_1, n_2, n_3) \in (\mathbb{N}^{*})^3,\
	\forall M \in M_{n_1, n_2}(\mathbb{R}),\
	\forall v \in \mathbb{R}^{n_3},\
	\forall Q \in M_{n_2, n_3}(\mathbb{R}),
	\\
	\begin{array}{ll}
		M\ Q\ v
		& =
		M
		\left(
			\displaystyle{\sum_{i_3 = 1}^{n_3}}
			(Q)_{i_2, i_3} (v)_{i_3}
		\right)_{i_2 \in [\![1, n_2]\!]}
		\\
		& =
		\left(
			\displaystyle{\sum_{i_2 = 1}^{n_2}}
			\left(
				(M)_{i_1, i_2}
				\left(
					\displaystyle{\sum_{i_3 = 1}^{n_3}}
					(Q)_{i_2, i_3} (v)_{i_3}
				\right)
			\right)
		\right)_{i_1 \in [\![1, n_1]\!]}
		\\
		& =
		\left(
			\displaystyle{\sum_{i_2 = 1}^{n_2}}\
			\displaystyle{\sum_{i_3 = 1}^{n_3}}
			(M)_{i_1, i_2} (Q)_{i_2, i_3} (v)_{i_3}
		\right)_{i_1 \in [\![1, n_1]\!]}
		\\
		& =
		\left(M \otimes v^T\right) Q
		\\
	\end{array}
\end{array}
\renewcommand{\arraystretch}{1.0}
\end{equation}

Thanks to all the elements introduced above, we can express the linear contribution to outputs estimator as the inverse Laplace of $\check{Y}$ on the step size $\dt^{[N]}$ as this is the final time of the time-shifted system \eqref{eq:Lin_on_a_step_time_shifted}.

\begin{equation}
\label{eq:y_L_estimation}
\renewcommand{\arraystretch}{1.5}
\begin{array}{lcl}
	\m \hat{y}_L^{[N+1]} \\
	=
		\mathcal{L}^{-1}\!\!\left( \check{Y}\right)(\dt^{[N]})
		& \phantom{a} & \\
	=
		\mathcal{L}^{-1}\!\!\left( G\check{U} + P\tilde{x}^{[N]} + R\tilde{f}_C^{[N]}\right)(\dt^{[N]})
		&& \text{from \eqref{eq:transfer_system_Laplace}} \\
	=
		\mathcal{L}^{-1}\!\!\left( G\check{U} \right)(\dt^{[N]}) + \mathcal{L}^{-1}\!\!\left(P\tilde{x}^{[N]}+R\tilde{f}_C^{[N]}\right)(\dt^{[N]})
		&& \text{from}\ \mathcal{L}^{-1}\ \text{linearity} \\
	=
		\mathcal{L}^{-1}\!\!\left( G\check{\Xid}\bar{U} \right)(\dt^{[N]}) + \mathcal{L}^{-1}\!\!\left(P\tilde{x}^{[N]}+R\tilde{f}_C^{[N]}\right)(\dt^{[N]})
		&& \text{from \eqref{eq:split_U_and_coeffs}} \\
	=
		\mathcal{L}^{-1}\!\!\left( (G\otimes\bar{U}^T) \check{\Xid} \right)(\dt^{[N]}) + \mathcal{L}^{-1}\!\!\left(P\tilde{x}^{[N]}+R\tilde{f}_C^{[N]}\right)(\dt^{[N]})
		&& \text{from \eqref{eq:reordering_a_MMv_product}} \\
	=
		\mathcal{L}^{-1}\!\!\left( G\otimes\bar{U}^T \right)(\dt^{[N]})\ \check{\Xid} + \mathcal{L}^{-1}\!\!\left(P\tilde{x}^{[N]}+R\tilde{f}_C^{[N]}\right)(\dt^{[N]})
		&& \text{from}\ \mathcal{L}^{-1}\ \text{linearity} \\
	=
		\mathcal{G}_V\ \check{\Xid} + \mathcal{P}_V\ \tilde{x}^{[N]} + \mathcal{R}_V\ \tilde{f}_C^{[N]}
		&& \text{from}\ \mathcal{L}^{-1}\ \text{linearity} \\
\end{array}
\renewcommand{\arraystretch}{1.0}
\end{equation}

with

\begin{equation}
\label{eq:GcalV_PcalV_RcalV_defs}
\renewcommand{\arraystretch}{1.5}
\begin{array}{ll}
	\mathcal{G}_V
	& = \mathcal{L}^{-1}\left( G\otimes\bar{U}^T \right)(\dt^{[N]}) \\
	\mathcal{P}_V
	& = \mathcal{L}^{-1}\left(P\right)(\dt^{[N]}) \\
	\mathcal{R}_V
	& = \mathcal{L}^{-1}\left(R\right)(\dt^{[N]})
\end{array}
\renewcommand{\arraystretch}{1.0}
\end{equation}

Analogously, we can write the expression of $\hat{\dot{y}}_L^{[N+1]}$ the estimator of the time-derivative of the outputs:

\begin{equation}
\label{eq:dot_y_L_estimation}
\renewcommand{\arraystretch}{2.5}
\begin{array}{lll}
	\m\hat{\dot{y}}_L^{[N+1]}
	& = &
	\dspfrac{d\Big(\mathcal{L}^{-1}\left( G\otimes\bar{U}^T \right) \check{\Xid} + \mathcal{L}^{-1}\left(P\right) \tilde{x}^{[N]} + \mathcal{L}^{-1}\left(R\right) \tilde{f}_C^{[N]}\Big)}{dt} (\dt^{[N]})
	\\
	& = &
	\dspfrac{d\Big(\mathcal{L}^{-1}\left( G\otimes\bar{U}^T \right)\Big)}{dt} (\dt^{[N]})
	\ \check{\Xid}
	+
	\dspfrac{d\Big(\mathcal{L}^{-1}\left(P\right)\Big)}{dt} (\dt^{[N]})
	\ \tilde{x}^{[N]}
	\\
	&&
	+
	\dspfrac{d\Big(\mathcal{L}^{-1}\left(R\right)\Big)}{dt} (\dt^{[N]})
	\ \tilde{f}_C^{[N]}
	\\
	& = &
	\mathcal{G}_D\ \check{\Xid}
	+ \mathcal{P}_D\ \tilde{x}^{[N]}
	+ \mathcal{R}_D\ \tilde{f}_C^{[N]}
	\\
\end{array}
\renewcommand{\arraystretch}{1.0}
\end{equation}

with

\begin{equation}
\label{eq:GcalD_PcalD_RcalD_defs}
\renewcommand{\arraystretch}{2.5}
\begin{array}{ll}
	\mathcal{G}_D
	& = \dspfrac{d\Big(\mathcal{L}^{-1}\left( G\otimes\bar{U}^T \right)\Big)}{dt} (\dt^{[N]}) \\
	\mathcal{P}_D
	& = \dspfrac{d\Big(\mathcal{L}^{-1}\left(P\right)\Big)}{dt} (\dt^{[N]}) \\
	\mathcal{R}_D
	& = \dspfrac{d\Big(\mathcal{L}^{-1}\left(R\right)\Big)}{dt} (\dt^{[N]})
\end{array}
\renewcommand{\arraystretch}{1.0}
\end{equation}

\subsubsection{Linear part numerical evaluation}
\label{subsubsection:linear_part_numerical_evaluation}

The remaining problem lies in the evaluation of the quantities $\mathcal{G}_V$, $\mathcal{P}_V$, $\mathcal{R}_V$, $\mathcal{G}_D$, $\mathcal{P}_D$ and  $\mathcal{R}_D$. Indeed, the inverse Laplace transform of a function $F$ of the Laplace variable $s$ can be evaluated at a given time $t$ with several numerical methods \cite{Abate2006}. Our implementation, to generate the results of section \ref{section:examples_and_test_cases}, used the Gaver-Stehfest method \cite{Jacquot1983}, based on \cite{Stehfest1970}.

First of all, the inverse Laplace transform of a matrix (in our case $P$ and $R$) or a tensor of order $3$ (in our case: $G\otimes\bar{U}^T$) is the matrix (or a tensor of order $3$, respectively) of the inverse Laplace transforms of every element. The resulting matrix (or tensor of order $3$, respectively) has the same size as the one in the Laplace domain.

Let $F$ either denote a matrix or tensor function of the Laplace variable $s$. The numerical computation of $\mathcal{L}^{-1}(F)(t)$ for a given time $t$ requires several evaluations of $F$ at different values of $s$.

In our case, any evaluation of $G\otimes\bar{U}^T$, $P$ or $R$ requires, among others, a matrix inversion (see expressions \eqref{eq:P_G_R}) and several matrix products.

In case $A^{[N]}$ is diagonalizable, it can be written as $K \Delta K^{-1}$ where $\Delta$ is diagonal. The matrices $P$, $R$ and $G$ can then be written as follows:

\begin{equation}
\label{eq:P_when_A_is_diagonalizable}
\begin{array}{ll}
	P(s)
	& = C^{[N]}(sI-A^{[N]})^{-1} \\
	& = C^{[N]}(sI-K\Delta K^{-1})^{-1} \\
	& = C^{[N]}(sKIK^{-1}-K\Delta K^{-1})^{-1} \\
	& = C^{[N]}\big(K(sI-\Delta)K^{-1}\big)^{-1} \\
	& = C^{[N]}K(sI-\Delta)^{-1}K^{-1} \\
\end{array}
\end{equation}

\begin{equation}
\label{eq:G_when_A_is_diagonalizable}
\begin{array}{ll}
	G(s)
	& = P(s)B^{[N]}+D^{[N]} \\
	& = C^{[N]}K(sI-\Delta)^{-1}K^{-1}B^{[N]}+D^{[N]} \\
\end{array}
\end{equation}

\begin{equation}
\label{eq:R_when_A_is_diagonalizable}
\begin{array}{ll}
	R(s)
	& = \frac{1}{s} P(s) \\
	& = \frac{1}{s} C^{[N]}K(sI-\Delta)^{-1}K^{-1} \\
\end{array}
\end{equation}

In that case, the products $C^{[N]}K$ and $K^{-1}B^{[N]}$ can be computed only once when $A^{[N]}$ is known, that is to say, once for each iteration of the time-loop. Indeed, they do not depend on $s$. Moreover, the inversion $(sI-\Delta)^{-1}$ is immediate.

\begin{equation}
\label{eq:sI_m_D_inv}
(sI-\Delta)^{-1} =
\left(
	\begin{array}{ccc}
		\left(\frac{1}{s-(\Delta)_{1, 1}}\right) & & 0 \\
		& \ddots & \\
		0 & & \left(\frac{1}{s-(\Delta)_{\nst, \nst}}\right) \\
	\end{array}
\right)
\end{equation}

However, this approach has two main drawbacks:

\begin{itemize}
\ite in case $A^{[N]}$ is not diagonalizable this cannot be done (for instance: in case $A^{[N]}$ is nilpotent but not zero), and
\ite in case an evaluation at a given value $s$ corresponding to an eigenvalue of $A^{[N]}$ is required, $(sI-\Delta)^{-1}$ cannot be computed (in \eqref{eq:sI_m_D_inv}, at least one of the diagonal coefficients becomes $\nicefrac{1}{0}$).
\end{itemize}

Consequently, the approach we used in our implementation is the explicit computation of every coefficient in the matrices $G$, $P$ and $R$. Indeed, every element of $G$ is a rational function in terms of $s$, and every coefficient of the rational functions of every element of $G$ can be computed with the \textbf{Patel \& Misra} method \cite{Misra1987}.

Regarding matrix $P$, we can notice that, despite it is not strictly speaking a transfer function, it would be the same as $G$ in the following conditions:

\begin{itemize}
\ite if $\nin$ were equal to $\nst$,
\ite if $B^{[N]}$ were $I_{\nst\times\nst}$ the identity matrix of size $\nst \times \nst$, and
\ite if $D^{[N]}$ were $\textbf{0}_{\nout\times\nst}$ the null matrix of size $\nout \times \nst$.
\end{itemize}

Hence, running the Patel \& Misra method on the fake state-space system made of matrices $A^{[N]}$, $I_{\nst\times\nst}$, $C^{[N]}$ and $\textbf{0}_{\nout\times\nst}$, we obtain all the coefficients of the rational functions of every element of $P$.

Finally, regarding matrix $R$, it can easily be computed from $P$ as elements of the latter are rational fractions, and $R:s\mapsto\frac{1}{s} P(s)$ (see \eqref{eq:P_G_R}).

The tensor $\mathcal{G}_V$ and the matrices $\mathcal{P}_V$ and $\mathcal{R}_V$ can finally be computed as every coefficient of $G(s)$, $P(s)$, $R(s)$ and $\bar{U}(s)$ (see \eqref{eq:U_bar}) are known rational functions of $s$. Regarding $\mathcal{G}_V = \mathcal{L}^{-1}(G\otimes\bar{U}^T)$ in particular, the outer product $G\otimes\bar{U}^T$ works as defined in \eqref{eq:outer_product_Mv}. However, due to the indexing of $\check{\Xid}$ and $\bar{U}$ (see \eqref{eq:Xi_check} and \eqref{eq:U_bar}, respectively), we consider the indexing \eqref{eq:G_otimes_UT} for the outer product $G\otimes\bar{U}^T$.

\begin{equation}
\label{eq:G_otimes_UT}
\renewcommand{\arraystretch}{1.5}
\begin{array}{ll}
	G\otimes\bar{U}^T
	& =
	\left(
		(G)_{ij} (\bar{U})_k
	\right)_{\substack
	{
		i \in \Inout \\
		j \in \Inin \\
		k \in [\![0, n]\!]
	}}
	\\
	& =
	\left(
		(G\otimes\bar{U}^T)_{ijk}
	\right)_{\substack
	{
		i \in \Inout \\
		j \in \Inin \\
		k \in [\![0, n]\!]
	}}
	\\
\end{array}
\renewcommand{\arraystretch}{1.0}
\end{equation}

The Gaver-Stehfest method being applied element-wise, we obtain:

\begin{equation}
\label{eq:Gaver_Stehfest_elementwise}
\begin{array}{lclcl}
	\mathcal{G}_V
	& = &
	\big(
		(\mathcal{G}_V)_{i, j, k}
	\big)_{\substack{
		i\in\Inout\\
		j\in\Inin\\
		k\in[\![0, n]\!]
	}}
	& = &
	\Big(
		\mathcal{L}^{-1}\big((G\otimes\bar{U}^T)_{i, j, k}\big)\big(\dt^{[N]}\big)
	\Big)_{\substack{
		i\in\Inout\\
		j\in\Inin\\
		k\in[\![0, n]\!]
	}}
	\\
	\mathcal{P}_V
	& = &
	\big(
		(\mathcal{P}_V)_{i, j}
	\big)_{\substack{
		i\in\Inout\\
		j\in\Inst\\
	}}
	& = &
	\Big(
		\mathcal{L}^{-1}\big((P)_{i, j}\big)\big(\dt^{[N]}\big)
	\Big)_{\substack{
		i\in\Inout\\
		j\in\Inst
	}}
	\\
	\mathcal{R}_V
	& = &
	\big(
		(\mathcal{R}_V)_{i, j}
	\big)_{\substack{
		i\in\Inout\\
		j\in\Inst\\
	}}
	& = &
	\Big(
		\mathcal{L}^{-1}\big((R)_{i, j}\big)\big(\dt^{[N]}\big)
	\Big)_{\substack{
		i\in\Inout\\
		j\in\Inst
	}}
\end{array}
\end{equation}

In case the Gaver-Stehfest method requires an evaluation with $s$ being a pole of $P$, $R$ or $G$, the co-simulation step will diverge and $t^{[N+1]}$ will change leading to a different value of $\dt^{[N]}$ making the Gaver-Stehfest method evaluate $G$, $P$ and $R$ at different values of $s$.

If an estimation of the time-derivatives of the outputs is required, $\mathcal{G}_D$, $\mathcal{P}_D$ and $\mathcal{R}_D$ must be computed in order to evaluate $\m \hat{\dot{y}}_L^{[N+1]}$ (see \eqref{eq:dot_y_L_estimation} and \eqref{eq:GcalD_PcalD_RcalD_defs}). Additional evaluations of the inverse Laplace of $G\otimes\bar{U}^T$, $P$ and $R$ at different times than $\dt^{[N]}$ can help compute these quantities.

Let $h$ denote a small strictly positive quantity with respect to $\dt^{[N]}$. A simple first-order estimation of $\mathcal{G}_D$ and $\mathcal{P}_D$ can be achieved by an additional inverse Laplace computation for each of these quantities.

\begin{equation}
\label{eq:GcalD_PcalD_RcalD_FO}
\renewcommand{\arraystretch}{2.5}
\begin{array}{ll}
	\mathcal{G}_{D, FO}
	& =
	\dspfrac{1}{h} \left(\mathcal{G}_V - \mathcal{L}^{-1}\left( G\otimes\bar{U}^T \right) (\dt^{[N]} - h)\right)
	\\
	\mathcal{P}_{D, FO}
	& =
	\dspfrac{1}{h} \left(\mathcal{P}_V - \mathcal{L}^{-1}\left( P \right) (\dt^{[N]} - h)\right)
	\\
	\mathcal{R}_{D, FO}
	& =
	\dspfrac{1}{h} \left(\mathcal{R}_V - \mathcal{L}^{-1}\left( R \right) (\dt^{[N]} - h)\right)
	\\
\end{array}
\renewcommand{\arraystretch}{1.0}
\end{equation}

Higher order estimations can also be used, for instance the implementation that has been used to present the results in \ref{section:examples_and_test_cases} uses a $2$\up{nd} order Richardson (2Rich) approximation \eqref{eq:GcalD_PcalD_RcalD_2Rich} with $h=0.2\ \dt^{[N]}$.

\begin{equation}
\label{eq:GcalD_PcalD_RcalD_2Rich}
\begin{small}
\begin{array}{ll}
	\mathcal{G}_{D, 2Rich}
	& =
	\dspfrac
	{
		\mathcal{L}^{-1}\left( G\otimes\bar{U}^T \right) (\dt^{[N]} - h) - 4 \mathcal{L}^{-1}\left( G\otimes\bar{U}^T \right) (\dt^{[N]} - \frac{h}{2}) + 3 \mathcal{G}_V
	}
	{h}
	\\ \\
	\mathcal{P}_{D, 2Rich}
	& =
	\dspfrac
	{
		\mathcal{L}^{-1}\left( P \right) (\dt^{[N]} - h) - 4 \mathcal{L}^{-1}\left( P \right) (\dt^{[N]} - \frac{h}{2}) + 3 \mathcal{P}_V
	}
	{h}
	\\ \\
	\mathcal{R}_{D, 2Rich}
	& =
	\dspfrac
	{
		\mathcal{L}^{-1}\left( R \right) (\dt^{[N]} - h) - 4 \mathcal{L}^{-1}\left( R \right) (\dt^{[N]} - \frac{h}{2}) + 3 \mathcal{R}_V
	}
	{h}
	\\
\end{array}
\end{small}
\end{equation}

Indeed, these estimations have a consistency order of $2$. Let's take an arbitrary regular enough real scalar function $f$. Taylor formula of $f(t-h)$ and $f(t-\frac{h}{2})$ give:

\begin{equation}
\label{eq:Richardson_2Rich_order_proof_3}
\renewcommand{\arraystretch}{1.5}
\begin{array}{ll}
	\dspfrac{f(t-h) - 4 f(t-\frac{h}{2}) + 3 f(t)}{h}
	& =
	f^{\prime}(t)-\dspfrac{h^2}{12}f^{\prime\prime\prime}(t)+o(h^2)
	\\
	& =
	f^{\prime}(t) + O(h^2)
\end{array}
\renewcommand{\arraystretch}{1.0}
\end{equation}

\noindent which shows the $2$\up{nd} order of consistence of the 2Rich formula used in \eqref{eq:GcalD_PcalD_RcalD_2Rich}.

\newpage 

\subsection{Estimator update}
\label{subsection:estimator_update}

Let's recap the intermediate computations required the get the estimators of table \ref{table:estimators_at_tnp} from the known quantities of table \ref{table:known_quantities_at_tn}. Figure \ref{fig:dependencies} shows the intermediate quantities, and the quantities they require to be computed.

\begin{center}
\includegraphics[scale=0.55]{\figuresdir/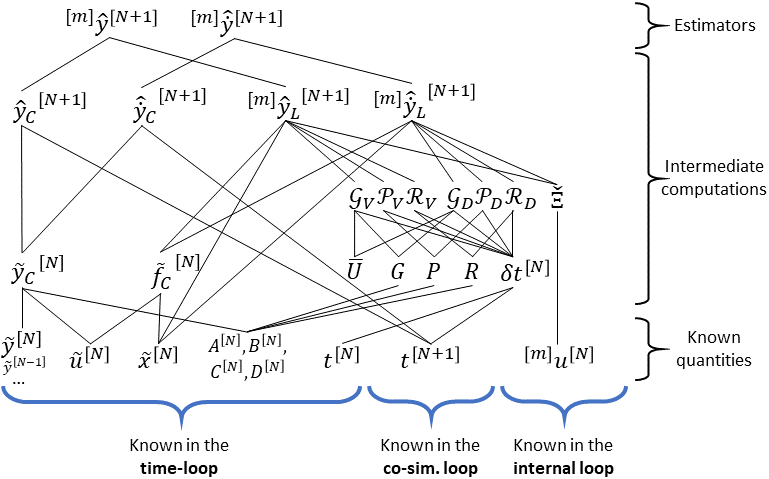}
\captionof{figure}{Quantity dependencies to get the output estimators}
\label{fig:dependencies}
\end{center}

We notice, on figure \ref{fig:dependencies}, that some intermediate quantities can be computed before the moment where the estimators are needed. Therefore, instead of computing the estimators when required (algorithm \ref{alg:single_worker_stepestim_icsa}), a \textit{lazy} update version of COSTARICA usage in a co-simulation worker program can be achieved. This lazy version avoids re-computing intermediate quantities at stages where they are not supposed to change from the previous computations. This is presented in algorithm \ref{alg:single_worker_lazyCOSTARICA}.

Algorithm \ref{alg:single_worker_lazyCOSTARICA} enables the usage of COSTARICA in an optimized way. Every time the co-simulation method needs to iterate on a given macro-step (internal loop), the cost of the estimation of the outputs is:

\begin{itemize}
\ite the $\check{\Xid}$ computation \eqref{eq:a_check_def},
\ite the tensor-matrix product $\mathcal{G}_V\check{\Xid}$,
\ite the matrix-vector product $\mathcal{P}_V\tilde{x}^{[N]}$,
\ite the matrix-vector product $\mathcal{R}_V\tilde{f}_C^{[N]}$, and
\ite three vectors sums.
\end{itemize}

This might be multiplied by $2$ in case the time-derivatives of the outputs are required as well.

Other computations are required, yet they can be done outside of the internal loop, which enables the co-simulation method to iterate a lot internally in order to get a high convergence in case the internal part is due to an iterative numerical method (for instance \cite{Eguillon2021IfosmondiJFM}).

Finally, the computational cost of the estimators only depends on the sizes $\nin$, $\nout$, $\nst$ and $n$. Unlike a minimal solver being responsible for the resolution of the linearized system \eqref{eq:Lin_on_a_step}, the dynamics of the linearized system do not affect the computational cost.

\begin{footnotesize}
\begin{algorithm}[H]
\caption{Lazy COSTARICA injection in single worker process in an iterative co-simulation method}
\label{alg:single_worker_lazyCOSTARICA}
{\setstretch{1.3}
$N:=0$\;
$t^{[0]}:=\tinit$\;
\While(\tcp*[h]{Time loop}){$t^{[N]} < \tend$}
{
	Ask system to provide $A^{[N]}, B^{[N]}, C^{[N]}, D^{[N]}$\tcp*{update 1}
	Ask system to provide $\tilde{x}^{[N]}$ and $\tilde{f}^{[N]}$ (see table \ref{table:known_quantities_at_tn}) \tcp*{update 1}
	From these quantities and $t^{[N]}$, $\tilde{y}^{[N]}$ and $\tilde{u}^{[N]}$ (known), \hspace{10cm}compute $\tilde{y}_C^{[N]}$, $\tilde{f}_C^{[N]}$ and coefficients of rational functions \hspace{10cm}at each element of $s:\mapsto\left(G\otimes\bar{U}^T\right)(s)$, of $s:\mapsto P(s)$ \hspace{10cm}and of $s:\mapsto R(s)$ \tcp*{update 1}
	\While(\tcp*[h]{Co-sim. step loop}){Step starting on $t^{[N]}$ is not converged}
	{
		$m:=0$\;
		Method (re)computes $t^{[N+1]}$\;
		From $t^{[N]}$, $t^{[N+1]}$, $\m\tilde{y}_C^{[N]}$ and eventually  $(\tilde{y}_C^{[N-r]})_{r=1, 2, ...}$, \hspace{10cm}compute $\hat{y}_C^{[N+1]}$ and $\hat{\dot{y}}_C^{[N+1]}$\tcp*{update 2}
		Compute $\dt^{[N]} := t^{[N+1]} - t^{[N]}$\tcp*{update 2}
		From expressions of $s:\mapsto\left(G\otimes\bar{U}^T\right)(s)$, $s:\mapsto P(s)$ \hspace{10cm}and $s:\mapsto R(s)$, and as we now know $\dt^{[N]}$, compute \hspace{10cm}several inverse Laplace transforms to get $\mathcal{G}_V$, $\mathcal{P}_V$, \hspace{10cm}$\mathcal{R}_V$, $\mathcal{G}_D$, $\mathcal{P}_D$ and $\mathcal{R}_D$\tcp*{update 2}
		
		\While(\tcp*[h]{Internal loop}){True}
		{
			Method computes $\m u^{[N]}$\;
			\sout{Compute $\m \tilde{y}^{[N+1]}$ by step integration $S^{[N]}$ with inputs $\m u^{[N]}$}\;
			From $\m u^{[N]}$, compute matrix $\check{\Xid}$\tcp*{update 3}
			\fbox{Estimate outputs: $\m \hat{y}^{[N+1]}:=\mathcal{G}_V\check{\Xid}+\mathcal{P}_V\tilde{x}^{[N]}+\mathcal{R}_V\tilde{f}_C^{[N]}+\hat{y}_C^{[N+1]}$\;}\\
			\fbox{Estimate outputs: $\m \hat{\dot{y}}^{[N+1]}:=\mathcal{G}_D\check{\Xid}+\mathcal{P}_D\tilde{x}^{[N]}+\mathcal{R}_D\tilde{f}_C^{[N]}+\hat{\dot{y}}_C^{[N+1]}$\;}\\
			From $\m \hat{y}^{[N+1]}$ and outputs of other systems on other workers, \hspace{10cm}and eventually the time-derivative estimations too, \hspace{10cm}method decides if the step is \textbf{converged}, \textbf{to-be-redone} or \textbf{rejected}\;
			\uIf{Step $[t^{[N]}, t^{[N+1]}[$ is \textbf{to-be-redone}}
			{
				$m:=m+1$\;
			}
			\Else
			{
				\tcp{either step is converged or has been rejected}
				Break\;
			}
		}
	}
	$\mmax(N)=m$\;
	{Compute $\!{}^{[\mmax(N)]} \tilde{y}^{[N+1]}$ by step integration $S^{[N]}\!$ with inputs $\!{}^{[\mmax(N)]} u^{[N]}$}\;
	$N := N+1$\;
}
} 
\end{algorithm}
\end{footnotesize}

\section{Relevance of the linearization over time}
\label{section:Relevance_of_the_linearization_over_time}

As the COSTARICA estimator is based on the linearization \eqref{eq:Lin_on_a_step}, its quality directly depends on the quality of this linearization. Moreover, this linearization is not complete in the sense that partial derivatives of $f$ and $g$ functions in $x$ and $u$ variables contribute to the linearization, but the $t$ variable is missing. This directly comes from the fact that there is no standardized ways to get partial derivatives of $f$ and $g$ across $t$ in practice (namely in the FMI standard). The numerical effect of the choice (non-complete linearization) made in the development of the numerical technique (COSTARICA estimator based on this very linearization) is studied hereafter. This analysis of the relevance of the non-complete linearization over time shows that the error order of this linearization directly characterizes the quality of the COSTARICA estimatior.

\subsection{Error analysis regarding the state response}
\label{subsection:Error_analysis_regarding_the_state_response}

In order to perform an analysis of the error of the linearization, we first need some prerequisites derived from the Landau notation $o$. The latter are shown in \ref{subsubsection:Prerequisite_asymptotic_order_of_states_and_inputs}. Then, in \ref{subsubsection:Theoretical_error_order_due_to_the_linearization}, the error order calculations are given.

\subsubsection{Prerequisite: asymptotic order of states and inputs}
\label{subsubsection:Prerequisite_asymptotic_order_of_states_and_inputs}

Landau's notation $o$, when applied to the state variables, bring the existence of a function $\ex$ with a limit of $0$ in $0$ so that we can write:

\begin{equation}
\label{eq:o_x_prep}
\renewcommand{\arraystretch}{1.5}
\begin{array}{l}
	o\big((x(\tau)-\tildexN)^2\big) \\
	=
		\ex(x(\tau)-\tildexN)\cdot(x(\tau)-\tildexN)^2 \\
	=
		\ex(x(\tau)-\tildexN) \cdot \Big(\tildexN + \tildexdNplus\cdot(\tau-\tN) + \half\tildexddNplus\cdot(\tau-\tN)^2 + o((\tau-\tN)^2) - \tildexN\Big)^2 \\
	=
		\ex(x(\tau)-\tildexN) \cdot \Big(\tildexdNplus\cdot(\tau-\tN) + \half\tildexddNplus\cdot(\tau-\tN)^2 + o((\tau-\tN)^2) \Big)^2 \\
	=
		\ex(x(\tau)-\tildexN) \cdot \Big(\tildexdNplus\cdot(\tau-\tN) + \half\tildexddNplus\cdot(\tau-\tN)^2 + \ext(\tau-\tN) \cdot (\tau-\tN)^2 \Big)^2 \\
	=
		\ex(x(\tau)-\tildexN) \cdot \Big((\tildexdNplus)^2\cdot(\tau-\tN)^2 + \fourth(\tildexddNplus)^2\cdot(\tau-\tN)^4 + \ext(\tau-\tN)^2 \cdot (\tau-\tN)^4 \\
	\h\h +
		\tildexdNplus\tildexddNplus(\tau-\tN)^3 + 2 \tildexdNplus\ext(\tau-\tN)\cdot(\tau-\tN)^3 + \tildexddNplus\ext(\tau-\tN)^4 \Big) \\
	=
		\ex(x(\tau)-\tildexN) \cdot \Big((\tildexdNplus)^2 + \fourth(\tildexddNplus)^2\cdot(\tau-\tN)^2 + \ext(\tau-\tN)^2 \cdot (\tau-\tN)^2 \\
	\h\h +
		\tildexdNplus\tildexddNplus(\tau-\tN) + 2 \tildexdNplus\ext(\tau-\tN)\cdot(\tau-\tN) + \tildexddNplus\ext(\tau-\tN)^2 \Big) \cdot (\tau-\tN)^2 \\
\end{array}
\renewcommand{\arraystretch}{1.0}
\end{equation}

As we have a finite limit \eqref{eq:o_x_why1} of a term in the last expression of \eqref{eq:o_x_prep}, we can write the zero limit of the product \eqref{eq:o_x_why2}.

\begin{equation}
\label{eq:o_x_why1}
\begin{array}{l}
	\dsp{\lim_{\substack{\tau \rightarrow \tN\\\tau>\tN}}}
	\Big((\tildexdNplus)^2 + \fourth(\tildexddNplus)^2\cdot(\tau-\tN)^2 + \ext(\tau-\tN)^2 \cdot (\tau-\tN)^2
	\\
	\hspace{25mm} +
		\tildexdNplus\tildexddNplus(\tau-\tN) + 2 \tildexdNplus\ext(\tau-\tN)\cdot(\tau-\tN) + \tildexddNplus\ext(\tau-\tN)^2 \Big) = (\tildexdNplus)^2
\end{array}
\end{equation}

\begin{equation}
\label{eq:o_x_why2}
\begin{array}{l}
	\dsp{\lim_{\substack{\tau \rightarrow \tN\\\tau>\tN}}}
	\bigg(
		\ex(x(\tau)-\tildexN) \cdot \Big((\tildexdNplus)^2 + \fourth(\tildexddNplus)^2\cdot(\tau-\tN)^2 + \ext(\tau-\tN)^2 \cdot (\tau-\tN)^2
		\\
		\hspace{25mm} +
			\tildexdNplus\tildexddNplus(\tau-\tN) + 2 \tildexdNplus\ext(\tau-\tN)\cdot(\tau-\tN) + \tildexddNplus\ext(\tau-\tN)^2 \Big)
	\bigg) =	0
\end{array}
\end{equation}

By defining an epsilon function $\epsilon_{xt}$ by \eqref{eq:o_x_why3}, we dispose of a function with a limit of zero in zero \eqref{eq:o_x_why4} and we can thus rewrite the last expression of \eqref{eq:o_x_prep} as \eqref{eq:o_x}, showing that an order $2$ error in the states corresponds to an order $2$ error in time.

\begin{equation}
\label{eq:o_x_why3}
\begin{array}{ll}
	\epsilon_{xt}:\xi\mapsto
	&
	\ex(x(\xi+\tN)-\tildexN)
	\\
	&
	\cdot \Big((\tildexdNplus)^2 + \fourth(\tildexddNplus)^2\cdot\xi^2 + \ext(\xi)^2 \cdot (\xi)^2
	+
	\tildexdNplus\tildexddNplus\xi + 2 \tildexdNplus\ext(\xi)\cdot\xi + \tildexddNplus\ext(\xi)^2 \Big)
\end{array}
\end{equation}

\begin{equation}
\label{eq:o_x_why4}
\lim_{\substack{\tau\rightarrow\tN\\\tau>\tN}} \epsilon_{xt}(\tau-\tN) = 0
\end{equation}

\begin{equation}
\label{eq:o_x}
\renewcommand{\arraystretch}{1.5}
\begin{array}{lll}
	o\big((x(\tau)-\tildexN)^2\big)
	& \multicolumn{2}{l}
		{
			=
			\ex(x(\tau)-\tildexN) \cdot \Big((\tildexdNplus)^2 + \fourth(\tildexddNplus)^2\cdot(\tau-\tN)^2 + \ext(\tau-\tN)^2 \cdot (\tau-\tN)^2
		} \\
	& \multicolumn{2}{l}
		{
			\h +
			\tildexdNplus\tildexddNplus(\tau-\tN) + 2 \tildexdNplus\ext(\tau-\tN)\cdot(\tau-\tN) + \tildexddNplus\ext(\tau-\tN)^2 \Big)
		} \\
	& \multicolumn{2}{l}
		{
			\h \cdot (\tau-\tN)^2
		} \\
	& =
		\epsilon_{xt}(\tau-\tN) \cdot (\tau-\tN)^2
		\h \phantom{.}
		&
			\text{see \eqref{eq:o_x_why3}} \\
	& =
		o\big((\tau-\tN)^2\big)
		&
			\text{because of \eqref{eq:o_x_why4}} \\
\end{array}
\renewcommand{\arraystretch}{1.0}
\end{equation}

Analogously, an order $2$ error in the inputs corresponds to an order $2$ error in time \eqref{eq:o_u}.

\begin{equation}
\label{eq:o_u}
o\big((u(\tau)-\tildeuNplus)^2\big)
=
o\big((\tau-\tN)^2\big)
\end{equation}

\subsubsection{Theoretical error order due to the linearization}
\label{subsubsection:Theoretical_error_order_due_to_the_linearization}

Let's consider the states at a given time $t$ within the macro-step $[\tN, \tNp[$. Its theoretical expression is \eqref{eq:x_exact_1}.

\begin{equation}
\label{eq:x_exact_1}
x(t)
=
\tildexN + \dsp{\int_{\tN}^{t}} f(\tau, x(\tau), u(\tau)) d\tau
\end{equation}

The limited development of \eqref{eq:x_exact_1} can be done, leading to \eqref{eq:x_exact_2}.

\begin{equation}
\label{eq:x_exact_2}
\renewcommand{\arraystretch}{1.8}
\arraycolsep=2pt
\begin{array}{l}
	x(t)
	= \\
	\tildexN + \dsp{\int_{\tN}^{t}}
	\Big(
		\tildefNplus + \tildepfNplus{t} (\tau-\tN) + \tildepfNplus{x} (x(\tau)-\tildexN) + \tildepfNplus{u} (u(\tau)-\tildeuNplus)
		\\ \hsd
		+ \half \tildepfNplus{tt} (\tau-\tN)^2 + \half \tildepfNplus{tx} (\tau-\tN)(x(\tau)-\tildexN) + \half \tildepfNplus{tu} (\tau-\tN)(u(\tau)-\tildeuNplus)
		\\ \hsd
		+ \half \tildepfNplus{xt} (x(\tau)-\tildexN)(\tau-\tN) + \half \tildepfNplus{xx} (x(\tau)-\tildexN)^2 + \half \tildepfNplus{xu} (x(\tau)-\tildexN)(u(\tau)-\tildeuNplus)
		\\ \hsd
		+ \half \tildepfNplus{ut} (u(\tau)-\tildeuNplus)(\tau-\tN) + \half \tildepfNplus{ux} (u(\tau)-\tildeuNplus)(x(\tau)-\tildexN) + \half \tildepfNplus{uu} (u(\tau)-\tildeuNplus)^2
		\\ \hsd
		+ o\big((\tau-\tN)^2\big) + o\big((x(\tau)-\tildexN)^2\big) + o\big((u(\tau)-\tildeuNplus)^2\big)
	\Big)
	d\tau
	\\
\end{array}
\renewcommand{\arraystretch}{1.0}
\end{equation}

\newpage 

The limited development of the states and the inputs in \eqref{eq:x_exact_2} can also be done, leading to \eqref{eq:x_exact_3}.

\begin{equation}
\label{eq:x_exact_3}
\renewcommand{\arraystretch}{1.8}
\arraycolsep=2pt
\begin{array}{ll}
	x(t)
	= &
		\tildexN + \dsp{\int_{\tN}^{t}}
		\Big(
			\tildefNplus + \tildepfNplus{t} (\tau-\tN) + \tildepfNplus{x} (x(\tau)-\tildexN) + \tildepfNplus{u} (u(\tau)-\tildeuNplus)
			\\ & \h
			+ \halftildepfNplus{tt} (\taumtn)^2 + \halftildepfNplus{tx} (\taumtn)\big(\tildexdNplus(\taumtn)+\halftildexddNplus(\taumtn)^2+o((\taumtn)^2)\big)
			\\ & \h\h
			+ \halftildepfNplus{tu} (\taumtn)\big(\tildeudNplus(\taumtn)+\halftildeuddNplus(\taumtn)^2+o((\taumtn)^2)\big)
			\\ & \h
			+ \halftildepfNplus{xt} \big(\tildexdNplus(\taumtn)+\halftildexddNplus(\taumtn)^2+o((\taumtn)^2)\big)(\taumtn)
			\\ & \h\h
			+ \halftildepfNplus{xx} \big(\tildexdNplus(\taumtn)+\halftildexddNplus(\taumtn)^2+o((\taumtn)^2)\big)^2
			\\ & \h\h
			+ \halftildepfNplus{xu} \big(\tildexdNplus(\taumtn)+\halftildexddNplus(\taumtn)^2+o((\taumtn)^2)\big)
			\\ & \h\h
					\cdot\big(\tildeudNplus(\taumtn)+\halftildeuddNplus(\taumtn)^2+o((\taumtn)^2)\big)
			\\ & \h
			+ \halftildepfNplus{ut} \big(\tildeudNplus(\taumtn)+\halftildeuddNplus(\taumtn)^2+o((\taumtn)^2)\big)(\taumtn)
			\\ & \h\h
			+ \halftildepfNplus{ux} \big(\tildeudNplus(\taumtn)+\halftildeuddNplus(\taumtn)^2+o((\taumtn)^2)\big)
			\\ & \h\h
					\cdot\big(\tildexdNplus(\taumtn)+\halftildexddNplus(\taumtn)^2+o((\taumtn)^2)\big)
			\\ & \h\h
			+ \halftildepfNplus{uu} \big(\tildeudNplus(\taumtn)+\halftildeuddNplus(\taumtn)^2+o((\taumtn)^2)\big)^2
			\\ & \h
			+ o\big((\taumtn)^2\big) + o\big((x(\tau)-\tildexN)^2\big) + o\big((u(\tau)-\tildeuNplus)^2\big)
		\Big)
		d\tau
		\\
\end{array}
\renewcommand{\arraystretch}{1.0}
\end{equation}

Gathering all the $o$ terms and merging them thanks to the properties described in the prerequisites \ref{subsubsection:Prerequisite_asymptotic_order_of_states_and_inputs}, we get \eqref{eq:x_exact_4}.

\begin{equation}
\label{eq:x_exact_4}
\renewcommand{\arraystretch}{1.8}
\arraycolsep=2pt
\begin{array}{ll}
	x(t)
	= &
		\tildexN + \dsp{\int_{\tN}^{t}}
		\Big(
			\tildefNplus + \tildepfNplus{t} (\tau-\tN) + \tildepfNplus{x} (x(\tau)-\tildexN) + \tildepfNplus{u} (u(\tau)-\tildeuNplus)
			\\ & \h
			+ \halftildepfNplus{tt} (\taumtn)^2 + \halftildepfNplus{tx} (\taumtn)\big(\tildexdNplus(\taumtn)+\halftildexddNplus(\taumtn)^2\big)
			\\ & \h\h
			+ \halftildepfNplus{tu} (\taumtn)\big(\tildeudNplus(\taumtn)+\halftildeuddNplus(\taumtn)^2\big)
			\\ & \h
			+ \halftildepfNplus{xt} \big(\tildexdNplus(\taumtn)+\halftildexddNplus(\taumtn)^2\big)(\taumtn)
			\\ & \h\h
			+ \halftildepfNplus{xx} \big(\tildexdNplus(\taumtn)+\halftildexddNplus(\taumtn)^2\big)^2
			\\ & \h\h
			+ \halftildepfNplus{xu} \big(\tildexdNplus(\taumtn)+\halftildexddNplus(\taumtn)^2\big)\big(\tildeudNplus(\taumtn)+\halftildeuddNplus(\taumtn)^2\big)
			\\ & \h
			+ \halftildepfNplus{ut} \big(\tildeudNplus(\taumtn)+\halftildeuddNplus(\taumtn)^2\big)(\taumtn)
			\\ & \h\h
			+ \halftildepfNplus{ux} \big(\tildeudNplus(\taumtn)+\halftildeuddNplus(\taumtn)^2\big)\big(\tildexdNplus(\taumtn)+\halftildexddNplus(\taumtn)^2\big)
			\\ & \h\h
			+ \halftildepfNplus{uu} \big(\tildeudNplus(\taumtn)+\halftildeuddNplus(\taumtn)^2\big)^2
			\\ & \h
			+ o\big((\taumtn)^2\big)
		\Big)
		d\tau
		\\
\end{array}
\renewcommand{\arraystretch}{1.0}
\end{equation}

\newpage 

Developping \eqref{eq:x_exact_4} and absorbing the terms in $(\taumtn)^k$ where $k \geqslant 3$ leads to \eqref{eq:x_exact_5}

\begin{equation}
\label{eq:x_exact_5}
\renewcommand{\arraystretch}{1.8}
\arraycolsep=2pt
\begin{array}{ll}
	x(t)
	= &
		\tildexN + \dsp{\int_{\tN}^{t}}
		\Big(
			\tildefNplus + \tildepfNplus{t} (\tau-\tN) + \tildepfNplus{x} (x(\tau)-\tildexN) + \tildepfNplus{u} (u(\tau)-\tildeuNplus)
			\\
	& \h
			+ \halftildepfNplus{tt} (\taumtn)^2 + \halftildepfNplus{tx} \tildexdNplus(\taumtn)^2 + \halftildepfNplus{tu} \tildeudNplus(\taumtn)^2
			\\
	& \h
			+ \halftildepfNplus{xt} \tildexdNplus(\taumtn)^2 + \halftildepfNplus{xx} (\tildexdNplus)^2(\taumtn)^2 + \halftildepfNplus{xu} \tildexdNplus\tildeudNplus(\taumtn)^2
			\\
	& \h
			+ \halftildepfNplus{ut} \tildeudNplus(\taumtn)^2 + \halftildepfNplus{ux} \tildeudNplus\tildexdNplus(\taumtn)^2 + \halftildepfNplus{uu} (\tildeudNplus)^2(\taumtn)^2
			\\
	& \h
			+ o\big((\taumtn)^2\big)
		d\tau
		\\
\end{array}
\renewcommand{\arraystretch}{1.0}
\end{equation}

Factorizing the elements of \eqref{eq:x_exact_5} leads to \eqref{eq:x_exact_6}.

\begin{equation}
\label{eq:x_exact_6}
\renewcommand{\arraystretch}{1.8}
\arraycolsep=2pt
\begin{array}{ll}
	x(t)
	= &
		\tildexN + \dsp{\int_{\tN}^{t}}
		\bigg(
			\tildefNplus + \tildepfNplus{t} (\tau-\tN) + \tildepfNplus{x} (x(\tau)-\tildexN) + \tildepfNplus{u} (u(\tau)-\tildeuNplus)
			\\
	& \h
			+ \half \Big(\tildepfNplus{tt} + \tildepfNplus{tx} \tildexdNplus + \tildepfNplus{tu} \tildeudNplus\Big)(\taumtn)^2
			\\
	& \h
			+ \half \Big(\tildepfNplus{xt} \tildexdNplus + \tildepfNplus{xx} (\tildexdNplus)^2 + \tildepfNplus{xu} \tildexdNplus\tildeudNplus\Big)(\taumtn)^2
			\\
	& \h
			+ \half \Big(\tildepfNplus{ut} \tildeudNplus + \tildepfNplus{ux} \tildeudNplus\tildexdNplus + \tildepfNplus{uu} (\tildeudNplus)^2\Big)(\taumtn)^2
			\\
	& \h
			+ o\big((\taumtn)^2\big)
		\bigg)
		d\tau
		\\
\end{array}
\renewcommand{\arraystretch}{1.0}
\end{equation}

We can now regroup terms of \eqref{eq:x_exact_6} so that we get the expression \eqref{eq:x_exact_7}.

\begin{equation}
\label{eq:x_exact_7}
\renewcommand{\arraystretch}{1.8}
\arraycolsep=2pt
\begin{array}{ll}
	x(t)
	= &
		\tildexN + \dsp{\int_{\tN}^{t}}
		\bigg(
			\tildefNplus + \tildepfNplus{t} (\tau-\tN) + \tildepfNplus{x} (x(\tau)-\tildexN) + \tildepfNplus{u} (u(\tau)-\tildeuNplus)
			\\
	& \h
			+
			\Big(
				\half \big(\tildepfNplus{tt} + \tildepfNplus{xx} (\tildexdNplus)^2 + \tildepfNplus{uu} (\tildeudNplus)^2 \big)
				\\ & \h\h
				+ \big(\tildepfNplus{xt} \tildexdNplus + \tildepfNplus{xu} \tildexdNplus\tildeudNplus + \tildepfNplus{ut} \tildeudNplus \big)
			\Big)
			\\
	& \h
			\cdot(\taumtn)^2
			+ o\big((\taumtn)^2\big)
		\bigg)
		d\tau
		\\
\end{array}
\renewcommand{\arraystretch}{1.0}
\end{equation}

Expression \eqref{eq:x_exact_7} can be written in the clearer way \eqref{eq:x_exact_8} with a term $\DeltaN$ defined in \eqref{eq:DeltaN}.

\begin{equation}
\label{eq:x_exact_8}
\renewcommand{\arraystretch}{1.8}
\arraycolsep=2pt
\begin{array}{ll}
	x(t)
	= &
		\tildexN + \dsp{\int_{\tN}^{t}}
		\bigg(
			\tildefNplus + \tildepfNplus{t} (\tau-\tN) + \tildepfNplus{x} (x(\tau)-\tildexN) + \tildepfNplus{u} (u(\tau)-\tildeuNplus)
			\\
	& \h
			+
			\DeltaN (\taumtn)^2
			+ o\big((\taumtn)^2\big)
		\bigg)
		d\tau
		\\
\end{array}
\renewcommand{\arraystretch}{1.0}
\end{equation}

\begin{equation}
\label{eq:DeltaN}
\renewcommand{\arraystretch}{1.8}
\arraycolsep=2pt
\begin{array}{ll}
  \DeltaN
  = &
    \half \big(\tildepfNplus{tt} + \tildepfNplus{xx} (\tildexdNplus)^2 + \tildepfNplus{uu} (\tildeudNplus)^2 \big)
    \\ &
    + \big(\tildepfNplus{xt} \tildexdNplus + \tildepfNplus{xu} \tildexdNplus\tildeudNplus + \tildepfNplus{ut} \tildeudNplus \big)
\end{array}
\renewcommand{\arraystretch}{1.0}
\end{equation}

On the other side, the states estimated by the COSTARICA estimator are computed by solving the linearized system. Their theoretical expression is thus \eqref{eq:xL}.

\begin{equation}
\label{eq:xL}
x_L(t)
=
\tildexN + \dsp{\int_{\tN}^{t}} \tildefNplus + \tildepfNplus{x} (x_L(\tau) - \tildexN) + \tildepfNplus{u} (u(\tau) - \tildeuNplus) d\tau
\end{equation}

The error between the estimator \eqref{eq:xL} and the theoretical states (using the full system instead of a linearization) \eqref{eq:x_exact_8} is denoted as $e$ and its expression is given in \eqref{eq:e}.

\begin{equation}
\label{eq:e}
\renewcommand{\arraystretch}{2.0}
\arraycolsep=2pt
\begin{array}{ll}
	e(t)
	& = x(t) - x_L(t)
	\\
	& =
		\dsp{\int_{\tN}^{t}}
		\bigg(
			\tildefNplus + \tildepfNplus{t} (\tau-\tN) + \tildepfNplus{x} (x(\tau)-\tildexN) + \tildepfNplus{u} (u(\tau)-\tildeuNplus)
			\\
	& \h
			+
			\DeltaN (\taumtn)^2
			+ o\big((\taumtn)^2\big)
			\\
	& \h
			-
			\Big(
				\tildefNplus + \tildepfNplus{x} (x_L(\tau) - \tildexN) + \tildepfNplus{u} (u(\tau) - \tildeuNplus)
			\Big)
		\bigg)
		d\tau
		\\
	& =		
		\dsp{\int_{\tN}^{t}}
		\bigg(
			\tildepfNplus{t} (\tau-\tN) + \tildepfNplus{x} (x(\tau)-x_L(\tau))
			+
			\DeltaN (\taumtn)^2
			+ o\big((\taumtn)^2\big)
		\bigg)
		d\tau
		\\
	& =
		\dsp{\int_{\tN}^{t}}
		\bigg(
			\tildepfNplus{t} (\tau-\tN) + \tildepfNplus{x} e(\tau)
			+
			\DeltaN (\taumtn)^2
			+ o\big((\taumtn)^2\big)
		\bigg)
		d\tau
		\\
\end{array}
\renewcommand{\arraystretch}{1.0}
\end{equation}

The value of $e$ at $\tN$ is zero. Indeed, \eqref{eq:e} clearly shows this (the integral becomes null as the real quantity under it is integrated on a punctual set if $t=\tN$), and it can also been seen as both the states and the estimated states start with the same value at $\tN$ \eqref{eq:e_iv}.

\begin{equation}
\label{eq:e_iv}
e(\tN) = x(\tN)-x_L(\tN) = \tildexN - \tildexN = 0
\end{equation}

From \eqref{eq:e} and \eqref{eq:e_iv}, we can write the ODE \eqref{eq:ed}.

\begin{equation}
\label{eq:ed}
\renewcommand{\arraystretch}{1.2}
\left\{
\begin{array}{lcl}
	\ed(t)
	& = &
		\tildepfNplus{t} (t-\tN) + \tildepfNplus{x} e(t)
		+
		\DeltaN (t-\tN)^2
		+ o\big((t-\tN)^2\big)
	\\
	e(\tN) & = & 0 \\
\end{array}
\right.
\renewcommand{\arraystretch}{1.0}
\end{equation}

To study the error, let's consider a scalar version of \eqref{eq:ed}. The scalar analog of $\tildepfNplus{x}$ is denoted by $\lambda$ (one may consider that it is a relevant eigenvalue of the Jacobian $\tildepfNplus{x}$ of the system), the scalar analog of $\tildepfNplus{t}$ is denoted by $\sigma$, the scalar analog of $\DeltaN$ is denoted as $\Delta$, and the scalar analog of $e$ is denoted as $\varepsilon$. Finally, the scalar equivalent of ODE \eqref{eq:ed} is \eqref{eq:ed_scal}.

\begin{equation}
\label{eq:ed_scal}
\renewcommand{\arraystretch}{1.2}
\left\{
\begin{array}{lcl}
	\dot{\varepsilon}(t)
	& = &
	\lambda
	\varepsilon(t)
	+ \sigma (t-\tN)
	+ \Delta (t-\tN)^2
	+ o\big((t-\tN)^2\big)
	\\
	\varepsilon(\tN) & = & 0 \\
\end{array}
\right.
\renewcommand{\arraystretch}{1.0}
\end{equation}

To simplify the study, let's introduce the affine change of variable \eqref{eq:eps_def_var_change}.

\begin{equation}
\label{eq:eps_def_var_change}
\epsilon : \tau \mapsto \varepsilon(\tau + \tN)
\end{equation}

As $t\in[\tN, \tNp[$, we have $\tau\in [0, \dtN[$ with $\tau=t-\tN$. Hence, we can write $\varepsilon$ in terms of $\epsilon$ and a similar relation for the derivative as the change of variable is affine, see \eqref{eq:eps_prop_var_change}.

\begin{equation}
\label{eq:eps_prop_var_change}
\arraycolsep=2pt
\begin{array}{lcl}
  \varepsilon(\tau) & = & \epsilon(\tau - \tN) \\
  \dot{\varepsilon}(\tau) & = & \dot{\epsilon}(\tau - \tN) \\
\end{array}
\end{equation}

The ODE \eqref{eq:ed_scal} can be rewritten in terms of the unknown $\epsilon$ \eqref{eq:ed_scal_var_change_t}.

\begin{equation}
\label{eq:ed_scal_var_change_t}
\renewcommand{\arraystretch}{1.2}
\left\{
\begin{array}{lcl}
	\dot{\epsilon}(t-\tN)
	& = &
	\lambda
	\epsilon(t-\tN)
	+ \sigma (t-\tN)
	+ \Delta (t-\tN)^2
	+ o\big((t-\tN)^2\big)
	\\
	\epsilon(0) & = & 0 \\
\end{array}
\right.
\renewcommand{\arraystretch}{1.0}
\end{equation}

\newpage 

Moreover, we can use the new variable $\tau$ (affine time-shift of $t$) to write \eqref{eq:ed_scal_var_change_tau}.

\begin{equation}
\label{eq:ed_scal_var_change_tau}
\renewcommand{\arraystretch}{1.2}
\left\{
\begin{array}{lcl}
	\dot{\epsilon}(\tau)
	& = &
	\lambda
	\epsilon(\tau)
	+ \sigma \tau
	+ \Delta\ \tau^2
	+ o(\tau^2)
	\\
	\epsilon(0) & = & 0 \\
\end{array}
\right.
\renewcommand{\arraystretch}{1.0}
\end{equation}

Equation \eqref{eq:ed_scal_var_change_tau} can be solved by removing its asymptotic error term and the solution is \eqref{eq:epsilon_final_solution}.

\begin{equation}
\label{eq:epsilon_final_solution}
\epsilon:\tau\mapsto \frac{-1}{\lambda^3}
\Big(
	\lambda^2 \Delta\ \tau^2
	+
	(\lambda \sigma + 2 \Delta) (-\exp(\lambda \tau)+\lambda \tau + 1)
\Big)
\end{equation}

Around $\tau=0$, we can write the limited development \eqref{eq:DT_epsilon} of the solution \eqref{eq:epsilon_final_solution}.

\begin{equation}
\label{eq:DT_epsilon}
\renewcommand{\arraystretch}{2.2}
\begin{array}{ll}
	\epsilon(\tau)
	& =
		\dfrac{-1}{\lambda^3}
		\Bigg(
			\lambda^2 \Delta\ \tau^2
			+
			(\lambda \sigma + 2 \Delta) \bigg(-\Big(1+\lambda \tau + \dfrac{\lambda^2 \tau^2}{2} + \dfrac{\lambda^3 \tau^3}{6} + o(\tau^3)\Big)+\lambda \tau + 1\bigg)
		\Bigg)
		\\
	& =
		\dfrac{-1}{\lambda^3}
		\Bigg(
			\lambda^2 \Delta\ \tau^2
			+
			(\lambda \sigma + 2 \Delta) \bigg(- \dfrac{\lambda^2 \tau^2}{2} - \dfrac{\lambda^3 \tau^3}{6} + o(\tau^3)\bigg)
		\Bigg)
		\\
	& =
		\dfrac{-1}{\lambda^3}
		\Bigg(
			\lambda^2 \Delta\ \tau^2
			+
			\lambda \sigma \bigg(- \dfrac{\lambda^2 \tau^2}{2} - \dfrac{\lambda^3 \tau^3}{6} + o(\tau^3)\bigg)
			+
			2 \Delta \bigg(- \dfrac{\lambda^2 \tau^2}{2} - \dfrac{\lambda^3 \tau^3}{6} + o(\tau^3)\bigg)
		\Bigg)
		\\
	& =
		\dfrac{-1}{\lambda^3}
		\Bigg(
			\lambda^2 \Delta\ \tau^2
			-
			\dfrac{\lambda^3 \sigma \tau^2}{2}
			-
			\dfrac{\lambda^4 \sigma \tau^3}{6}
			-
			\dfrac{2 \Delta\lambda^2 \tau^2}{2}
			-
			\dfrac{2 \Delta\lambda^3 \tau^3}{6}
			+
			o(\tau^3)
		\Bigg)
		\\
	& =
		\dfrac{- \Delta\ \tau^2}{\lambda}
		+
		\dfrac{\sigma \tau^2}{2} + \dfrac{\lambda \sigma \tau^3}{6}
		+
		\dfrac{2 \Delta\ \tau^2}{2 \lambda} + \dfrac{2 \Delta\ \tau^3}{6}
		+ o(\tau^3)
		\\
	& =
		\sigma \dfrac{\tau^2}{2}
		+
		\left(
			\lambda\sigma + 2 \Delta
		\right)
		\dfrac{\tau^3}{6}
		+ o(\tau^3)
		\\
\end{array}
\renewcommand{\arraystretch}{1.0}
\end{equation}

The last line of \eqref{eq:DT_epsilon} is the result we are interested in, here. Indeed, is shows that we can expect \textbf{an error of order two or more} between the states of the linearized system and the states of the full system. A closer look at the final expression of $\epsilon$ in \eqref{eq:DT_epsilon} even shows that this error is \textbf{of order three or more when the derivatives function of the system does not explicitly depend on the time variable}. Indeed, $\sigma$ was defined as the scalar version of $\tildepfNplus{t}$, which means that a null sigma is implied by a null partial derivative in terms of time. This can be locally true if $f$ only has a term in $t^2$, for instance, but it is clearly implied if $t$ does not even appear in the expression of the function. Please note that this result is namely a consequence of the fact that the linearization \eqref{eq:Lin_on_a_step}, on which the COSTARICA estimator is based, takes into account the partial derivatives in the directions of the states and the input but not the time. This limitation is directly related to the industrial limitation of available partial derivatives in the systems (see table \ref{table:capabilities_related_to_COSTARICA} in the first chapter).

Higher orders can be reached with $\sigma=0$ and the extra condition $\Delta=0$ that is far from being straightforward to be verified, in practice. However, we can notice the particular case of a state-space system where $f$ is of the form $f:t,x,u\mapsto Ax+Bu$ for given $A$ and $B$ matrices. In this case, the linearization \eqref{eq:Lin_on_a_step} is exact and we expect to have no error. Is is consistent with our analysis as, in this case, the partial derivative of $f$ is null in the direction of time, and $\DeltaN$, which scalar version is $\Delta$, is null as well as $f$ has all its second order derivatives being null (see definition of $\DeltaN$ \eqref{eq:DeltaN}: it is a sum of second order derivatives of $f$).

\newpage 

\subsection{Corroborative empirical observations}
\label{subsection:Corroborative_empirical_observations}

In order to illustrate the error order determined in \ref{subsection:Error_analysis_regarding_the_state_response}, empirical observations of error across macro-step size are made here. The measured errors will all be given by comparing the states at the end \underline{of a single} macro-step in this section. For errors over a whole co-simulation, section \ref{section:examples_and_test_cases} will treat entire co-simulations on test-cases.

First, the models will be chosen and described in \ref{subsubsection:Model_choice_and_design} in order to highlight the result found in \ref{subsection:Error_analysis_regarding_the_state_response}. Then, the difference of behavior between a non-linear system and its linearized version will be shown in \ref{paragraph:Ideal_error_due_to_the_linearization_only}; and when the linearized version is not computed with an integration but estimated (over the macro-step) with the COSTARICA estimator, the difference between this estimation and the integration of the non-linear system is presented in \ref{paragraph:Practical_error_on_the_linearization_based_numerical_estimation}.

\subsubsection{Model choice and design}
\label{subsubsection:Model_choice_and_design}

As shown in \ref{subsubsection:Theoretical_error_order_due_to_the_linearization}, the error order depends on the $f$ function of a system: in case the latter does not depend on the time variable, the error order increases. Hence, the academic Lotka-Volterra model \cite{Volterra1928} will be used in a decoupled version where each species is represented by one system, and a modified version with a time-dependency on $f$ will be introduce to compare the error in these two approaches.

\paragraph{Classical Lotka-Volterra}
\label{paragraph:Classical_Lotka_Volterra}

The classical Lotka-Volterra test-case will be used as non-linear test model here. Figure \ref{fig:analysis_classical_LV_isolating_prey} shows how the prey system is extracted to be the system on which the observations will be made. 

\begin{center}
\includegraphics[scale=0.48]{\figuresdir/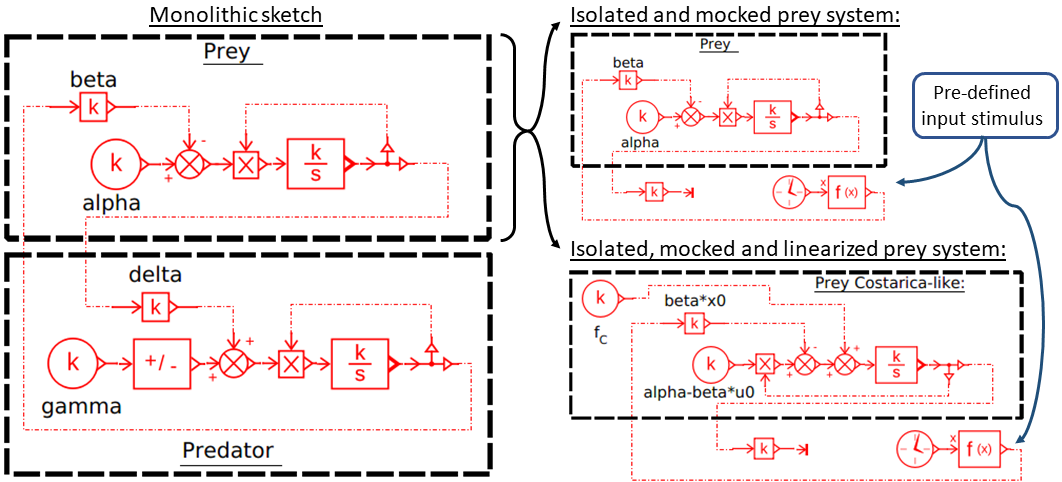}
\vspace{-2mm} 
\captionof{figure}{Isolation of the prey system in a classical Lotka-Volterra model in Simcenter Amesim}
\label{fig:analysis_classical_LV_isolating_prey}
\end{center}

The equation of the isolated prey system is given in \eqref{eq:isolated_prey_classical_LV} where the state is also the output, corresponding to the amount of prey.

\begin{equation}
\label{eq:isolated_prey_classical_LV}
\begin{array}{lcl}
  \dot{x}(t) & = & x(t) (\alpha - \beta u(t)) \\
  y(t) & = & x(t) \\
\end{array}
\end{equation}

In \eqref{eq:isolated_prey_classical_LV}, we observe that the underlying function $f$ of the system is $f:t,x,u\mapsto x(\alpha - \beta u)$ and \textbf{does not depend on the time}. Parameters $\alpha$ and $\beta$ are real fixed parameters. The input $u$ is expected to be the amount of predators.

The linearized version of \eqref{eq:isolated_prey_classical_LV} around $\tinit$ in the way linearizations are done in COSTARICA, that is to say with repect to $x$ and $u$ but not $t$ (see \eqref{eq:Lin_on_a_step}), is presented in \eqref{eq:isolated_prey_classical_LV_linearized}.

\begin{equation}
\label{eq:isolated_prey_classical_LV_linearized}
\begin{array}{lcl}
  \dot{x}(t) & = & (\alpha - \beta u(\tinit)) x(t) - \beta x(\tinit) u(t) + \beta x(\tinit) u(\tinit) \\
  y(t) & = & x(t) \\
\end{array}
\end{equation}

We want to study the error on the prey system only, so our tests will feed both systems \eqref{eq:isolated_prey_classical_LV} and \eqref{eq:isolated_prey_classical_LV_linearized} with a stimulus signal as input that corresponds to a polynomial approximation of the amount of predator obtained by a preliminary monolithic simulation, as shown on figure \ref{fig:analysis_classical_LV_stimulus}.

\begin{center}
\includegraphics[scale=0.53]{\figuresdir/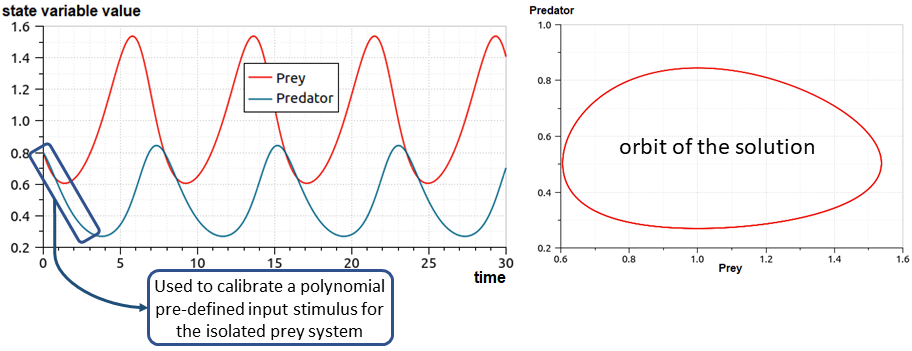}
\vspace{-2mm} 
\captionof{figure}{Prelimirary monolithic simulation of the classical Lotka-Volterra model to determine the predator stimulus for prey system}
\label{fig:analysis_classical_LV_stimulus}
\end{center}

The parameters of the model and corresponding input predator stimulus signal for the prey system are given in table \ref{tab:isolated_prey_classical_LV_params}, where namely $x(\tinit)$ denotes the initial amount of prey, and $u(\tinit)$ the initial amount of predators. The input predator stimulus signal $u$ for the prey system is also given in table \ref{tab:isolated_prey_classical_LV_params}.

\begin{center}
\captionof{table}{Parameter set of the classical Lotka-Volterra model}
\label{tab:isolated_prey_classical_LV_params}
\renewcommand{\arraystretch}{1.3}
\begin{tabularx}{8cm}{ |>{\centering\arraybackslash}X|>{\centering\arraybackslash}X| }
  \hline
  \multicolumn{2}{|c|}
  {
    Full model:
  } \\
  \hline
  \hline
  $\alpha = 0.67$ & $\beta = 1.33$ \\
  \hline
  $\gamma = 1$ & $\delta = 1$ \\
  \hline
  \hline
  \multicolumn{2}{|c|}
  {
    Isolated prey:
  } \\
  \hline
  \hline
  $x(\tinit) = 0.8$ & $u(\tinit) = 0.8$
  \\
  \hline
  \multicolumn{2}{|c|}
  {
    Input stimulus $u:t\mapsto 0.8 - 0.16 \ t - 0.11008 \ t^2$
  } \\
  \hline
\end{tabularx}
\renewcommand{\arraystretch}{1.0}
\end{center}

\paragraph{Lotka-Volterra with time dependency}
\label{paragraph:Lotka_Volterra_with_time_dependency}

A modification of the classical Lotka-Volterra test-case will be used to introduce a dependency of the $f$ function of the prey system to the time. The interactions between the prey and the predator are, in this modified Lotka-Volterra model, conditioned by the time of the day. The non-linear terms are multiplied by a sinusoid signal $s(t)$ going from $10\%$ to $100\%$ with a period of $2.4$s (which makes it analogous, with a second-to-ten-hours change of variable, to a full day). As we study the linearization at the initial time, this sinusoid signal is considered as having no phase delay (so that its time-derivative is non-null at $t=0$). Figure \ref{fig:analysis_LV_with_time_isolating_prey} shows this modification on the model.

\begin{center}
\includegraphics[scale=0.45]{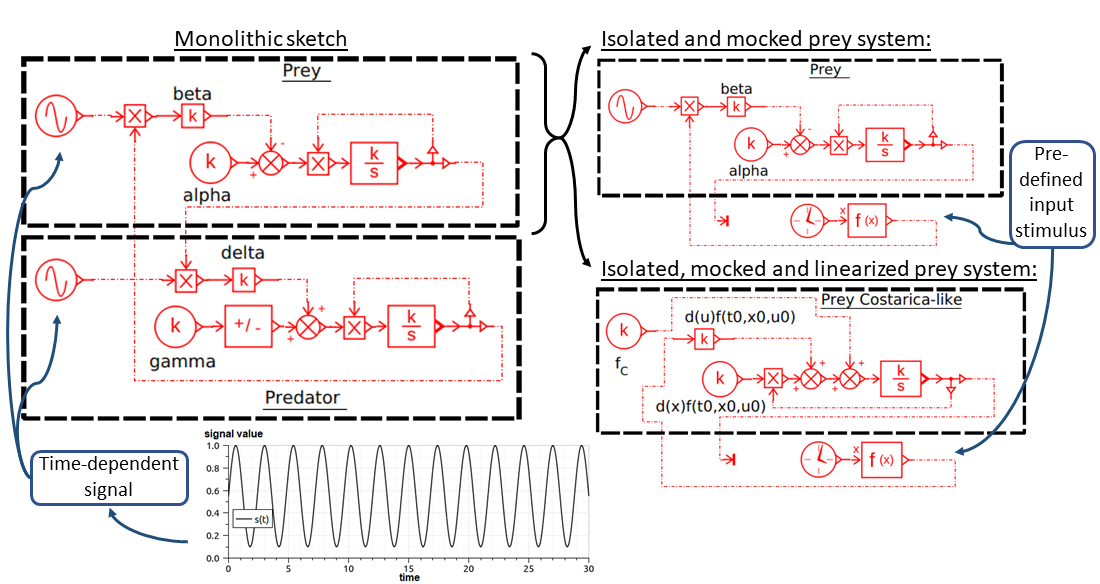}
\vspace{-2mm} 
\captionof{figure}{Isolation of the prey system in a modified Lotka-Volterra model in Simcenter Amesim}
\label{fig:analysis_LV_with_time_isolating_prey}
\end{center}

The equation of the isolated prey system is given in \eqref{eq:isolated_prey_LV_with_time} where the state is also the output, corresponding to the amount of prey.

\begin{equation}
\label{eq:isolated_prey_LV_with_time}
\begin{array}{lcl}
  \dot{x}(t) & = & x(t) (\alpha - s(t)\beta u(t)) \\
  y(t) & = & x(t) \\
\end{array}
\end{equation}

In \eqref{eq:isolated_prey_LV_with_time}, we observe that the underlying function $f$ of the system is $f:t,x,u\mapsto x(\alpha - s(t) \beta u)$ and \textbf{does depend on the time}, contrary to the one of the prey system of the classical Lotka-Volterra model. Parameters $\alpha$ and $\beta$ are the same real fixed parameters as in the non-modified model. The input $u$ is still expected to be the amount of predators, however this input is not the same than in the previous case as far as the model modification affects the monolithic simulation as well.

The linearized version of \eqref{eq:isolated_prey_LV_with_time} around $\tinit$, still in the way linearizations are done in COSTARICA, is presented in \eqref{eq:isolated_prey_LV_with_time_linearized}.

\begin{equation}
\label{eq:isolated_prey_LV_with_time_linearized}
\begin{array}{lcl}
  \dot{x}(t) & = & (\alpha - s(\tinit) \beta u(\tinit)) x(t) - \beta s(\tinit) x(\tinit) u(t) + \beta s(\tinit) x(\tinit) u(\tinit) \\
  y(t) & = & x(t) \\
\end{array}
\end{equation}

The stimulus input signal aiming at feeding both \eqref{eq:isolated_prey_classical_LV} and \eqref{eq:isolated_prey_classical_LV_linearized} and coming from a preliminary monolithic simulation differs from the prvious case. Figure \ref{fig:analysis_classical_LV_stimulus} shows the results of the modified Lotka-Volterra model.

\begin{center}
\includegraphics[scale=0.55]{\figuresdir/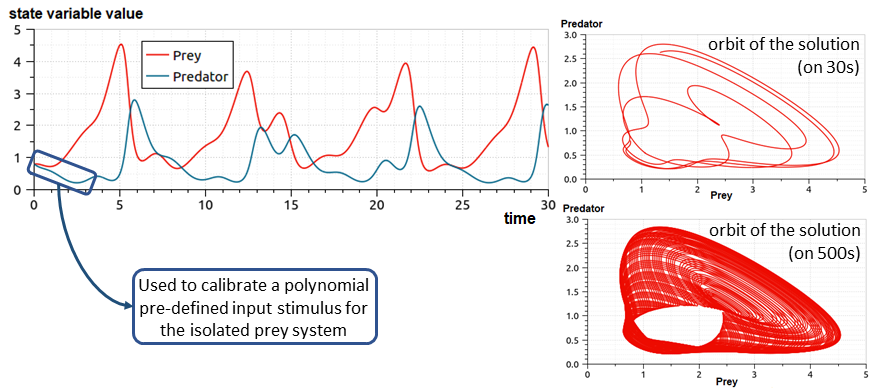}
\vspace{-2mm} 
\captionof{figure}{Prelimirary monolithic simulation of the modified Lotka-Volterra model to determine the predator stimulus for prey system}
\label{fig:analysis_LV_with_time_stimulus}
\end{center}

The parameters of the model (including the sinusoid signal $s$) and corresponding input predator stimulus signal $u$ for the prey system are given in table \ref{tab:isolated_prey_LV_with_time_params}.

\begin{center}
\captionof{table}{Parameter set of the modified Lotka-Volterra model}
\label{tab:isolated_prey_LV_with_time_params}
\renewcommand{\arraystretch}{1.3}
\begin{tabularx}{9cm}{ |>{\centering\arraybackslash}X|>{\centering\arraybackslash}X| }
  \hline
  \multicolumn{2}{|c|}
  {
    Full model:
  } \\
  \hline
  \hline
  $\alpha = 0.67$ & $\beta = 1.33$ \\
  \hline
  $\gamma = 1$ & $\delta = 1$ \\
  \hline
  \multicolumn{2}{|c|}
  {
    $s:t\mapsto 0.55 + 0.45\ \sin(2 \pi t / 2.4)$
  }
  \\
  \hline
  \hline
  \multicolumn{2}{|c|}
  {
    Isolated prey:
  } \\
  \hline
  \hline
  $x(\tinit) = 0.8$ & $u(\tinit) = 0.8$ \\
  \hline
  \multicolumn{2}{|c|}
  {
    Input stimulus $u:t\mapsto 0.8 - 0.448 \ t + 0.5173559185 \ t^2$
  } \\
  \hline
\end{tabularx}
\renewcommand{\arraystretch}{1.0}
\end{center}

\subsubsection{Observed error order}
\label{subsubsection:Observed_error_order}

\paragraph{Ideal error due to the linearization only}
\label{paragraph:Ideal_error_due_to_the_linearization_only}

The ideal error is the difference between the simulation realized with the non-linear version of the isolated prey systems and the linearization in terms of $x$ and $u$. In other words, it would be the error in the response to a given stimulus over a single macro-step \textit{in case the only approximation made were to use the linearized system instead of the non-linear real one}.

This error is expected to be of order $3$ and $2$ for the classical Lotka-Volterra model and its modified version, respectively (object of the analysis in \ref{subsection:Error_analysis_regarding_the_state_response}). Hence, four simulations of the isolated and mocked systems of figure \ref{fig:analysis_classical_LV_isolating_prey} and \ref{fig:analysis_LV_with_time_isolating_prey} have been conducted:

\begin{enumerate}
\item the mocked prey isolated from the classical Lotka-Volterra model,
\item the linearization (in $x$ and $u$ only) of the mocked prey isolated from the classical Lotka-Volterra model,
\item the mocked prey isolated from the modified Lotka-Volterra model, and
\item the linearization (in $x$ and $u$ only) of the mocked prey isolated from the modified Lotka-Volterra model.
\end{enumerate}

In each modeling version (classical and modified Lotka-Volterra), the difference between the non-linear and linear versions of the mocked prey system, submitted to the same predator stimulus, have been computed at each micro-step. These results, shown on figure \ref{fig:analysis_Ideal_error_dueTo_Lin}, show the error we could expect if the macro-step size were of the size of the corresponding ascissa. A micro-step size of $10^{-5}$ has been chosen here. The error a macro-step of size $\dt$ would be, in case the only approximation made were to use the linearized system instead of the non-linear real one, the y-value of the point in the corresponding curve at the abscissa $\dt$ in figure \ref{fig:analysis_Ideal_error_dueTo_Lin}. On the latter, "LV with time" denotes the modified Lotka-Volterra.

Figure \ref{fig:analysis_Ideal_error_dueTo_Lin} clearly shows that the error has the expected order: quadratic in the modified Lotka-Volterra case, due to the dependency of $f$ on the time variable, and cubic in the classical Lotka-Volterra case as the independency of $f$ on the time variable cancels the order $2$ term in the error asymptotic expression \eqref{eq:DT_epsilon}.

This error shown in figure \ref{fig:analysis_Ideal_error_dueTo_Lin} is \textit{ideal} in the sense that it characterizes the error due to the linearization process only. However, over a macro-step estimated with COSTARICA, numerical calculations are made (numerical Laplace inverse, numerical computation of the $G$, $P$ and $R$ matrices using Misra \& Patel method \cite{Misra1987}, including numerical Hessenberg decomposition, \etc). Therefore, a similar error observation with numerical COSTARICA estimations have to be conducted in order to measure a \textit{practical} error.

\begin{center}
\includegraphics[scale=0.55]{\figuresdir/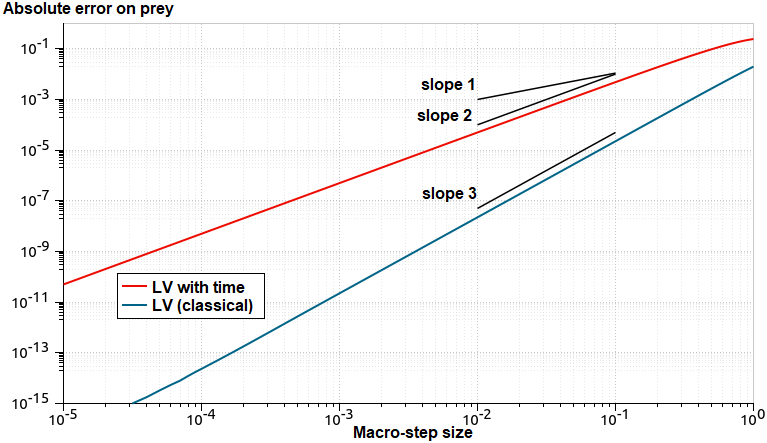}
\captionof{figure}{Error between the non-linear and linearized isolated preys for the two Lotka-Volterra models}
\label{fig:analysis_Ideal_error_dueTo_Lin}
\end{center}

\paragraph{Practical error on the linearization-based numerical estimation}
\label{paragraph:Practical_error_on_the_linearization_based_numerical_estimation}

The process to measure the \textit{practical} error is slightly different than simulations comparisons used to highlight the ideal error. Indeed, instead of conducting four simulations and comparing the results at each micro-step, we want to get the error of the COSTARICA estimation over macro-steps of various sizes. Hence, one COSTARICA estimation is realized for each macro-step size and compared to the real integration of the corresponding isolated prey in its real non-linear form.

The difference of amount of prey at the end of the macro-steps have been compared and the results for different macro-step sizes are presented in figure \ref{fig:analysis_Error_on_COSTARICA_on_prey_of_classical_LV} for the classical Lotka-Volterra case, and on figure \ref{fig:analysis_Error_on_COSTARICA_on_prey_of_LV_with_time} for the modified Lotka-Volterra case.

\begin{center}
\includegraphics[scale=0.35]{\figuresdir/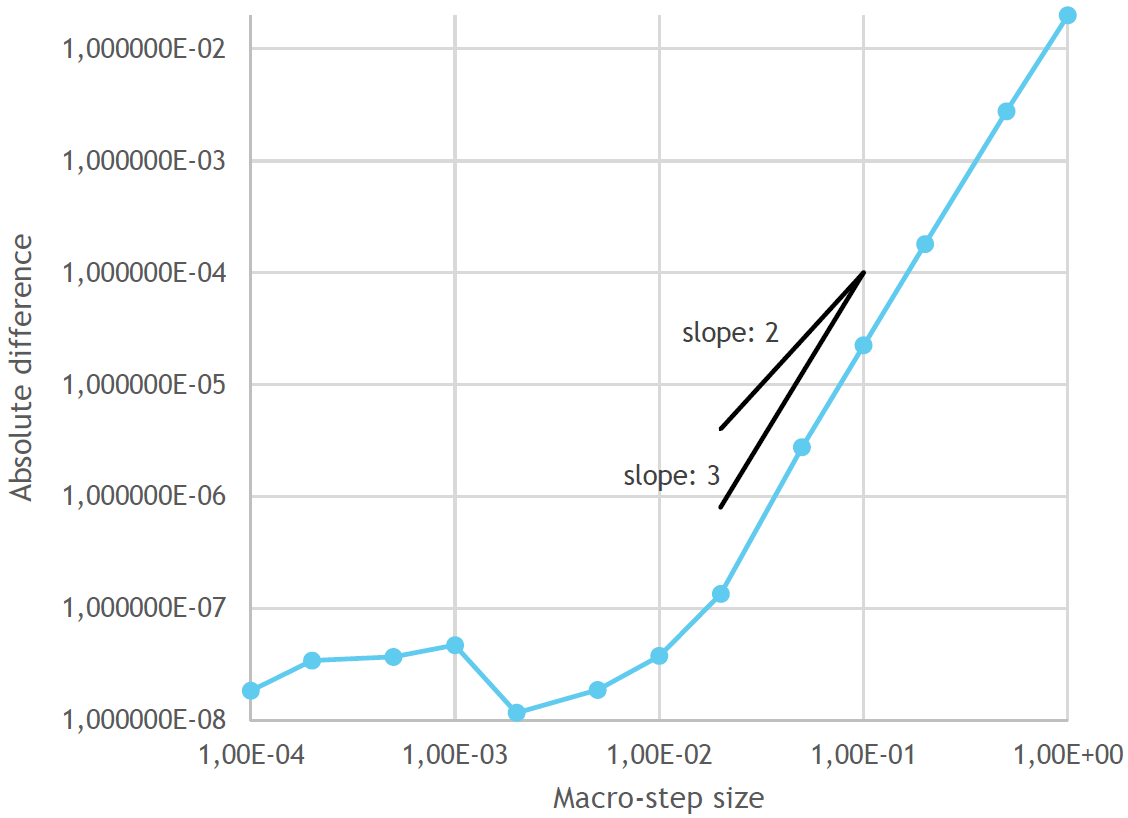}
\captionof{figure}{Error on numerical COSTARICA estimation or the prey of a classical Lotka-Volterra model}
\label{fig:analysis_Error_on_COSTARICA_on_prey_of_classical_LV}
\end{center}

\begin{center}
\includegraphics[scale=0.35]{\figuresdir/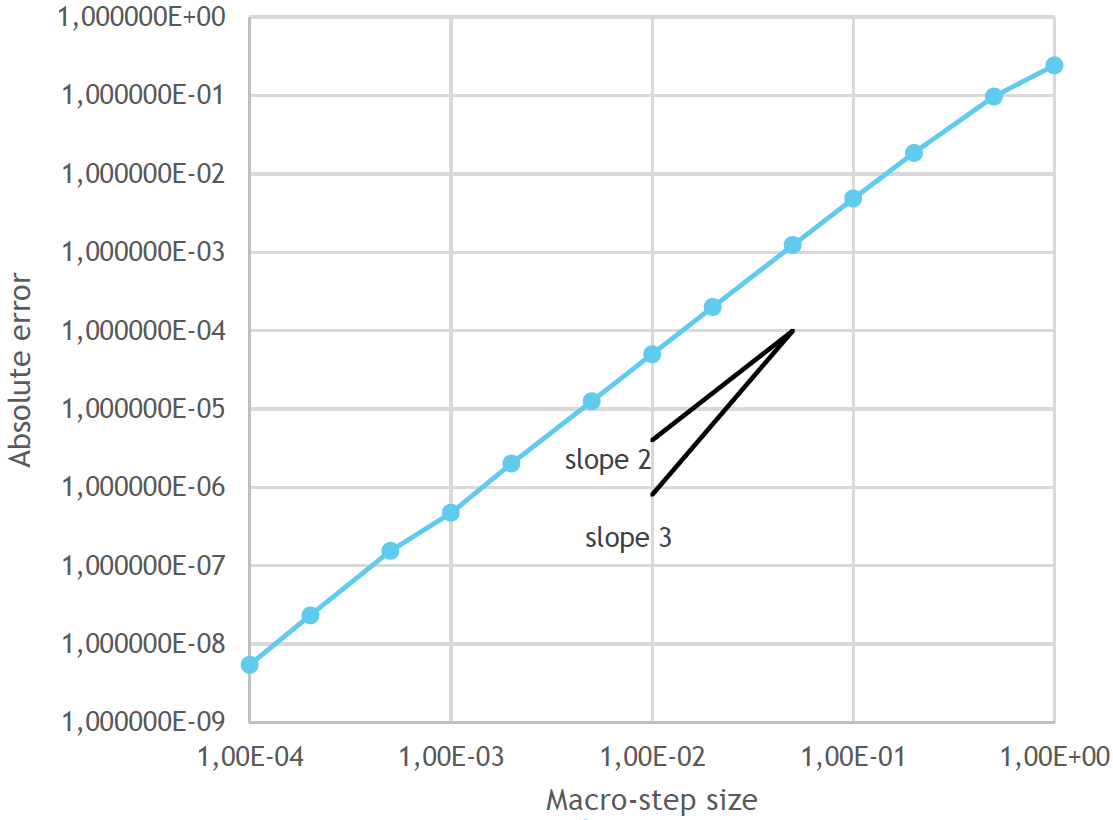}
\captionof{figure}{Error on numerical COSTARICA estimation or the prey of a Lotka-Volterra-with-time model}
\label{fig:analysis_Error_on_COSTARICA_on_prey_of_LV_with_time}
\end{center}

The theoretical error orders of $3$ for the classical Lotka-Volterra case and $2$ for the modified Lotka-Volterra case are matched by the measurements presented on figures \ref{fig:analysis_Error_on_COSTARICA_on_prey_of_classical_LV} and \ref{fig:analysis_Error_on_COSTARICA_on_prey_of_LV_with_time}, respectively. This means that the numerical processes involved in the COSTARICA estimator computation do preserve the error order driven by the linearization process. However, when the macro-step becomes too small, these numerical effects can still deteriorate the error decreasing, see namely the left of the curve of figure \ref{fig:analysis_Error_on_COSTARICA_on_prey_of_classical_LV}.

\section{Examples and test cases}
\label{section:examples_and_test_cases}

This section presents the use of COSTARICA on concrete cases. First of all, a simple generic equation is used to show the behavior of COSTARICA on a tough case where the linearization is not representative of the system. Then, co-simulations on modular models will be presented. The iterative co-simulation algorithm on which the COSTARICA process is injected is IFOSMONDI-JFM \cite{Eguillon2021IfosmondiJFM}.

One of these modular models is a simple mechanical model made of two systems connected by a force and velocity/displacement coupling. This model has been introduced in \cite{Eguillon2022F3ornits} and is strongly inspired by test-cases in \cite{Busch2016} and \cite{Meyer2021}, and it enables us to show the different terms of the estimators. It will be presented in subsection \ref{subsection:mechanical_bodies_with_behavior_change}. The second modular model, presented in subsection \ref{subsection:lotka_volterra_predation}, consists in the non-linear Lotka-Volterra model \cite{Volterra1928} decoupled in two non-linear systems. It is the co-simulation model (modular system) corresponding to the one from which the prey system has been isolated for the analysis in \ref{paragraph:Classical_Lotka_Volterra}.

For the sake of reproducibility, the equations of the latter models will be detailed.

\textbf{Important remark 1:} please note that, for the modular models presented in \ref{subsection:mechanical_bodies_with_behavior_change} and \ref{subsection:lotka_volterra_predation}, the inputs $u(t)$ of a system corresponds to the outputs $y(t)$ of the other system and vice-versa (the output of a system is connected to the input of the other one). The formalism of this paper was centered around one co-simulation system, however the modular models are made of several such systems connected to one another. In order to remove ambiguity, indices have been introduced on $u$, $y$ and $x$ quantities in subsections \ref{subsection:mechanical_bodies_with_behavior_change} and \ref{subsection:lotka_volterra_predation}.

\textbf{Important remark 2:} regarding the modular models of subsections \ref{subsection:mechanical_bodies_with_behavior_change} and \ref{subsection:lotka_volterra_predation}, co-simulations with IFOSMONDI-JFM using rollback and with replacement of the rollback by the COSTARICA process have been run and compared to the monolithic system (acting as reference results) simulated with Simcenter Amesim. Please note that, in the context of industrial cases, such monolithic references can usually not be assembled as the systems are black-boxes. The models presented in this paper have been designed on purpose so that error measurements can be done.

\subsection{Tough case: time-only dependent terms}
\label{subsection:tough_case_time_only_dependent_terms}

The COSTARICA estimator is based on the linearization of the considered system. The assumption is made that, in case the system is not a linear ODE, it behaves similarly to its linearization in a neighborhood of the currently reached time. The co-simulation step size must therefore be small enough to stay in an acceptable neighborhood.

An example of model involving such non-linear systems is presented further in \ref{subsection:lotka_volterra_predation}. However, tough cases can arise from very simple systems without non-linearities: namely when the ODE contains, among others, a term that is completely independent of the state variable $x$ and the input $u$.

Let $a$ and $b$ be real scalar functions of times. Let's consider the simple following system:

\begin{equation}
\label{eq:test_case_TimeOnlyTerm_ODE}
\left\{
	\begin{array}{lcl}
		\dspfrac{dx}{dt} & = & a(t) \\
		y & = & x + b(t)
	\end{array}
\right.
\end{equation}

This system is not sensitive to any input. We can either consider the system as inputs-less (with $\nin=0$), or as having unused inputs (with $\nin=1$ for instance) to avoid degenerated matrices. Fortunately, these two choices will generate the same results.

Let's consider the case where there is one unused input $\nin=1$, so that we can detail here all the involved matrices. An equivalent version of system \eqref{eq:test_case_TimeOnlyTerm_ODE} is:

\begin{equation}
\label{eq:test_case_TimeOnlyTerm_ODE_v2}
\left\{
	\begin{array}{lcl}
		\dspfrac{dx}{dt} & = & 0 \cdot x + 0 \cdot u + a(t) \\
		y & = & 1\cdot x + 0 \cdot u + b(t)
	\end{array}
\right.
\end{equation}

This system can be sketched in a modelling and simulation software: an example is shown on figure \ref{fig:test_case_TimeOnlyTerm_sketch}.

\begin{center}
\includegraphics[scale=0.4]{\figuresdir/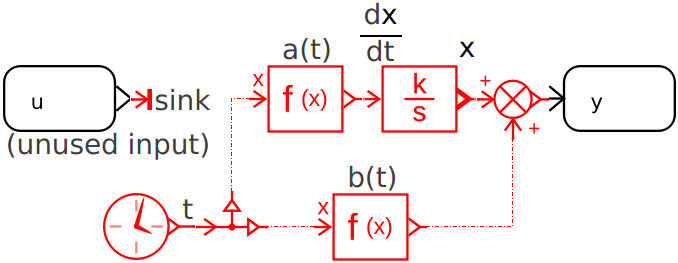}
\captionof{figure}{Tough system with time-only dependent terms in Simcenter Amesim}
\label{fig:test_case_TimeOnlyTerm_sketch}
\end{center}

At any point in time, the linearization of \eqref{eq:test_case_TimeOnlyTerm_ODE} is the same (\textit{i.e.} the directional derivatives are the same regardless of the time index), so let's use the notations $A$, $B$, $C$ and $D$ instead of $A^{[N]}$, $B^{[N]}$, $C^{[N]}$ and $D^{[N]}$.

The state-space version (application of \eqref{eq:SSR_on_a_step}) of \eqref{eq:test_case_TimeOnlyTerm_ODE} around a time $\tN$ is

\begin{equation}
\label{eq:test_case_TimeOnlyTerm_SSRized_ODE}
\left\{
	\begin{array}{lcl}
		\dspfrac{dx}{dt} & = & 0 \\
		y_L & = & x
	\end{array}
\right.
\end{equation}

\noindent because the directional derivatives are

\begin{equation}
\label{eq:test_case_TimeOnlyTerm_ABCD}
\begin{array}{lll}
	A & = 0 & \in M_{1, 1}(\mathbb{R}) \\
	B & = 0 & \in M_{1, 1}(\mathbb{R}) \\
	C & = 1 & \in M_{1, 1}(\mathbb{R}) \\
	D & = 0 & \in M_{1, 1}(\mathbb{R}) \\
\end{array}
\end{equation}

The linearized version (application of \eqref{eq:Lin_on_a_step_alter}) of \eqref{eq:test_case_TimeOnlyTerm_ODE} around a time $\tN$ is

\begin{equation}
\label{eq:test_case_TimeOnlyTerm_linearized_ODE}
\left\{
	\begin{array}{lcl}
		\dspfrac{dx}{dt} & = & a(\tN) \\
		y_L & = & x
	\end{array}
\right.
\end{equation}

We notice that the state-space system \eqref{eq:test_case_TimeOnlyTerm_SSRized_ODE} has a significant lack of information compared to \eqref{eq:test_case_TimeOnlyTerm_ODE}, due to the terms that are transparent in the linearization regarding the directional derivatives. On its side, the linearized system \eqref{eq:test_case_TimeOnlyTerm_linearized_ODE} also has a lack of information (dynamical behavior of $a$ is hidden and $b$ completely disappeared).

Let's detail the COSTARICA estimator on a general co-simulation step $[t^{[N]}, t^{[N+1]}[$ to see the consequences of this information loss.

First, we have:

\begin{equation}
\label{eq:test_case_TimeOnlyTerm_GP}
\left\{
	\begin{array}{lccl}
		G(s) & = & C(sI-A)^{-1}B+D & = s^{-1} \cdot 0 + 0 = 0 \\
		P(s) & = & C(sI-A)^{-1} & = s^{-1} \\
		R(s) & = & C(sI-A)^{-1}/s & = s^{-2}
	\end{array}
\right.
\end{equation}

\noindent so, regarding the inverse Laplace matrices, we have:

\begin{equation}
\label{eq:test_case_TimeOnlyTerm_GcalPcal}
\left\{
	\renewcommand{\arraystretch}{2.0}
	\begin{array}{lccl}
		\mathcal{G}_V & = & \mathcal{L}^{-1}(G\otimes\bar{U}^T)(\dt^{[N]}) & = 0 \\
		\mathcal{P}_V & = & \mathcal{L}^{-1}(P)(\dt^{[N]}) & = 1 \\
		\mathcal{R}_V & = & \mathcal{L}^{-1}(R)(\dt^{[N]}) & = \dt^{[N]} \\
		\mathcal{G}_D & = & \dspfrac{d(\check{t}\mapsto 0)}{dt}(\dt^{[N]}) & = 0 \\
		\mathcal{P}_D & = & \dspfrac{dH}{dt}(\dt^{[N]}) & = 0 \\
		\mathcal{R}_D & = & \dspfrac{d(\check{t}\mapsto \check{t}H(\check{t}))}{dt}(\dt^{[N]}) & = 1 \\
	\end{array}
	\renewcommand{\arraystretch}{1.0}
\right.
\end{equation}

\noindent where $H$ denotes the Heaviside function, and where the size of $\mathcal{G}_V$ and $\mathcal{G}_D$ is $1\times 1\times n$ where $n$ does not matter as it is the maximum polynomial degree of an unused input. Whatever the value of $n$ is, $\mathcal{G}_V$ and $\mathcal{G}_D$ are filled with zeros.

Finally, the expression of the linear parts of the estimators for the output value and derivative are:

\begin{equation}
\label{eq:test_case_TimeOnlyTerm_yL_ydotL}
\forall m\in [\![0, \mmax(N)]\!],\
\left\{
	\renewcommand{\arraystretch}{1.5}
	\begin{array}{lcl}
		\m \hat{y}_L^{[N+1]} & = & \tilde{x}^{[N]} + \dt^{[N]}\ a(\tN) \\
		\m \hat{\dot{y}}_L^{[N+1]} & = & a(\tN) \\
	\end{array}
	\renewcommand{\arraystretch}{1.0}
\right.
\end{equation}

In other words, the linear part of the outputs estimator exactly acts as a first-order hold estimator (we would have obtained a zero-order hold with a state-space representation \eqref{eq:test_case_TimeOnlyTerm_SSRized_ODE} instead of \eqref{eq:test_case_TimeOnlyTerm_linearized_ODE}, for instance if we wouldn't have had access to the state derivatives).

Regarding the control part of the estimator of the output, if a zero-order hold is used (see \eqref{eq:control_part_ZOH_val}), the estimator is:

\begin{equation}
\label{eq:test_case_TimeOnlyTerm_yC}
\begin{array}{lcl}
	\hat{y}_C^{[N+1]}
	& = & \tilde{y}_C^{[N]} \\
	& = & \tilde{y}^{[N]} - \left(C \tilde{x}^{[N]} + D\ {}^{[\mmax(N-1)]} \tilde{u}^{[N-1]}(t^{[N]})\right) \\
	& = & \tilde{y}^{[N]} - \tilde{x}^{[N]} \\
	& = & \tilde{x}^{[N]} + b(t^{[N]}) - \tilde{x}^{[N]} \\
	& = & b(t^{[N]}) \\
\end{array}
\end{equation}

Finally, summing the terms (see \eqref{eq:estimators_terms}), the COSTARICA estimator for the output is:

\begin{equation}
\label{eq:test_case_TimeOnlyTerm_ynp}
\begin{array}{lcl}
	\m \hat{y}^{[N+1]}
	& = & \m \hat{y}_L^{[N+1]} + \hat{y}_C^{[N+1]} \\
	& = & \tilde{x}^{[N]} + \dt^{[N]}\ a(\tN) + b(t^{[N]}) \\
\end{array}
\end{equation}

\noindent which exactly corresponds to a first-order hold estimation (recall: in \eqref{eq:test_case_TimeOnlyTerm_ODE}, we have: $y=x+b(t)$) where the slope is only drived by the slope of the state. Using a higher order estimator for the control part would have catched a slope made of one contribution by the state's slope, and one given by the first-order approximation of $b$ function's slope over the co-simulation step.

Regarding the estimation of the derivative of the output, either the control term $\hat{\dot{y}}_C^{[N+1]}$ will be added to the linear term $\m\hat{\dot{y}}_L^{[N+1]} = a(\tN)$ (see \eqref{eq:test_case_TimeOnlyTerm_yL_ydotL}), or, in case the state-spare representation was used only, the control term will be the only one that matters (as $\m \hat{\dot{y}}_L^{[N+1]}$ would be $0$ in that case). In the first case (normal case, all required capabilities in table \ref{table:capabilities_related_to_COSTARICA} are satisfied and thus a real linearization can be used instead of a state-space approximation only, the estimator of the derivative of the output will be able to take into account both the $a(\tN)$ part (instantaneous derivative of the state), and the derivative of the $b(t)$ part in case the estimation is done with a few past values $\tilde{y}_C^{[N]}$, $\tilde{y}_C^{[N-1]}$, ...

This tough case shows that, even when the system is different from its linearization due to lost information (difference between \eqref{eq:test_case_TimeOnlyTerm_ODE} and \eqref{eq:test_case_TimeOnlyTerm_linearized_ODE}), the COSTARICA estimator is not worse than a first-order hold estimation (as it reproduces it). Even on a state-space representation, where the lack of information is even more significant (difference between \eqref{eq:test_case_TimeOnlyTerm_ODE} and \eqref{eq:test_case_TimeOnlyTerm_SSRized_ODE}), the COSTARICA estimator is not worse than a zero-order hold estimation (as it reproduces it).

Please note that, while on the one hand the main drawback of such estimations is that the inputs are not taken into account ($m$ does not appear in final expression of \eqref{eq:test_case_TimeOnlyTerm_ynp}), on the other hand this example has no inputs (or an unused one, equivalently) so it does not matter in this particular case. If the system had inputs (and used them), the $B$ and $D$ matrices wouldn't have been null, and thus, the estimator would have taken them into account.

\subsection{Mechanical bodies with behavior change}
\label{subsection:mechanical_bodies_with_behavior_change}

This model is a uni-dimensional linear mechanical model. Two bodies (inertias) of $10\ 000$ kg are interconnected to one another and to zero-speed points (walls) with springs and dampers. After the instantaneous transition time $t=100$ s, the body on the right vanishes and the corresponding system spontaneously acts as a constant pulling force of $1000$ N. A representation of this model, introduced in \cite{Eguillon2022F3ornits}, is presented in figure \ref{fig:test_case_Meca_sketch_monolithic_scheme}. The corresponding sketch in Simcenter Amesim is shown in figure \ref{fig:test_case_Meca_sketch_monolithic_Amesim}.

\begin{center}
$
\begin{small}
\begin{array}{c|lcl|}
	\cline{2-4}
	\multirow{13}{*}
	{
		\includegraphics[scale=0.87]{\figuresdir/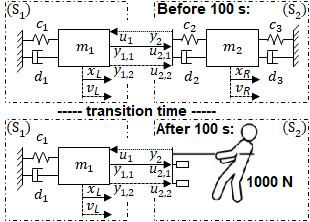}
	}
	& c_1 & = &   1\ \text{kN/m} \\
	&
	c_2 & = &   1\ \text{kN/m} \\
	&
	c_3 & = &   1\ \text{kN/m} \\
	&
	d_1 & = &   1\ \text{kN/(m/s)} \\
	&
	d_2 & = &   0\ \text{kN/(m/s)} \\
	&
	d_3 & = &   1\ \text{kN/(m/s)} \\
	&
	m_1 & = & 10000\ \text{kg} \\
	&
	m_2 & = & 10000\ \text{kg} \\
	&
	x_1(\tinit) & = & -1\ \text{m} \\
	&
	x_2(\tinit) & = & 0\ \text{m} \\
	&
	v_1(\tinit) & = & 0\ \text{m/s} \\
	&
	v_2(\tinit) & = & 0\ \text{m/s} \\
	&
	\multicolumn{3}{|l|}
	{
		[\tinit, \tend] = [0\ \text{s},\  200\ \text{s}]
	} \\
	\cline{2-4}
\end{array}
\end{small}
$
\vspace{-2mm}
\captionof{figure}{Mechanical bodies with behavior change: linear uni-dimensional co-simulation test case}
\label{fig:test_case_Meca_sketch_monolithic_scheme}
\end{center}

\begin{center}
\includegraphics[scale=0.23]{\figuresdir/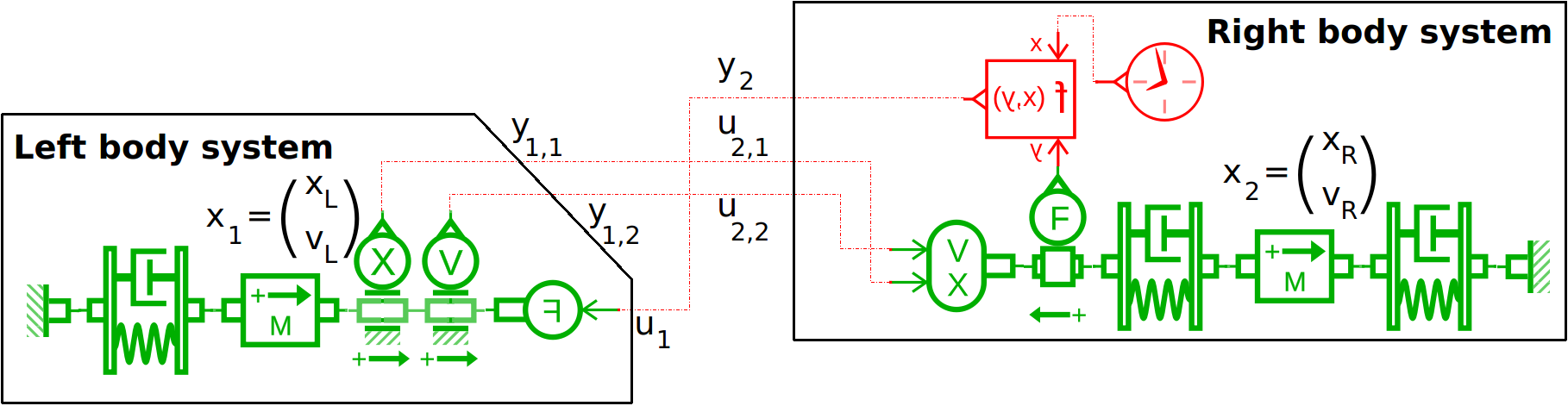}
\vspace{-2mm}
\captionof{figure}{Mechanical bodies with behavior change in Simcenter Amesim}
\label{fig:test_case_Meca_sketch_monolithic_Amesim}
\end{center}

Let's detail the equations of the left body system and denote it by $(S_1)$. The left body system for co-simulation, as represented in figure \ref{fig:test_case_Meca_sketch_Left_body_Amesim}, has $\nin=1$ input (the force coming from the right), $\nout=2$ outputs (the displacement and velocity of the left body), and $\nout=2$ state variables (the position and speed of the body, modelled as a mechanical inertia).

\begin{center}
\includegraphics[scale=0.2]{\figuresdir/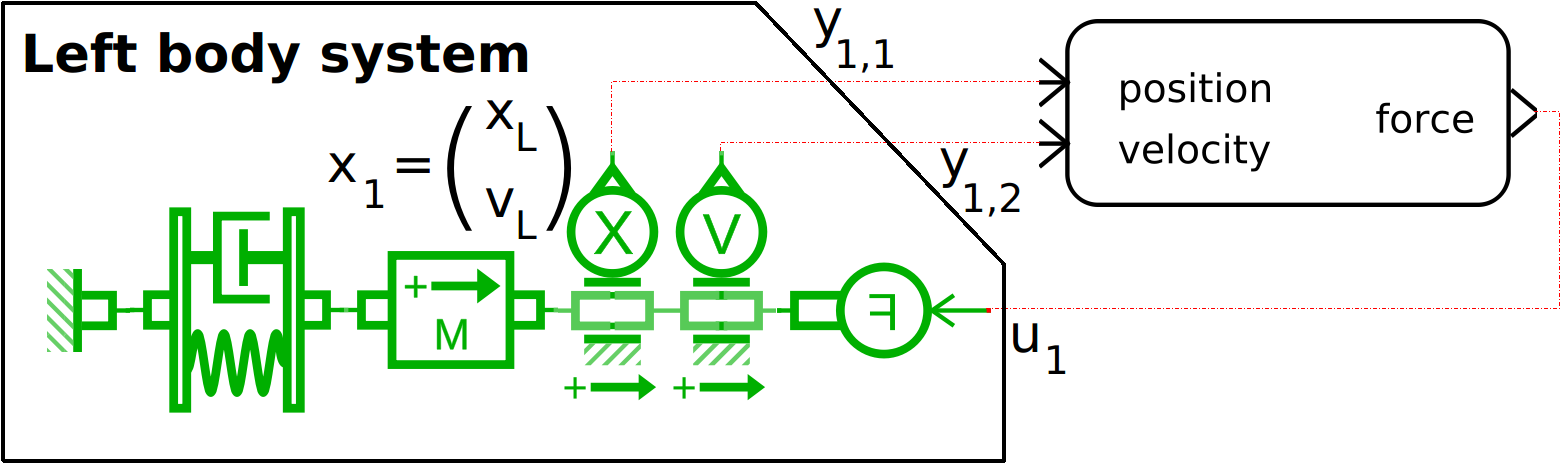}
\vspace{-2mm}
\captionof{figure}{Left body system $(S_1)$ in Simcenter Amesim}
\label{fig:test_case_Meca_sketch_Left_body_Amesim}
\end{center}

The state variables are denoted by $x_1=(x_L, v_L)^T$, the input force by $u_1$ (input) and the outputs by $y_{1, 1}$ and $y_{1, 2}$ respectively. The spring of rate $c_1$ and the damper of rate $d_1$ (see figure \ref{fig:test_case_Meca_sketch_monolithic_scheme}) in the left of the body generate forces $-c_1 x_L$ and $-d_1 v_L$ respectively (measured positively from left to right). Second Newton's law gives: $m_1\dot{v}_L = -c_1 x_L - d_1 v_L - u_1$ as the force at the interface is measured positively from right to left. Finally, the equations of system $(S_1)$ are given in \eqref{eq:test_case_Meca_S1_ODE}.

\begin{equation}
\label{eq:test_case_Meca_S1_ODE}
(S_1):\
\left\{
\begin{array}{ccccccc}
	\left(
		\begin{array}{c}
			\dot{x}_L \\
			\dot{v}_R \\
		\end{array}
	\right)
	&
	=
	&
	\left(
		\begin{array}{cc}
			0 & 1 \\
			\frac{-c_1}{m_1} & \frac{-d_1}{m_1} \\
		\end{array}
	\right)
	&
	\left(
		\begin{array}{c}
			x_L \\
			v_L \\
		\end{array}
	\right)
	&
	+
	&
	\left(
		\begin{array}{c}
			0 \\
			\frac{-1}{m_1} \\
		\end{array}
	\right)
	&
	u_1
	\\ \\
	\left(
		\begin{array}{c}
			y_{1, 1} \\
			y_{1, 2} \\
		\end{array}
	\right)
	&
	=
	&
	\left(
		\begin{array}{P{4mm}P{4mm}}
			1 & 0 \\
			0 & 1 \\
		\end{array}
	\right)
	&
	\left(
		\begin{array}{c}
			x_L \\
			v_L \\
		\end{array}
	\right)
	&
	+
	&
	\left(
		\begin{array}{c}
			0 \\
			0 \\
		\end{array}
	\right)
	&
	u_1
\end{array}
\right.
\end{equation}

Let's detail the equations of the right body system and denote it by $(S_2)$. It has $\nin=2$ inputs (the position and the velocity of the body in $(S_1)$), $\nout=1$ output (the force at the left of the spring-damper component on the left of figure \ref{fig:test_case_Meca_sketch_Left_body_Amesim}) and $\nin=2$ state variables (the position and speed of the body). As shown on the figure \ref{fig:test_case_Meca_sketch_Right_body_Amesim}, the force interface variable produced by $(S_2)$ (output of it) and used by $(S_1)$ (input of it) is measured positively from right to left. The other quantities (forces of the springs and dampers around the body, position and velocity of the body) will be measured positively from left to right.

\begin{center}
\includegraphics[scale=0.2]{\figuresdir/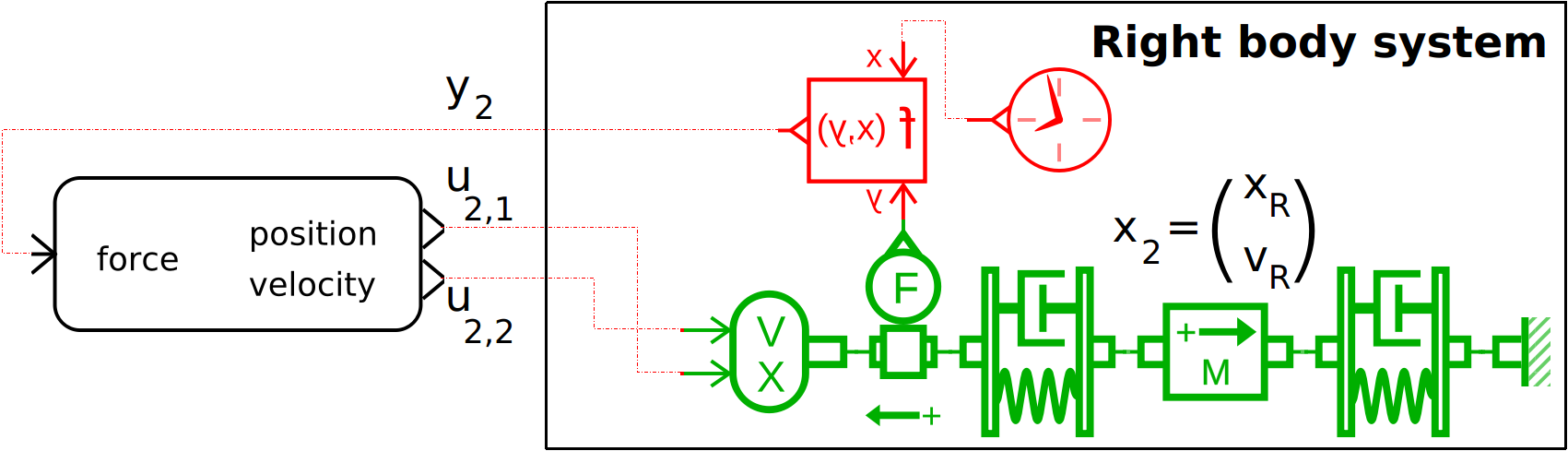}
\captionof{figure}{Right body system $(S_2)$ in Simcenter Amesim}
\label{fig:test_case_Meca_sketch_Right_body_Amesim}
\end{center}

The body has a mass $m_2$ and is modelled as a mechanical inertia which position and speed are the state variables denoted by $x_2=(x_R, v_R)^T$. It is submitted to 4 forces (measured positively from left to right): the spring of rate $c_2$, the damper with a damper rating of $d_2$, the spring of rate $c_3$ and the damper with a damper rating of $d_3$ (see figure \ref{fig:test_case_Meca_sketch_monolithic_scheme}). Let's focus on this situation which corresponds to the behavior of $(S_2)$ before the transition time $t=100$ s.

The forces coming from the left of the body in $(S_2)$ come from a spring and a damper and are respectively $-c_2(x_R-u_{2,1})$ and $-d_2(v_R-u_{2,2})$. They indeed depend on the left body's position (input $u_{2, 1}$) and the left body's velocity (input $u_{2, 2}$). The forces coming from the right of the right body also come a spring and a damper, and they are respectively $-c_3x_R$ and $-d_3v_R$. Second Newton's law gives: $m_2\dot{v}_R = -c_2(x_R-u_{2,1})-d_2(v_R-u_{2,2})-c_3x_R-d_3v_R$.

\newpage 

Finally, the equations of system $(S_2)$ are:

\vspace{-2mm} 

\begin{equation}
\label{eq:test_case_Meca_S2_ODE}
\begin{small}
(S_2):\
\left\{
\begin{array}{ccccccl}
	\left(
		\begin{array}{c}
			\dot{x}_R \\
			\dot{v}_R \\
		\end{array}
	\right)
	&
	=
	&
	\left(
		\begin{array}{cc}
			0 & 1 \\
			\frac{-c_2-c_3}{m_2} & \frac{-d_2-d_3}{m_2} \\
		\end{array}
	\right)
	&
	\left(
		\begin{array}{c}
			x_R \\
			v_R \\
		\end{array}
	\right)
	&
	+
	&
	\left(
		\begin{array}{cc}
			0 & 0 \\
			\frac{c_2}{m_2} & \frac{d_2}{m_2} \\
		\end{array}
	\right)
	&
	\left(
		\begin{array}{c}
			u_{2, 1} \\
			u_{2, 2} \\
		\end{array}
	\right)
	\\ \\
	\multirow{2}{*}{$y_2$}
	&
	\multirow{2}{*}{$=$\Bigg\{}
	&
	\left(
		\begin{array}{cc}
			-c_2\ &\ -d_2 \\
		\end{array}
	\right)
	&
	\left(
		\begin{array}{c}
			x_R \\
			v_R \\
		\end{array}
	\right)
	&
	+
	&
	\left(
		\begin{array}{cc}
			c_2\ &\ d_2 \\
		\end{array}
	\right)
	&
	\left(
		\begin{array}{c}
			u_{2, 1} \\
			u_{2, 2} \\
		\end{array}
	\right)
	\hspace{3mm} \text{if}\ t\in[0, 100[
	\\
	&
	&
	\multicolumn{5}{l}
	{
		\hspace{5mm} -1000 \hspace{3mm} \text{otherwise}
	}
\end{array}
\right.
\end{small}
\end{equation}

As explained in the remark of the introduction of section \ref{section:examples_and_test_cases}, co-simulations with IFOSMONDI-JFM using rollback and with replacement of the rollback by the COSTARICA process have been run and compared to the reference monolithic system presented in figure \ref{fig:test_case_Meca_sketch_monolithic_Amesim}.

A fixed-step version of the IFOSMONDI-JFM method has been used, so that errors can be computed for various values of the co-simulation step size. The following results use the "Anderson" version of IFOSMONDI-JFM, with an epsilon of $1\cdot 10^{-6}$ (see \cite{Eguillon2021IfosmondiJFM}).

\vspace{-2mm} 

\begin{center}
\includegraphics[scale=0.58]{\figuresdir/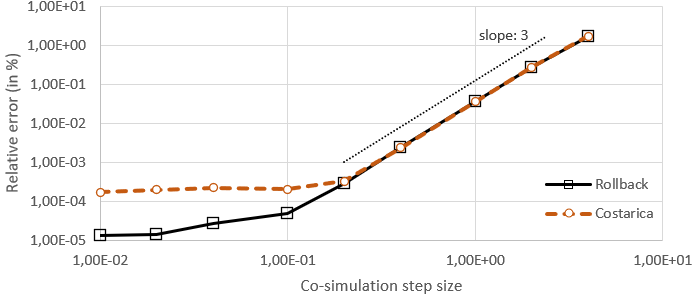}
\vspace{-3mm} 
\captionof{figure}{Comparison of the convergence graphs of IFOSMONDI-JFM method on the mechanical test case depending on the way to iterate on the co-simulation steps}
\label{fig:test_case_Meca_results}
\end{center}

Figure \ref{fig:test_case_Meca_results} shows that both the rollback and the COSTARICA process make IFOSMONDI-JFM method reach an order $3$ of convergence on the characteristic variable of the modular model: $x_R$. Regardless of the numerical effects for small values of the co-simulation step size, the estimators involved in the COSTARICA process are very accurate which enables the IFSOMONDI-JFM co-simulation method to be unaffected by the rollback avoidance. Indeed, both systems $(S_1)$ and $(S_2)$ (see \eqref{eq:test_case_Meca_S1_ODE} and \eqref{eq:test_case_Meca_S2_ODE}) are linear. This makes their linearizations \eqref{eq:Lin_on_a_step} exact (and, in this particular case, their state-space representation \eqref{eq:SSR_on_a_step} do represent them exactely as well).

The left body's displacement (position) is indeed very close to the monolithic reference in both cases (rollback and COSTARICA), as shown on figure \ref{fig:test_case_Meca_curves_dt=2e-1}.

\vspace{-2mm} 

\begin{center}
\includegraphics[scale=0.41]{\figuresdir/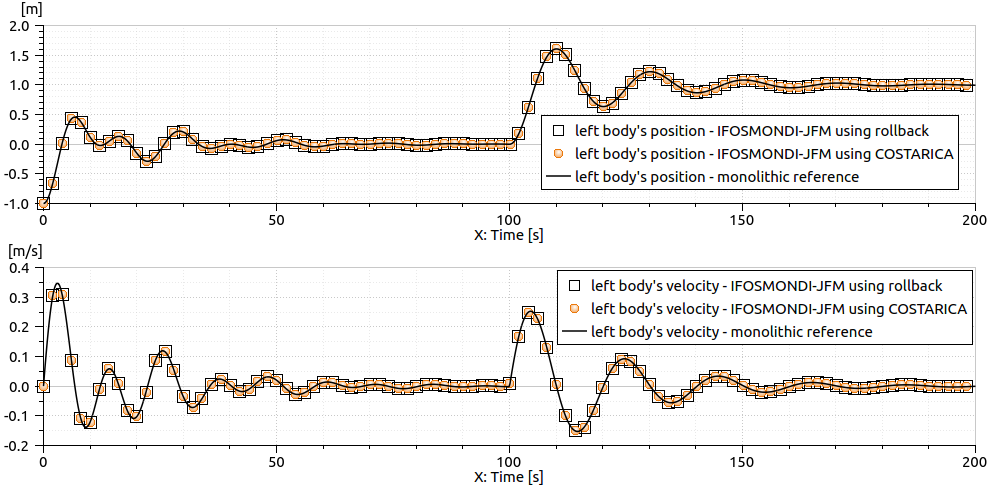}
\vspace{-3mm} 
\captionof{figure}{Left body's motion comparison depending on the (co-)simulation method: co-simulation with IFOSMONDI-JFM using rollback (co-simulation step size: $2\cdot 10^{-1}$), co-simulation with IFOMSONDI-JFM replacing the need for rollback by the COSTARICA process (co-simulation step size: $2\cdot 10^{-1}$) and monolithic simulation (reference)}
\label{fig:test_case_Meca_curves_dt=2e-1}
\end{center}

Therefore, co-simulations on this test case with an iterative co-simulation method could have been performed even if the systems were not capable of rollback, thanks to the COSTARICA process, and without loss of accuracy.

\subsection{Lotka-Volterra predation (non-linear)}
\label{subsection:lotka_volterra_predation}

This model is a decoupled version of the classical two species Lotka-Volterra predation equations \cite{Volterra1928} also presented in \ref{paragraph:Classical_Lotka_Volterra}. Despite it has been described in this paragraph, the full model (including equations of both parts) is presented here as we aim at conducting a full co-simulation in this subsection (and not only a single macro-step on only one system, as in \ref{subsection:Corroborative_empirical_observations}). The monolithic system is shown on figure \ref{fig:test_case_LV_sketch_monolithic} and the two systems of the corresponding modular model for co-simulation are presented on figures \ref{fig:test_case_LV_sketch_prey} and \ref{fig:test_case_LV_sketch_predator}.

\vspace{-2mm} 

\begin{center}
$
\begin{small}
\begin{array}{c|lcl|}
	\cline{2-4}
	\multirow{7}{*}
	{
		\includegraphics[scale=0.28]{\figuresdir/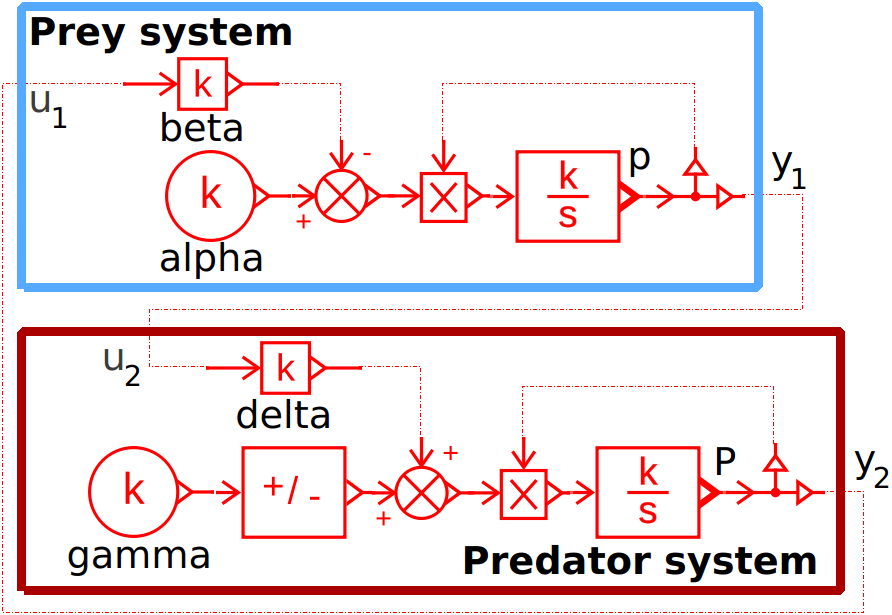}
	}
	&
	\alpha & = & 0.67 \\
	&
	\beta & = & 1.33 \\
	&
	\gamma & = &   1 \\
	&
	\delta & = &   1 \\
	&
	p(\tinit) & = & 1 \\
	&
	P(\tinit) & = & 1 \\
	&
	\multicolumn{3}{|l|}
	{
		[\tinit, \tend] = [0\ \text{s},\  20\ \text{s}]
	} \\
	\cline{2-4}
	\multicolumn{4}{c}{\vspace{16mm}}
\end{array}
\end{small}
$
\vspace{-4mm} 
\captionof{figure}{Lotka-Volterra predation model in Simcenter Amesim}
\label{fig:test_case_LV_sketch_monolithic}
\end{center}

\vspace{-2mm} 

System $(S_1)$ represents the prey. A sketch of it is presented on figure \ref{fig:test_case_LV_sketch_prey}. The population of prey, denoted by $p$, is the single state variable of $(S_1)$. It is also the output. The single input to this system is the population of predator. The $\alpha$ and $\beta$ parameters are the natural birth rate and the rate of predation upon the prey respectively. The (non-linear) equations of $(S_1)$ are given in \eqref{eq:test_case_LV_S1_ODE}.

\vspace{-2mm} 

\begin{center}
\includegraphics[scale=0.26]{\figuresdir/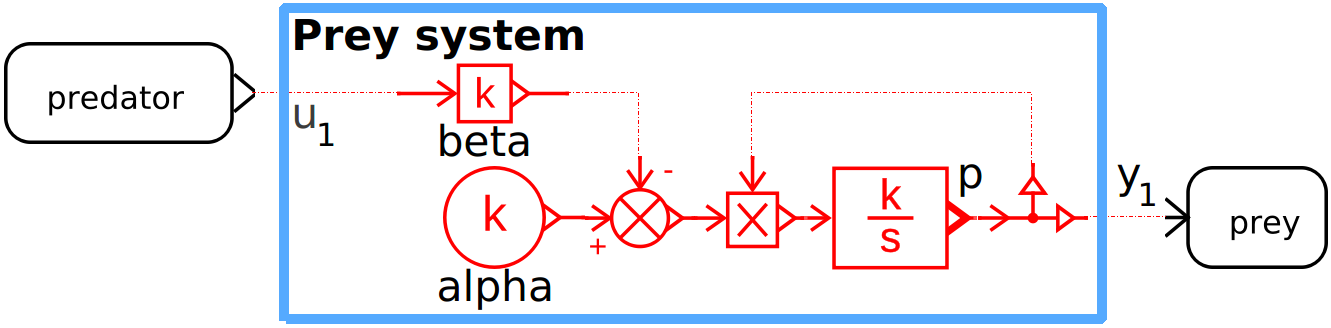}
\vspace{-4mm} 
\captionof{figure}{Prey system $(S_1)$ in Simcenter Amesim}
\label{fig:test_case_LV_sketch_prey}
\end{center}

\vspace{-4mm} 

\begin{equation}
\label{eq:test_case_LV_S1_ODE}
(S_1):\
\left\{
\begin{array}{ccc}
	\dot{p}
	&
	=
	&
	p(\alpha - \beta u_1)
	\\
	y_1
	&
	=
	&
	1\ p + 0\ u_1
\end{array}
\right.
\end{equation}

System $(S_2)$ represents the predator. A sketch of it is presented on figure \ref{fig:test_case_LV_sketch_predator}. The population of predator, denoted by $P$, is the single state variable of $(S_2)$. It is also the output. The single input to this system is the population of prey. The $\gamma$ and $\delta$ parameters are the natural death rate and the growth rate upon predator (due to predation) respectively. The (non-linear) equations of $(S_2)$ are given in \eqref{eq:test_case_LV_S2_ODE}.

\vspace{-2mm} 

\begin{center}
\includegraphics[scale=0.24]{\figuresdir/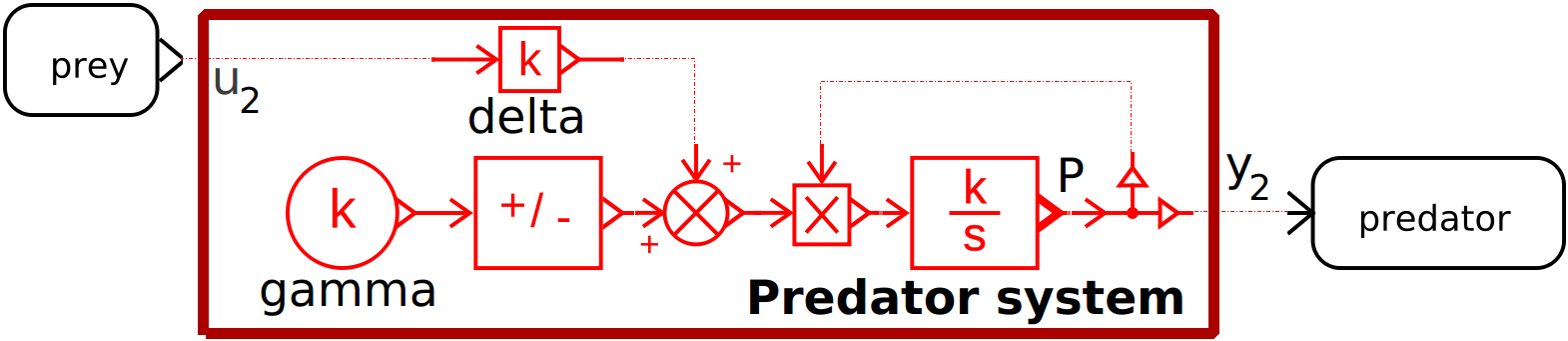}
\vspace{-4mm} 
\captionof{figure}{Predator system $(S_2)$ in Simcenter Amesim}
\label{fig:test_case_LV_sketch_predator}
\end{center}

\vspace{-4mm} 

\begin{equation}
\label{eq:test_case_LV_S2_ODE}
(S_2):\
\left\{
\begin{array}{ccc}
	\dot{P}
	&
	=
	&
	P(\delta u_2 - \gamma)
	\\
	y_2
	&
	=
	&
	1\ P + 0\ u_2
\end{array}
\right.
\end{equation}

As for the previous test case, IFOSMONDI-JFM has been used as co-simulation method to compare the COSTARICA process to the rollback usage. The reference results are the ones obtained from the simulation of the monolithic system of figure \ref{fig:test_case_LV_sketch_monolithic}.

A fixed-step version of the IFOSMONDI-JFM method has been used, as well as the "Anderson" version of the algorithm. The epsilon parameter (see \cite{Eguillon2021IfosmondiJFM}) has been set to $1\cdot 10^{-6}$.

Figure \ref{fig:test_case_LV_results} shows that the COSTARICA process injected in IFOSMONDI-JFM does not achieve the same accuracy than the real usage of the rollback. The fact that the estimators used in COSTARICA are based on the systems' linearizations \eqref{eq:Lin_on_a_step} explains the higher error than the cases where the successive iterations of IFOSMONDI-JFM are done with the real systems' integrations. The COSTARICA on the state-space representation (called "COSTARICA on SSR only" on the figure) is what is obtained when the systems are not able to provide the time-derivatives of the state variables (assumption \eqref{eq:SSR_erroneous_version} thus has to be made, and state-space representation \eqref{eq:SSR_on_a_step} has to be used instead of full linearization \ref{eq:Lin_on_a_step}).

\begin{center}
\includegraphics[scale=0.34]{\figuresdir/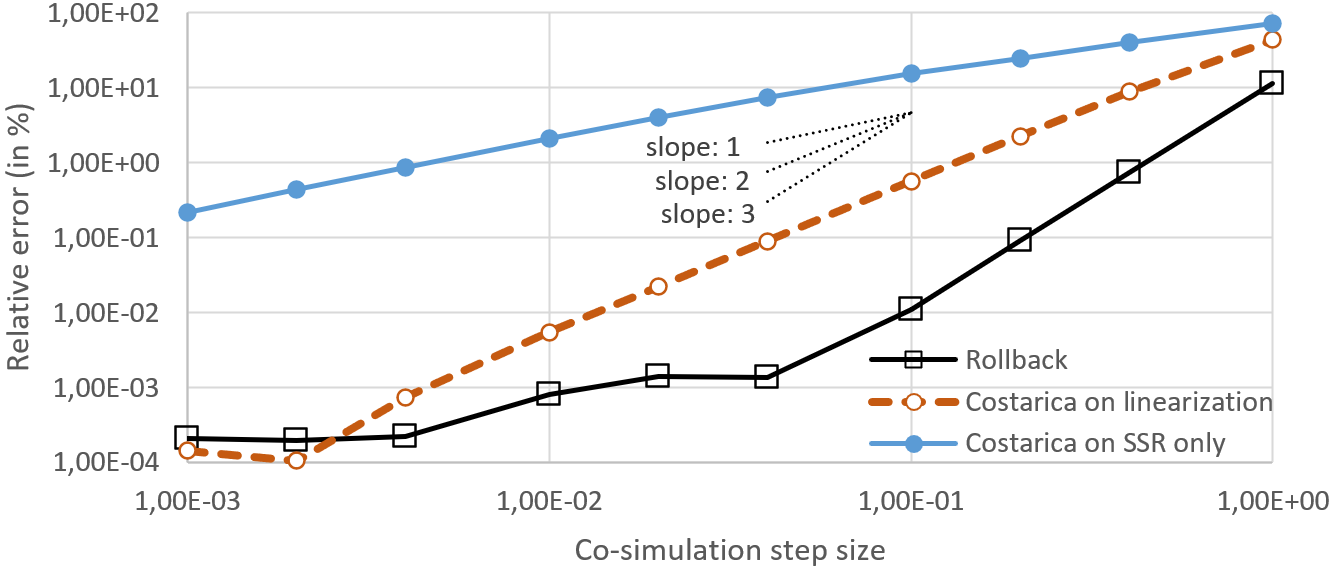}
\captionof{figure}{Comparison of the convergence graphs of IFOSMONDI-JFM method on the Lotka-Volterra test case depending on the way to iterate on the co-simulation steps}
\label{fig:test_case_LV_results}
\end{center}

Nevertheless, a method of order two (measured) is obtained, which allows any desired precision to be reached by simply refining the co-simulation step size. In case the systems are not capable of rollback, this example shows that the IFOSMONDI-JFM method (or any other iterative co-simulation method \cite{Eguillon2019Ifosmondi} \cite{Schweizer2016} \cite{Viot2018} \cite{Kraft2021} ...) can be used thanks to COSTARICA.

Regarding the results (proportion of prey and proportion of predator), figure \ref{fig:test_case_LV_curves_dt=1e-2} shows that the results are satisfactory even on the co-simulation replacing the rollback usage by the COSTARICA estimators. The differences between the different co-simulations can be observed more easily on the zoom on figure \ref{fig:test_case_LV_curves_dt=1e-2_zoom}.

\begin{center}
\includegraphics[scale=0.45]{\figuresdir/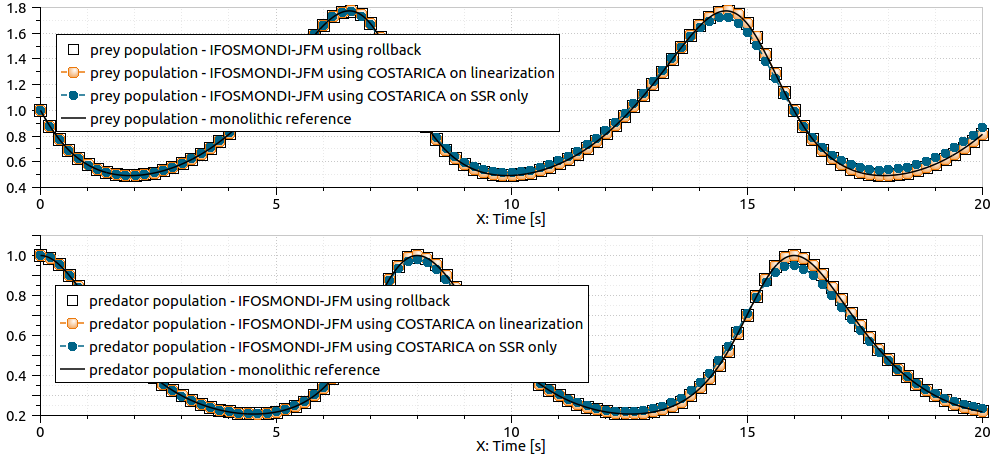}
\captionof{figure}{Prey and predator proportions depending on the (co-)simulation method: co-simulation with IFOSMONDI-JFM with a co-simulation step size of $10^{-2}$ with different methods to replay a co-simulation step, and monolithic simulation (reference)}
\label{fig:test_case_LV_curves_dt=1e-2}
\end{center}

\begin{center}
\includegraphics[scale=0.45]{\figuresdir/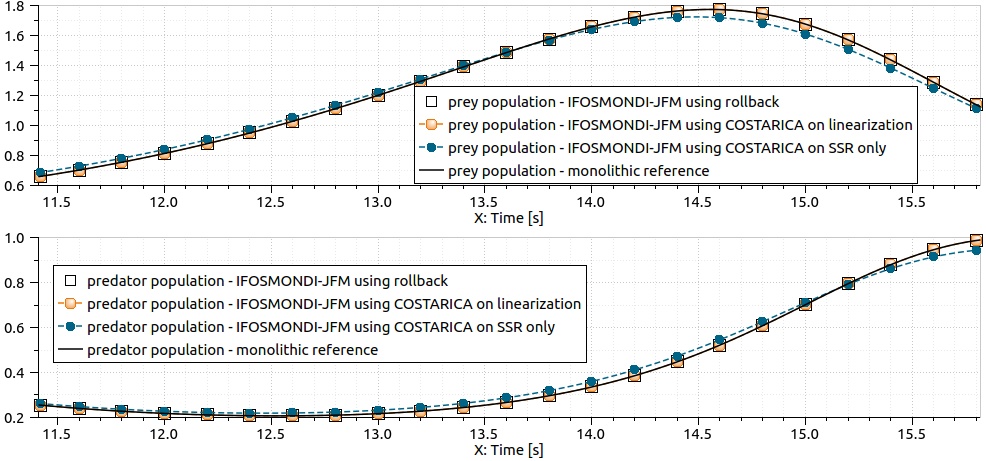}
\captionof{figure}{Zoom on the curves of figure \ref{fig:test_case_LV_curves_dt=1e-2} on $[11.5, 15.5]$}
\label{fig:test_case_LV_curves_dt=1e-2_zoom}
\end{center}

\section{Conclusion}
\label{section:conclusion}

The introduced COSTARICA process enables to use iterative co-simulation methods on modular models made of interconnected systems even in the case where not every of these systems is capable of rollback. Through the use of estimators of the local behavior of the systems, the iterative part of the co-simulation methods \textit{prepares} the final integration on every co-simulation step.

These estimators are based on the local linearizations of the non-rollback-capable systems at the lastly reached time, which might be more or less accurate depending on each system' nature. In case a system is linear or nearly linear, the estimators of COSTARICA can reach a precision similar to the one that would have been obtained by a real integration: in that case, the rollback replacement by the COSTARICA process does not (or slightly) affect the accuracy of the co-simulation. Otherwise, in case a system is non-linear, the estimators of COSTARICA only generate approximations of the behavior of the system, and the co-simulation step should not be too large in order to prevent this approximation from causing an unreasonable error. It is anyway worth it to use this process on non-rollback-capable systems as, in the worst case, the first-order hold behavior behavior is recreated (or zero-order hold in case the states time-derivatives cannot be retrieved), and in the better cases extra information about the behaviors of the systems will enable the iterative co-simulation method to compute a solution on each step that couldn't have been obtained due to the rollback requirement.

The analysis shows that, in the general (non-linear) case, the first-order dependence of the evaluation function of the systems ($f$ function) to the time variable influences the convergence order of the leading COSTARICA estimator.

Enhancement will be tacked in further work of the authors, such as the use of the final guess of the last iteration of the co-simulation method with the COSTARICA process in a comparison involving the result of the genuine integration. Such an error estimation could be used in a co-simulation step size controlling strategy, as the whole paper was written without supposing that the co-simulation step size was constant. This can be possible using an known and expected error order, and the latter would result from an extension of the analysis of section \ref{section:Relevance_of_the_linearization_over_time} to the output equation (involving the $g$ function). A criterion estimating the urgency to reevaluate the matrices of the linearizations of the systems can also be developed to save numerical computations.

\section*{Declarations of interest}

Authors Yohan EGUILLON and Bruno LACABANNE are currently Siemens Digital Industries Software employees.

The patent "Advanced cosimulation scheduler for dynamic system simulation" is currently pending to Siemens and includes the COSTARICA process (computation of the estimators and injection in a co-simulation algorithm).

\section*{Acknowledgements}

The authors would like to thank Siemens Digital Industries Software for supporting this work, as well as Institut Camille Jordan and Universit\'e de Lyon for supervising this research.

\renewcommand{\lstlistingname}{Listing}

\bibliographystyle{splncs04}
\bibliography{EguillonLacabanneTromeurDervout_COSTARICA}

\end{document}